\documentclass[11pt]{article}
\usepackage{amssymb,latexsym, amsmath}
\newcommand{\bfi}{\bfseries\itshape}
 
\newcommand{\singlespace}{\baselineskip 4.2333mm \parskip 4.2333mm}

\makeatletter
\@addtoreset{figure}{section}
\def\thefigure{\thesection.\@arabic\c@figure}
\def\fps@figure{h, t}
\@addtoreset{table}{bsection}
\def\thetable{\thesection.\@arabic\c@table}
\def\fps@table{h, t}
\@addtoreset{equation}{section}

\makeatother
\def\circledS{\boldsymbol{\bigcirc\hspace{-.33cm}{s}}}

\def\boxeq#1{\\ \nopagebreak
\framebox{\rule[-#1\baselineskip]{\textwidth}{0pt}
\rule[-#1\baselineskip]{0pt}{#1.5\baselineskip}}
\vspace{-#1\baselineskip}\vspace{-1\baselineskip}
}

\begin{document}

\title{Fluctuation effects on 3D Lagrangian mean\\
and Eulerian mean fluid motion}
\author{Darryl D. Holm
\\Theoretical Division and Center for Nonlinear Studies
\\Los Alamos National
Laboratory, MS B284
\\ Los Alamos, NM 87545
\\ email: dholm@lanl.gov}
\date{To appear, {\it Physica D} }

\maketitle

\begin{abstract}
We formulate equations for the slow time dynamics of fluid motion that self
consistently account for the effects of the variability upon the mean. 
The time-average effects of the fluctuations introduce nonlinear dispersion
that acts to spatially smooth the transport velocity of the mean flow relative
to its circulation or momentum velocity, by the inversion of a Helmholtz operator
whose length scale corresponds to the covariance of the fluctuations.

\end{abstract}
\clearpage
\tableofcontents


\section{Introduction}\label{intro-sec}

We seek equations for the slow time dynamics of fluid motion that self
consistently account for the effects of the variability upon the mean. In
formulating such equations one must choose a suitable decomposition of the flow
into its rapid and slowly varying components and determine a strategy for
applying the corresponding averaging procedure. We consider Reynolds type
decompositions of either the Lagrangian fluid trajectory, or the Eulerian fluid
velocity. These decompositions lead respectively to either the Lagrangian mean
(the time average following a fluid parcel), or the Eulerian mean (the time
average at a fixed position). Our strategy in seeking self consistent slow time
dynamics is to apply these decompositions and their corresponding averaging
procedures to Hamilton's principle for an ideal incompressible fluid flow. The
resulting Lagrangian mean and Eulerian mean equations are obtained in both
cases by using the same Euler-Poincar\'e variational
framework~\cite{HMR[1998a]},~\cite{HMR[1998b]}. Hence, these equations possess
conservation laws for energy and momentum, as well as a Kelvin-Noether
circulation theorem that establishes how the mean properties of the fluctuations
affect the circulation of the mean flow. 

Thus, we present two formulations of the mean equations we seek: a Lagrangian
mean theory; and an Eulerian mean theory. It turns out these theories possess a
certain duality. In particular, the Eulerian mean velocity appears as the
momentum, or circulation velocity in the Lagrangian mean dynamics, and {\it vice
versa}. The effect of the averaging in either case is to make the advection or
transport velocity smoother than the momentum, or circulation velocity, via the
inversion of a Helmholtz operator that relates the two velocities. The length
scale that appears in this Helmholtz operator is the covariance of the
fluctuations, which has its own dynamics in each case. Thus, the temporal
averaging in Hamilton's principle leads to a dynamical spatial filtering in the
resulting equations of motion. This is the main point of the paper. The two
formulations we present here complement each other and provide a flexible
unified basis for further investigation and analysis of the effects of
fluctuations on mean fluid dynamics.

We begin by introducing new equations that describe the Lagrangian mean
effects of advected fluctuations in 3D incompressible fluid motion. The
results include a new second moment closure model for 3D fluid turbulence.
This model describes {\it dynamically self-consistent} interaction between a
Lagrangian mean flow and a distribution of advecting rapid (or random)
fluctuations described in slow time (or statistically) by their Lagrangian mean
covariance tensor. Its derivation combines the Euler-Poincar\'e theory of fluid
dynamics, the Taylor hypothesis for advection of the fluctuations and the
Reynolds decomposition of the Lagrangian fluid trajectory. We also consider the
effects of rotation and stratification in this model, for the sake of its
potential geophysical applications. To help develop intuition about the solution
behavior of these Lagrangian mean models we discuss 1D and 2D subcases, as well.
Then, we compare the Lagrangian mean models with their Eulerian mean
counterparts and emphasize the duality between them.


\subsection{Lagrangian mean equations}

This Section introduces a new self-consistent dynamical model that describes
the Lagrangian mean effects of advected fluctuations on 3D incompressible
stratified fluid motion in a rotating frame. This model is based on the
Lagrangian fluid description of two standard assumptions: (1) Reynolds
decomposition of the Lagrangian fluid trajectory; and (2) the Taylor hypothesis,
that rapid fluctuations are advected by the mean flow. We substitute these two
assumptions into Hamilton's principle for the Euler equations of an ideal
incompressible fluid, and apply the Lagrangian mean before taking variations.
Dissipation is then introduced in the traditional semi-empirical fashion. The
results include a new second moment closure model for 3D fluid turbulence. The
latter is a development of the one point closure model of Chen et al.~\cite{Chen
etal[1998a]}--~\cite{Chen etal[1998c]}, based on the
viscous Camassa-Holm equations (VCHE), also known as the Navier-Stokes alpha
model.  The {\bfi Lagrangian mean Euler-Boussinesq (LMEB) model} for a
stratified incompressible fluid in a rotating reference frame is given by the
equations,
\smallskip\boxeq{6}
\begin{eqnarray}\label{LMEB-model-intro}
&&\frac{d}{dt}\mathbf{v} 
- \mathbf{u}\times{\rm curl}\,\mathbf{R}(\mathbf{x})
+ \boldsymbol\nabla p + gb\,\mathbf{\hat{z}}
= \nu\,\tilde\Delta\mathbf{v}
\,,\quad\hbox{with}\quad
\boldsymbol{\nabla\cdot\,}\mathbf{u} = 0 \,,
\\
&& 
\hbox{where  }
\frac{d}{dt} \equiv \Big(\frac{\partial}{\partial t} 
+ \mathbf{u}\boldsymbol{\,\cdot\nabla}\Big),
\quad
\mathbf{v} \equiv (1-\tilde\Delta)\mathbf{u}\,,
\quad
\tilde\Delta \equiv
\boldsymbol{\nabla\cdot\langle\xi\xi\rangle\cdot\nabla}\,,
\nonumber\\
&&\hspace{-.2in} \hbox{and}\quad
\frac{db}{dt}
= (\boldsymbol{\nabla\cdot\kappa}_S\,
\boldsymbol{\cdot\nabla})\,b
\quad\hbox{with}\quad
2\boldsymbol{\kappa}_S 
\equiv
\frac{d}{dt}\boldsymbol{\langle\xi\xi\rangle}
= \boldsymbol{\langle\xi\xi\rangle\cdot\nabla\mathbf{u}}
+ \boldsymbol{\nabla\mathbf{u}}^{\rm T}
\boldsymbol{\cdot\langle\xi\xi\rangle}.
\qquad\label{LMEB-model-defs-intro}
\end{eqnarray}
\smallskip
These LMEB equations {\it include} the standard Euler-Boussinesq (EB) equations 
as an invariant subsystem, for which $\boldsymbol{\langle\xi\xi\rangle}=0$. The
LMEB introduce the additional dynamics of the covariance 
$\boldsymbol{\langle\xi\xi\rangle}$ of the rapid fluid parcel displacement
fluctuations into the metric of the {\bfi dynamical Helmholtz operator},
$1-\tilde\Delta$. Because of the advective nature of the
$\boldsymbol{\langle\xi\xi\rangle}$ dynamics, this Helmholtz operator {\it
commutes} with the advective time derivative, $d/dt$ for the Lagrangian mean
velocity. 

The LMEB model introduces two {\it different} fluid velocities into the
average description. The velocities in the LMEB equations
(\ref{LMEB-model-intro}) -- (\ref{LMEB-model-defs-intro}) are
$\mathbf{u}$, defined to be the {\bfi Lagrangian mean velocity}; and
$\mathbf{v}=(1-\tilde\Delta)\mathbf{u}$, found later to be the {\bfi Eulerian
mean velocity}, to order $o(|\boldsymbol\xi|^2)$. It is also useful to think of
the velocity $\mathbf{u}$ as the particle, or transport velocity, while the
velocity $\mathbf{v}$ is the flow, or circulation velocity. Because of the
relation $\mathbf{v}=(1-\tilde\Delta)\mathbf{u}$, the transport velocity
$\mathbf{u}$ is {\it smoother} than the flow, or circulation velocity
$\mathbf{v}$, by the inversion of the dynamical Helmholtz operator,
$(1-\tilde\Delta)$. That is, in the advective time derivative 
${d}/{dt} = \Big({\partial}/{\partial{t}} +
\mathbf{u}\boldsymbol{\,\cdot\nabla}\Big)$, we have
$\mathbf{u}=(1-\tilde\Delta)^{-1}\mathbf{v}$. The difference
$\mathbf{u}-\mathbf{v} = \tilde\Delta\mathbf{u}$ is the {\bfi Stokes mean
drift velocity}, due to the presence of the rapid fluctuations with Lagrangian
mean covariance $\boldsymbol{\langle\xi\xi\rangle}$. The interpretation of its
effects will be a recurring theme in this paper.

The symmetric tensor $\boldsymbol{\kappa}_S = \frac{1}{2}d\boldsymbol
{\langle\xi\xi\rangle}/dt$ is the {\bfi Taylor diffusivity tensor} (defined
here without including any antisymmetric corrections due to rotation).
The displacement fluctuation $\boldsymbol\xi$ is defined by
$\boldsymbol\xi=\mathbf{X}^{\xi}-\mathbf{X}$  with $\mathbf{X} =
\langle\mathbf{X}^{\xi}\rangle$ for an averaging process (the Lagrangian mean)
taken at fixed Lagrangian label $\mathbf{a}$ and denoted
$\langle\,\boldsymbol{\cdot}\,\rangle$ with, e.g.,
\begin{equation}\label{avg-brkt-def}
\mathbf{X}(\mathbf{a},t)
=
\langle\mathbf{X}^{\xi}(\mathbf{a},t\,;\omega)\rangle
\equiv
\lim_{T\to\infty}\frac{1}{T}\int_{-T}^T
\mathbf{X}^{\xi}(\mathbf{a},t\,;\omega)\,d\omega\,.
\end{equation}
Thus, a displacement fluctuation satisfying
$\langle\boldsymbol\xi\rangle = 0$ is given by
\begin{equation}\label{xi-fluct-def}
\boldsymbol\xi(\mathbf{X},t\,;\omega)
\equiv
\mathbf{X}^{\xi}(\mathbf{a},t\,;\omega)
-
\mathbf{X}(\mathbf{a},t)
\,.
\end{equation}
This displacement fluctuation has covariance
\begin{equation}\label{xi-xi-cov-def}
\boldsymbol{\langle\xi\xi\rangle}
\equiv
\langle\mathbf{X}^{\xi}\mathbf{X}^{\xi}\rangle
-\mathbf{X}\mathbf{X}
\,.
\end{equation}
Here, $\mathbf{X}^{\xi}(\mathbf{a},t\,;\omega)$ is the spatial trajectory of
a fluid parcel with Lagrangian label $\mathbf{a}$ that is undergoing EB
dynamics. We assume this motion depends on both a slow time $t$ and a rapid
(or random) time variation $\omega$. In taking the Lagrangian mean of this
spatial trajectory, we average at constant fluid label $\mathbf{a}$ over its
rapid variation in $\omega$ during a time interval $T$ that is long
(denoted $\lim_{T\to\infty}$) compared to the rapid time scale for variation
in $\omega$, but during which the slow time variation in $t$ may be regarded
as fixed. In particular, we shall assume that the result of the averaging
operation $\langle\,\boldsymbol{\cdot}\,\rangle$ is independent of the magnitude
of the time interval $T$. Thus, in equation (\ref{xi-fluct-def}) one may think
of the Lagrangian trajectory $\mathbf{X}^{\xi}(\mathbf{a},t\,;\omega)$ as
following the original EB dynamics in a flow regime in which a separation of
time scales into slow ($t$) and fast ($\omega$) is possible, and think of
$\mathbf{X}(\mathbf{a},t)$ as following the approximate
slow time dynamics determined by the LMEB equations. We shall see that $p$ is
the Lagrangian mean pressure. The other notation in (\ref{LMEB-model-intro}) and
(\ref{LMEB-model-defs-intro}) is standard for fluid dynamics: $b$ is buoyancy;
$\nu$ is kinematic viscosity; $g$ is the constant acceleration of gravity; and
${\rm curl}\,\mathbf{R}=2\boldsymbol{\Omega}(\mathbf{x})$ is the Coriolis
parameter, which may depend on position. The boundary conditions of the
{\it dissipative} LMEB model are
\begin{equation}\label{LMEB-bc}
\mathbf{v}=0\,,
\quad
\mathbf{u}=0
\quad\hbox{and}\quad
\boldsymbol{\langle\xi\xi\rangle\cdot\hat{n}}=0\,
\quad\hbox{on a fixed boundary.}
\end{equation}

When dissipation, rotation and stratification are {\it absent}, the LMEB model
in (\ref{LMEB-model-intro}) -- (\ref{LMEB-model-defs-intro}) reduces to the 
{\bfi ideal Lagrangian mean motion (LMM) equations}:
\boxeq{6}
\begin{eqnarray}\label{ideal-LMM-eqns}\vspace{-.1in}
&&\frac{d}{dt}\mathbf{v} 
+ \boldsymbol\nabla p 
= 0
\,,\quad\hbox{with}\quad
\boldsymbol{\nabla\cdot}\mathbf{u} = 0 \,,
\\
&& 
\hbox{where  }
\frac{d}{dt} \equiv \Big(\frac{\partial}{\partial t} 
+ \mathbf{u}\boldsymbol{\,\cdot\nabla}\Big),
\quad
\mathbf{v} \equiv (1-\tilde\Delta)\mathbf{u}\,,
\quad
\tilde\Delta \equiv
\boldsymbol{\nabla\cdot\langle\xi\xi\rangle\cdot\nabla}\,,
\nonumber\\
&& \hbox{and}\quad
\frac{d}{dt}\boldsymbol{\langle\xi\xi\rangle}
= \boldsymbol{\langle\xi\xi\rangle\cdot\nabla\mathbf{u}}
+ \boldsymbol{\nabla\mathbf{u}}^{\rm T}
\boldsymbol{\cdot\langle\xi\xi\rangle}
\,.\label{LMM-cov-dyn}
\end{eqnarray}
The {\bfi boundary conditions} for this {\it ideal} LMM model are:
\begin{equation}\label{LMM-bc}
\mathbf{v}\boldsymbol{\cdot\hat{n}}=0\,,
\quad
\mathbf{u}=0\,,
\quad\hbox{and}\quad
\boldsymbol{\langle\xi\xi\rangle\cdot\hat{n}}=0
\quad\hbox{on a fixed boundary.}
\end{equation}
The LMM motion equation (\ref{ideal-LMM-eqns}) may be rewritten equivalently as
\begin{equation}\label{LMM-mot-eqn}
\Big(\frac{\partial}{\partial t} 
+ \mathbf{v}\boldsymbol{\,\cdot\nabla}\Big)\mathbf{v} 
+ 
\underbrace{
(\mathbf{u}-\mathbf{v})\boldsymbol{\,\cdot\nabla}\mathbf{v} }
_{\hbox{\large {\bf Stokes transport}}}
+ \boldsymbol\nabla p 
= 0
\,.
\end{equation}
Thus, perhaps not unexpectedly, the Stokes mean drift velocity
$\mathbf{u}-\mathbf{v} = \tilde\Delta\mathbf{u}$ contributes an {\bfi additional
transport term} in this motion equation for the Eulerian mean velocity
$\mathbf{v}$. On the invariant manifold $\boldsymbol{\langle\xi\xi\rangle}=0$
the LMEB and LMM equation sets recover their original forms. 

The ideal LMM model preserves the {\bfi total kinetic energy},
\begin{equation}\label{LMM-cons-erg}
E =  \frac{1}{2}\int d^{\,3}x \Big(|\mathbf{u}\,|^2 
+ \langle\xi^k\xi^l\rangle 
\mathbf{u}_{,k}\boldsymbol\cdot\mathbf{u}_{,l}\Big)
= \frac{1}{2}\int d^{\,3}x\ \mathbf{u}\boldsymbol\cdot\mathbf{v}
\,,
\end{equation}
in which the Lagrangian covariance of the fluctuations couples to the gradients
of the Lagrangian mean velocity. (Throughout the paper, we sum on repeated
indices.) Conservation of this energy provides $L^2$ control on
$|\boldsymbol{\nabla}\mathbf{u}|$, provided
$\boldsymbol{\langle\xi\xi\rangle}$ is bounded away from zero. In fact, for
incompressible Lagrangian mean velocity $\mathbf{u}$, the determinant
det$\boldsymbol{\langle\xi\xi\rangle}$ is conserved on fluid parcels. Hence, the
covariance does remain bounded away from zero, if it is initially so. This
statement still holds when order $O(|\boldsymbol\xi|^2)$ compressibility is
allowed. The ideal LMM model also conserves the domain integrated momentum,
$\int\mathbf{v}\, d^{\,3}x$. (With boundary conditions (\ref{LMM-bc}), this
conserved momentum is also equal to $\int\mathbf{u}\, d^{\,3}x$.)

The ideal LMM model possesses a {\bfi Kelvin-Noether
circulation theorem} showing how the Stokes drift velocity generates
circulation of $\mathbf{v}$. Namely,
\begin{equation}\label{LMM-KelThm-v}
\frac{ d}{dt}\oint_{\gamma(\mathbf{u})}\mathbf{v}
\boldsymbol{\cdot}d\mathbf{x} 
= -\int\int_{S(\mathbf{u})}
\Big[\boldsymbol{\nabla}\,\tilde\Delta{u}_j
\times\boldsymbol{\nabla}\,u^j\Big]
\boldsymbol{\cdot}d\mathbf{S}
\,,
\end{equation}
where the closed curve ${\gamma(\mathbf{u})}$ moves with the 
Lagrangian mean fluid velocity $\mathbf{u}$ and is the boundary of the
surface $S(\mathbf{u})$. By virtue of the operator commutation relation
\begin{equation}\label{remark-comm-rel}
\frac{ d}{dt}\tilde\Delta - \tilde\Delta\frac{ d}{dt}
\equiv\bigg[\frac{ d}{dt},\tilde\Delta\bigg]=0\,,
\end{equation}
we may also express the ideal LMM motion equation (\ref{ideal-LMM-eqns}) in
its {\bfi alternative LMM form},
\smallskip\boxeq{2}
\begin{equation}\label{LMM-mot-u-eqn}
\frac{\partial\mathbf{u}}{\partial t} 
+ \mathbf{u}\boldsymbol{\cdot\nabla}\mathbf{u}
= -\ (1-\tilde{\Delta})^{-1}\boldsymbol\nabla{p}\,,
\quad \boldsymbol{\nabla\cdot\mathbf{u}}=0\,.
\end{equation}
\smallskip

\noindent
In this form, the effect of the advected fluctuations is to smooth the
pressure gradient in an {\it adaptive fashion} depending on the velocity
shear through the evolution of $\tilde{\Delta}$. The divergence
of equation (\ref{LMM-mot-u-eqn}) yields an elliptic equation for the 
Lagrangian mean pressure, $p$. The Kelvin-Noether circulation theorem
corresponding to the form (\ref{LMM-mot-u-eqn}) of the LMM motion equation is
\begin{equation}\label{LMM-KelThm-u}
\frac{ d}{dt}\oint_{\gamma(\mathbf{u})}\mathbf{u}
\boldsymbol{\cdot}d\mathbf{x}
= - \int\int_{S(\mathbf{u})}
\Big[\boldsymbol{\nabla}\times
(1-\tilde{\Delta})^{-1}\boldsymbol{\nabla}{p}\Big]
\boldsymbol{\cdot}d\mathbf{S}
\,,
\end{equation}
which represents the circulation dynamics of $\mathbf{u}$, rather than
$\mathbf{v}$.

Thus, the presence of the fluctuations with Lagrangian mean covariance
$\boldsymbol{\langle\xi\xi\rangle}$ in the ideal LMM equations has five
main effects, relative to the Euler equations: 
\begin{enumerate}
\item  it smoothes the Lagrangian mean (transport)
velocity $\mathbf{u}$ relative to the Eulerian mean (momentum) velocity
$\mathbf{v}$; 
\item  it introduces an {\bfi additional transport} of the Eulerian mean
velocity $\mathbf{v}$ by the Stokes drift velocity $\mathbf{u}-\mathbf{v}$; 
\item  it controls $\|\boldsymbol{\nabla}\mathbf{u}\|_2$ in the $L^2$ norm
via energy conservation (or energy dissipation, when viscosity is included);
\item  it creates circulation of both Eulerian mean velocity $\mathbf{v}$
and Lagrangian mean fluid velocity $\mathbf{u}$ around closed loops advecting
with the fluid parcels; and
\item  it smoothes the pressure gradient in an adaptive fashion
depending on the evolution of the fluctuation covariance
$\boldsymbol{\langle\xi\xi\rangle}$, driven by the velocity shear
$\boldsymbol{\nabla\mathbf{u}}$.
\end{enumerate}

Most of this paper is devoted to deriving the LMM and LMEB equations by
approximating Taylor series expansions and averaging over the rapid time
dependence in Hamilton's principle for the Euler equations of an ideal
incompressible fluid. We average over ``fast time'' at fixed Lagrangian fluid
label in Hamilton's principle for the Euler equations. This averaging over the
rapid ``microscopic'' fluid motions allows us to extract an approximate
Lagrangian mean Hamilton's principle whose Euler-Lagrange equations describe the
remaining slow ``macroscopic'' fluid motions. After a discussion of these
Euler-Lagrange equations from the viewpoint of the Lagrangian fluid
parcel description, we use the Euler-Poincar\'e theory of Holm, Marsden and
Ratiu~\cite{HMR[1998a]},~\cite{HMR[1998b]}, to develop and analyze the ideal LMM
equations (\ref{ideal-LMM-eqns}) -- (\ref{LMM-cov-dyn}) in the Eulerian
description. We then add rotation and stratification to derive the ideal LMEB
equations.  Finally, we introduce dissipation according to the traditional
semi-empirical reasoning by requiring that the energy dissipation rate be
negative definite. This both adds the fluctuation-dependent viscosity in the
motion equation and introduces the dynamics for the Taylor diffusivity
tensor $\boldsymbol\kappa_S$ into the buoyancy equation.

A {\bfi second moment Lagrangian mean turbulence model} is obtained by adding
phenomenological viscosity $\nu\tilde\Delta\mathbf{v}$ and forcing $\mathbf{F}$
to the ideal LMM motion equation (\ref{ideal-LMM-eqns}), so that,
\begin{equation}\label{LMM-2pt-eqns}
\Big(\frac{\partial}{\partial t} 
+ \mathbf{u}\boldsymbol{\,\cdot\nabla}\Big)\mathbf{v} 
+ \boldsymbol\nabla p 
= \nu\,\tilde\Delta\mathbf{v} + \mathbf{F}\,,
\quad
\boldsymbol{\nabla\cdot}\mathbf{u}=0\,,
\end{equation}
with viscous boundary conditions (\ref{LMEB-bc}). Note that the Lagrangian mean
fluctuation covariance appears in the dissipation operator
$\tilde\Delta$. In the absence of the forcing $\mathbf{F}$, this viscous LMM
turbulence model dissipates the energy $E$ in equation (\ref{LMM-cons-erg})
according to
\begin{equation}\label{LMM-erg-dissip}
\frac{dE}{dt}
= -\,\nu \int d^{\,3}x
\Big[ {\rm tr}(\boldsymbol{\nabla}\mathbf{u}^T 
\boldsymbol{\cdot\,\langle\xi\xi\rangle\,\cdot}
\boldsymbol{\nabla}\mathbf{u}) +
\tilde\Delta\mathbf{u}\boldsymbol{\cdot}\tilde\Delta\mathbf{u}\,\Big]\,.
\end{equation}
This strictly negative energy dissipation rate is the reason for adding
viscosity with $\tilde\Delta$, instead of using the ordinary Laplacian operator.

The generalization of the Lagrangian mean motion model to Riemannian
manifolds -- to make it applicable in any coordinate system and to
elucidate its intrinsic geometrical structure -- will be discussed
elsewhere~\cite{Holm-Shkoller-inprep}. We will also defer the derivation of
the isopycnal and hydrostatic versions of this model to another time and
place~\cite{Holm[1998]}.


\subsection{Eulerian mean equations}

We shall also present a parallel development of these Lagrangian mean results,
but applied to Eulerian mean models. The {\bfi Eulerian mean Euler-Boussinesq
(EMEB)} equations are given by
\medskip
\boxeq{9}
\begin{eqnarray}\label{EMEB-model-intro}
&&\frac{d}{dt}\mathbf{V} 
+ V_j\boldsymbol\nabla \bar{U}^j
- \bar{\mathbf{U}}\times{\rm curl}\,\mathbf{R}(\mathbf{x})
+ \boldsymbol\nabla P_{tot}^E 
\nonumber\\
&&\hspace{1in}
+ gb\,\mathbf{\hat{z}}
+ \frac{1}{2}\,
\big(\bar{\mathbf{U}}_{,k}\boldsymbol\cdot\bar{\mathbf{U}}_{,l}\big)
\boldsymbol\nabla
\langle\zeta^k\zeta^l\rangle^E
= \nu\,\tilde\Delta^E\mathbf{V}
\,,\\
&& 
\hbox{where  }
\frac{d}{dt} \equiv \Big(\frac{\partial}{\partial t} 
+ \bar{\mathbf{U}}\boldsymbol{\,\cdot\nabla}\Big),
\quad
\mathbf{V} \equiv (1-\tilde\Delta^E)\bar{\mathbf{U}}\,,
\quad
\tilde\Delta^E \equiv
\boldsymbol{\nabla\cdot\langle\,\zeta\zeta\rangle}^E
\boldsymbol{\cdot\nabla}\,,
\nonumber\\
&&
\frac{db}{dt}
= \kappa\tilde\Delta^E b
\,,\quad
\frac{d}{dt}\boldsymbol{\langle\,\zeta\zeta\rangle}^E
= 0
\,,
\quad
 P_{tot}^E
\equiv 
 P  
- \frac{1}{2}|\bar{\mathbf{U}}\,|^2 
- \frac{1}{2}\langle\zeta^k\zeta^l\rangle^E 
\big(\bar{\mathbf{U}}_{,k}\boldsymbol\cdot\bar{\mathbf{U}}_{,l}\big)
\,,
\nonumber\\
&&
\quad\hbox{and}\quad
\boldsymbol{\nabla\cdot}\bar{\mathbf{U}} = 0 
\,,
\quad\hbox{where}\quad
\bar{\mathbf{U}}
\equiv
\langle\mathbf{U}\rangle^E
\,.
\label{EMEB-model-defs-intro}
\end{eqnarray}
Here $P$ is the Eulerian mean pressure, and $\nu$ and $\kappa$ are constants
representing viscosity and diffusivity, respectively.
The {\bfi boundary conditions} for this {\it viscous} EMEB model are:
\begin{equation}\label{EMEB-bc}
\mathbf{V} = 0\,,
\quad
\bar{\mathbf{U}} = 0\,,
\quad\hbox{and}\quad
\boldsymbol{\langle\,\zeta\zeta\rangle}^E
\boldsymbol{\cdot\hat{n}}=0
\quad\hbox{on a fixed boundary.}
\end{equation}

These equations correspond to the same level of approximation as those for
LMEB, but they are based on the Eulerian mean velocity $\bar{\mathbf{U}}$,
appearing in the {\bfi Reynolds fluid velocity decomposition},
\begin{equation}\label{Reynolds-Eul-mean-decomp-intro}
\mathbf{U}(\mathbf{x},t;\omega)
\equiv 
\bar{\mathbf{U}}(\mathbf{x},t) 
+ \mathbf{U}^{\,\prime}(\mathbf{x},t;\omega)
\,.
\end{equation}
The {\bfi Eulerian mean averaging process} at fixed position $\mathbf{x}$ is
denoted $\langle\,\boldsymbol{\cdot}\,\rangle^E$ with, e.g.,
\begin{equation}\label{Eul-mean-def-intro}
\bar{\mathbf{U}}(\mathbf{x},t)
=
\langle\mathbf{U}(\mathbf{x},t\,;\omega)\rangle^E
\equiv
\lim_{T\to\infty}\frac{1}{T}\int_{-T}^T
\mathbf{U}(\mathbf{x},t\,;\omega)\,d\omega\,.
\end{equation}
We shall show that the Eulerian velocity fluctuation $\mathbf{U}^{\,\prime}$ is
related to the Eulerian displacement fluctuation $\boldsymbol{\zeta}$ by
\begin{equation}\label{Eul-mean-vel-displ}
0
=
\boldsymbol{\zeta\cdot\nabla} \bar{\mathbf{U}}
+
\mathbf{U}^{\,\prime}(\mathbf{x},t\,;\omega)
\,.
\end{equation}
Consequently, the Eulerian mean kinetic energy due to the velocity fluctuation
satisfies
\begin{equation}\label{Eul-mean-fluct-KE-intro}
\langle\,|\mathbf{U}^{\,\prime}|^2\,\rangle^E 
= \langle\zeta^k\zeta^l\rangle^E 
\bar{\mathbf{U}}_{,k}\boldsymbol\cdot\bar{\mathbf{U}}_{,l}
\,.
\end{equation}
Note that the advection of the Eulerian
mean displacement fluctuation covariance
$\boldsymbol{\langle\,\zeta\zeta\rangle}^E$ by the Eulerian mean
velocity $\bar{\mathbf{U}}$ is componentwise: 
$d\boldsymbol{\langle\,\zeta\zeta\rangle}^E/dt = 0$, so each component
of this symmetric tensor is carried along with the Eulerian mean flow as if it
were a scalar function. Moreover, we shall show that the momentum
$\mathbf{V}$ appearing in the Euler-Poincar\'e formulation of these EMEB
equations is in fact the {\bfi Lagrangian mean velocity} for this theory. This
duality between Lagrangian mean and Eulerian mean theories is another theme of
the present paper.

In the absence of dissipation, rotation and stratification, the EMEB model
in (\ref{EMEB-model-intro}) -- (\ref{EMEB-model-defs-intro}) reduces to the 
{\bfi ideal Eulerian mean motion (EMM) equations}:
\smallskip\boxeq{5}
\begin{eqnarray}\label{ideal-EMM-eqns}\vspace{-.1in}
&&
\Big(\frac{\partial}{\partial t} 
+ \bar{\mathbf{U}}\boldsymbol{\,\cdot\nabla}\Big)
\mathbf{V} 
+ V_j\boldsymbol{\nabla}\bar{U}^j
+ \boldsymbol\nabla P_{tot}^E
+ \frac{1}{2}\,
\big(\bar{\mathbf{U}}_{,k}\boldsymbol\cdot\bar{\mathbf{U}}_{,l}\big)
\boldsymbol\nabla
\langle\zeta^k\zeta^l\rangle^E
=0
\,,
\\
&& 
\hbox{where  }
\quad
\mathbf{V} \equiv (1-\tilde\Delta^E)\bar{\mathbf{U}}\,,
\quad
\tilde\Delta^E \equiv
\boldsymbol{\nabla\cdot\langle\,\zeta\zeta\rangle}^E
\boldsymbol{\cdot\nabla}\,,
\nonumber\\
&& 
\boldsymbol{\nabla\cdot}\bar{\mathbf{U}} = 0 \,,
\quad\hbox{and}\quad
\Big(\frac{\partial}{\partial t} 
+ \bar{\mathbf{U}}\boldsymbol{\,\cdot\nabla}\Big)
\boldsymbol{\langle\,\zeta\zeta\rangle}^E
= 
0
\,.\label{EMM-cov-dyn}
\end{eqnarray}
The {\bfi boundary conditions} for this {\it ideal} EMM model are:
\begin{equation}\label{EMM-bc}
\mathbf{V}\boldsymbol{\cdot\hat{n}}=0\,,
\quad
\bar{\mathbf{U}}=0\,,
\quad\hbox{and}\quad
\boldsymbol{\langle\,\zeta\zeta\rangle\cdot\hat{n}}=0
\quad\hbox{on a fixed boundary.}
\end{equation}
The EMM motion equation (\ref{ideal-EMM-eqns}) may be rewritten equivalently as
\begin{equation}\label{EMM-mot-eqn}
\Big(\frac{\partial}{\partial t} 
+ \mathbf{V}\boldsymbol{\,\cdot\nabla}\Big)
\mathbf{V} 
+ 
\underbrace{
(\mathbf{V}-\bar{\mathbf{U}})
\boldsymbol{\times}{\rm curl}\,\mathbf{V}}
_{\hbox{\large {\bf Stokes vortex force}}}
+\
 \boldsymbol\nabla P 
- \frac{1}{2}
\langle\zeta^k\zeta^l\rangle^E 
\boldsymbol\nabla
\big(\bar{\mathbf{U}}_{,k}\boldsymbol\cdot\bar{\mathbf{U}}_{,l}\big)
= 0
\,.
\end{equation}
Thus, the Stokes mean drift velocity
$\mathbf{V}-\bar{\mathbf{U}} = \tilde\Delta^E\bar{\mathbf{U}}$ contributes an
{\bfi additional vortex force} in this motion equation for the Lagrangian mean
fluid velocity $\mathbf{V}$.

On the invariant manifold $\boldsymbol{\nabla\langle\,\zeta\zeta\rangle}^E=0$ the
EMM equation set reduces to the 3D ideal Camassa-Holm
model~\cite{HMR[1998a]},~\cite{HMR[1998b]}. The Eulerian mean system
(\ref{ideal-EMM-eqns}) -- (\ref{EMM-cov-dyn}) recovers the 3D ideal
Camassa-Holm model as an invariant subsystem for
$\langle\zeta^k\zeta^l\rangle^E=\alpha^2\delta^{kl}$, with $\alpha$ a constant
length scale.

The ideal EMM model preserves the {\bfi total kinetic energy},
\begin{equation}\label{EMM-cons-erg}
E =  \frac{1}{2}\int d^{\,3}x 
\Big(|\bar{\mathbf{U}}\,|^2 
+ \langle\zeta^k\zeta^l\rangle^E 
\bar{\mathbf{U}}_{,k}\boldsymbol\cdot\bar{\mathbf{U}}_{,l}\Big)
= \frac{1}{2}\int d^{\,3}x\ \bar{\mathbf{U}}\boldsymbol\cdot\mathbf{V}
\,,
\end{equation}
in which the Eulerian covariance of the fluctuations couples to the gradients
of the Eulerian mean velocity. Conservation of this energy provides $L^2$
control on $|\boldsymbol{\nabla}\bar{\mathbf{U}}|$, provided
$\boldsymbol{\langle\xi\xi\rangle}^E$ is bounded away from zero, which it will
be, if it is initially so. The ideal EMM model also conserves the domain
integrated momentum, $\int\mathbf{V}\, d^{\,3}x$. (With boundary conditions
(\ref{EMM-bc}), this conserved momentum is also equal to $\int\bar{\mathbf{U}}\,
d^{\,3}x$.)

The ideal EMM model possesses a {\bfi Kelvin-Noether
circulation theorem} showing how the Stokes drift velocity generates
circulation of $\mathbf{V}$. Namely,
\begin{equation}\label{EMM-KelThm-v}
\frac{ d}{dt}\oint_{\gamma(\bar{\mathbf{U}})}\mathbf{V}
\boldsymbol{\cdot}d\mathbf{x} 
= - \frac{1}{2}\int\int_{S(\bar{\mathbf{U}})}
\Big[
\boldsymbol{\nabla}
\big(\bar{\mathbf{U}}_{,k}\boldsymbol\cdot\bar{\mathbf{U}}_{,l}\big)
\boldsymbol{\times\nabla}
\langle\zeta^k\zeta^l\rangle^E \Big]
\boldsymbol{\cdot}d\mathbf{S}
\,,
\end{equation}
where the closed curve ${\gamma(\bar{\mathbf{U}})}$ moves with the 
Eulerian mean fluid velocity $\bar{\mathbf{U}}$ and is the boundary of the
surface $S(\bar{\mathbf{U}})$. 

Thus, the presence of the fluctuations with Eulerian mean covariance
$\boldsymbol{\langle\,\zeta\zeta\rangle}^E$ in the ideal EMM equations has four
main effects, relative to the Euler equations: 
\begin{enumerate}
\item  it smoothes the Eulerian mean (transport) velocity $\bar{\mathbf{U}}$
relative to the Lagrangian mean (momentum) velocity $\mathbf{V}$; 
\item  it introduces an additional {\bfi vortex force} into the Lagrangian mean
velocity equation due to the Stokes drift velocity
$\bar{\mathbf{U}}-\mathbf{V} = \tilde\Delta^E\bar{\mathbf{U}}$; 
\item  it controls $\|\boldsymbol{\nabla}\bar{\mathbf{U}}\|_2$ in the $L^2$ norm
via energy conservation (or energy dissipation, when viscosity is included); and
\item  it creates circulation of Lagrangian mean velocity $\mathbf{V}$
around closed loops advecting with the Eulerian mean velocity $\bar{\mathbf{U}}$.
\end{enumerate}

A {\bfi second moment Eulerian mean turbulence model} is obtained by adding
phenomenological viscosity $\nu\tilde\Delta^E\mathbf{V}$ and forcing
$\mathbf{F}$ to the ideal EMM motion equation (\ref{ideal-EMM-eqns}), so that,
\begin{eqnarray}\label{EMM-2pt-eqns-intro}
\Big(\frac{\partial}{\partial t} 
+ \bar{\mathbf{U}}\boldsymbol{\,\cdot\nabla}\Big)\mathbf{V}
+ V_j \boldsymbol\nabla \bar{U}^j
&+& \boldsymbol\nabla P^E_{tot} 
+\ \frac{1}{2}\,
\big(\bar{\mathbf{U}}_{,k}\boldsymbol\cdot\bar{\mathbf{U}}_{,l}\big)
\boldsymbol\nabla
\langle\zeta^k\zeta^l\rangle^E
\\
&=& \nu\,\tilde\Delta^E\mathbf{V} + \mathbf{F}\,,
\quad\hbox{where}\quad
\boldsymbol{\nabla\cdot}\bar{\mathbf{U}}=0\,,
\nonumber
\end{eqnarray}
with viscous boundary conditions (\ref{EMEB-bc}). Note that the Eulerian mean
fluctuation covariance appears in the dissipation operator
$\tilde\Delta^E$. In the absence of the forcing $\mathbf{F}$, this viscous EMM
turbulence model dissipates the energy $E$ in equation (\ref{EMM-cons-erg})
according to
\begin{equation}\label{EMM-erg-dissip-intro}
\frac{dE}{dt}
= -\,\nu \int d^{\,3}x
\Big[ {\rm tr}(\boldsymbol{\nabla}\bar{\mathbf{U}}^T 
\boldsymbol{\cdot\,\langle\,\zeta\zeta\rangle}^E\,
\boldsymbol{\cdot\nabla }\bar{\mathbf{U}}) +
\tilde\Delta^E\bar{\mathbf{U}}
\boldsymbol{\cdot}
\tilde\Delta^E\bar{\mathbf{U}}\,\Big]\,.
\end{equation}
This energy dissipation law justifies adding viscosity with
$\tilde\Delta^E$, instead of using the ordinary Laplacian operator.

\subsection{Outline of the paper}

In Section \ref{EP-intro-sec}, we shall recall the Euler-Poincar\'e equations
(\ref{EPeqn}) in the context of the ideal Euler fluid equations. In Section
\ref{avg-L-sec}, we shall introduce into Hamilton's principle for Euler's
equations the Reynolds decomposition of a Lagrangian fluid trajectory as
the sum of its mean and fluctuating parts. In the remainder of Section
\ref{avg-L-sec}, we shall: substitute this decomposition into the Lagrangian
for Euler's equations; transform the resulting rapid (or random) Lagrangian
$L(\omega)$ in equation (\ref{Lag-xi}) into the Eulerian description; make
an approximation of it based on truncating a Taylor expansion; take
its Lagrangian mean, denoted $\langle L\rangle$ in equation
(\ref{mean-Lag-approx}); and finally use the Euler-Poincar\'e theorem to
obtain approximate Lagrangian mean motion equations at order
$O(|\boldsymbol\xi|^2)$. In Section \ref{phys-interp-sec} we shall provide
the physical interpretations of the quantities arising during this modeling
procedure and compare the resulting equations with other models. In Section
\ref{1pt closure-sec} we shall discuss both ideal and viscous one point closure
models based on the Euler-Poincar\'e equation. We shall also compare these one
point closure models with Reynolds-averaged Navier-Stokes (RANS) models, Large
Eddy Simulation (LES) models, and the Leray regularization of the Navier-Stokes
equations.

In Section \ref{2pt Eul closure-sec} we shall augment the Euler-Poincar\'e
equation in  (\ref{EPeqn-Lbar}) to include dynamically varying correlations
of the rapid/random fluctuations and use this Lagrangian mean fluid theory to
derive a new approximate motion equation for the slow evolution of the
Lagrangian mean fluid velocity in Eulerian coordinates. This is the basis for
the Eulerian analysis of the Lagrangian mean motion (LMM) model. The LMM
equations in (\ref{EP-aug-motion2}) -- (\ref{xi-xi-eqn-redux}) will include
self-consistent forces caused by the correlations of the rapid/random
fluctuations and expressed in terms of their Lagrangian mean covariance
tensor, $\boldsymbol{\langle\xi\xi\rangle}$. The approximate dynamics of the
covariance tensor $\boldsymbol{\langle\xi\xi\rangle}$ itself will also be
determined, thereby producing a two point, or second moment, Lagrangian mean
closure of Euler's ideal incompressible fluid equations. Several properties
of this ideal second moment closure model are also derived in Section
\ref{2pt Eul closure-sec}, including its conservation laws for momentum and
energy, and its Kelvin-Noether circulation theorem, which are all inherited from
the Euler-Poincar\'e formulation. We note that the Lagrangian mean effects of the
rapid fluctuations are purely {\it dispersive} at this stage and, thus, are
energy conserving.

The Lie-Poisson Hamiltonian structure of the ideal LMM model is given
in Section \ref{Ham-LPB-sec}. Its 2D behavior is discussed briefly in Section
\ref{2D-ideal-version-sec}. Upon adding viscosity in Section
\ref{NS2pt closure-sec}, we shall introduce a second moment
Lagrangian mean closure of the Navier-Stokes equations corresponding to the
LMM model and compare this closure with the Reynolds averaged Navier-Stokes
equations. 

In Section \ref{LM-geo-appl-sec}, we shall discuss geophysical applications of
these ideas and use our approach via the Euler-Poincar\'e theory to derive the
Lagrangian mean Euler-Boussinesq (LMEB) equation set  (\ref{LMEB-model}) --
(\ref{LMEB-model-defs}) for the Lagrangian mean motion of a rotating stratified
turbulent incompressible fluid. Section \ref{LMSW-eqns-sec} considers lower
dimensional subcases in one and two dimensions, in an effort to help develop
intuition about the solution behavior of these Lagrangian mean models.

Section \ref{EMM-sec} studies the Eulerian mean counterpart of the LMM
model and emphasizes the duality and parallel mathematical structures shared
in the two approaches through their formulations as Euler-Poincar\'e equations.
Section \ref{EM-geo-appl-sec} adds rotation and stratification to the Eulerian
mean model, then discusses some of its aspects in fewer dimensions. Section
\ref{conclusions-sec} contains a summary and conclusions.


\section{Review of Hamilton's principle and Euler-Poincar\'e equations for
ideal fluids}  
\label{EP-intro-sec}

Consider the Lagrangian $L$ comprised of fluid kinetic energy
with volume preservation imposed by a Lagrange multiplier $P$ (the
pressure),
\begin{equation}\label{Lag-Lag}
L(\mathbf{X},\mathbf{\dot{X}})
= \int d^{\,3}a \left\{\frac{1}{2} 
  \left|\mathbf{\dot{X}}(\mathbf{a},t)\right|^2 
+ P(\mathbf{X}(\mathbf{a},t),t) 
  \Big(\det(\mathbf{X}^{\,\prime}_a) (\mathbf{a},t) - 1 \Big)\right\}.
\end{equation}
Here, $\mathbf{X}(\mathbf{a},t)$ is the Lagrangian fluid trajectory: 
that is, $\mathbf{x}=\mathbf{X}(\mathbf{a},t)$ is the current position of
the material point starting at initial position $\mathbf{a}$ at time
$t=0$.  We denote the derivatives of the function
$\mathbf{X}(\mathbf{a},t)$ by
$\mathbf{\dot{X}}=\partial\mathbf{X}/\partial{t}$ and
$\mathbf{X}^{\,\prime}_a=\partial\mathbf{X}/\partial\mathbf{a}$. 
After a brief calculation, {\bfi Hamilton's principle},
\begin{equation}\label{Ham-princ}
\delta\int dt\,L=0, 
\end{equation}
for the Lagrangian $L(\mathbf{X},\mathbf{\dot{X}})$ yields the
following {\bfi Euler-Lagrange  equations},
\begin{equation}\label{Eul-Lag-eqn}
\hspace{-.35in}
\delta\mathbf{X}:\qquad
\frac{\partial}{\partial t}\bigg|_{\mathbf{a}}
\frac{\delta L}{\delta\mathbf{\dot{X}}}
- \frac{\delta L}{\delta\mathbf{X}} = 0
\,,
\end{equation}
given explicitly by (dropping boundary and endpoint contributions),
\begin{eqnarray}\label{Eul-Lag.eqn}
\delta\mathbf{X}:&&
\mathbf{\ddot{X}}
+ \det(\mathbf{X}^{\,\prime}_a)\frac{\partial P}{\partial\mathbf{X}} = 0\,,
\\
\delta P:&&\det(\mathbf{X}^{\,\prime}_a) = 1
\,.
\end{eqnarray}
These are Euler's equations for the incompressible motion of an ideal
fluid in the Lagrangian description.

We obtain the Eulerian description of this motion by defining the 
Eulerian fluid velocity $\mathbf{U}(\mathbf{x},t)$ and volume element
$D(\mathbf{x},t)$ via the {\bfi basic kinematic relations},
\begin{equation}\label{Basic-def}
\mathbf{U}(\mathbf{x},t) = \mathbf{\dot{X}}(\mathbf{a},t)
\quad{\rm and}\quad 
D(\mathbf{x},t) = \Big(\det\mathbf{X}^{\,\prime}_a(\mathbf{a},t)\Big)^{-1}
\quad{\rm at}\quad 
\mathbf{x} = \mathbf{X}(\mathbf{a},t)\,.
\end{equation}
The volume element $D$ satisfies the continuity equation, 
\begin{equation}\label{conteqn}
\frac{\partial D}{\partial t} 
+ \boldsymbol{\nabla\cdot}\,(D\mathbf{U})
= 0\,.
\end{equation}
The volume element $D(\mathbf{x},t)$ is an {\bfi advected quantity},
(sometimes called ``frozen-in''). By this, we mean a quantity that is
expressible purely in terms of Lagrangian labels; so that it is invariant
along the Lagrangian mean fluid trajectory, or equivalently, satisfies a
certain Lie-derivative relation~\cite{HMR[1998a]}. For example, the volume
element $D(\mathbf{x},t)$ is expressible as the Jacobian for the
transformation from Lagrangian to Eulerian coordinates. Thus, the volume
element satisfies
\begin{equation}\label{vol-elmnt}
D\,d^{\,3}x=d^{\,3}a\,, 
\end{equation}
and the continuity equation (\ref{conteqn}) is implied by the corresponding
{\bfi invariance relation} for such an advected quantity, in this case,
\begin{equation}\label{conteqn-der}
0 = \frac{\partial}{\partial t}\Big|_{\mathbf{a}}(d^{\,3}a)
  = \bigg(\frac{\partial}{\partial t} + \pounds_{\mathbf{U}}\bigg) (D d^3x)
  = \bigg(\frac{\partial D}{\partial t} 
+ \boldsymbol{\nabla\cdot}\,(D\mathbf{U})\bigg) (d^3x)\,,
\end{equation}
where $\pounds_{\mathbf{U}}$ denotes {\bfi Lie derivative} with respect to the
vector field $\mathbf{U}(\mathbf{x},t)$, the Eulerian fluid velocity. 
The basic relations (\ref{Basic-def}) allow one to transform the 
Lagrangian (\ref{Lag-Lag}) into Eulerian variables as
\begin{equation}\label{Lag-Eul}
L = \int d^{\,3}x \left\{\frac{D}{2} |\mathbf{U}(\mathbf{x},t)|^2 
+ P(\mathbf{x},t) \Big(1- D(\mathbf{x},t) \Big)\right\}\;.
\end{equation}
Hamilton's principle $\delta\int dt\,L=0$ for a Lagrangian 
$L(\mathbf{U},D)$ defined in this way yields the following
{\bfi Euler-Poincar\'e equation} for $\mathbf{U}$, see Holm, Marsden
and Ratiu~\cite{HMR[1998a]},~\cite{HMR[1998b]},
\boxeq{2}
\begin{equation}\label{EPeqn}
\left(\frac{\partial}{\partial t} 
+ \mathbf{U}\boldsymbol{\,\cdot\nabla}\right)
\frac{1}{D}
\frac{\delta L}{\delta \mathbf{U}} 
+ \frac{1}{D}\frac{\delta L}{\delta U^j}\boldsymbol\nabla U^j 
- \boldsymbol\nabla\frac{\delta L}{\delta D} = 0\,,
\end{equation}
(throughout, we sum on repeated indices) and the
Lagrange multiplier $P$ imposes the constraint,
\begin{equation}\label{q-var}
\frac{\delta L}{\delta P} = 0\;.
\end{equation}
Substituting the variational derivatives of the Lagrangian (\ref{Lag-Eul}),
\begin{equation}\label{Vareqn}
\frac{1}{D}\frac{\delta L}{\delta \mathbf{U}} = \mathbf{U}
\,,\quad
\frac{\delta L}{\delta D} = \frac{1}{2}|\mathbf{U}\,|^2-P
\,,\quad
\frac{\delta L}{\delta P} =1 -D\,,
\end{equation}
into the Euler-Poincar\'e equation (\ref{EPeqn}) with constraint
(\ref{q-var}) and using continuity (\ref{conteqn}) yields Euler's
equations for the incompressible motion of an ideal fluid in the
Eulerian description, namely,
\begin{equation}\label{Euler-eqns}
\left(\frac{\partial}{\partial t} 
+ \mathbf{U}\boldsymbol{\,\cdot\nabla}\right)\mathbf{U} 
+ \boldsymbol\nabla P = 0
\,,\quad
\boldsymbol{\nabla\cdot}\mathbf{U} = 0 \;.
\end{equation}
For Lagrangians of the type (\ref{Lag-Lag}) --- specifically those
Lagrangians that are invariant under ``particle relabeling'' (by the
right-action of the volume preserving diffeomorphism group acting on the
tangent space of the fluid parcel trajectories) --- the Euler-Poincar\'e
equations (\ref{EPeqn}) are equivalent to the Euler-Lagrange equations
(\ref{Eul-Lag-eqn}). See Holm, Marsden and
Ratiu~\cite{HMR[1998a]},~\cite{HMR[1998b]} for more details, discussions and
proofs of this type of equivalence.

Of course, the Euler equations (\ref{Euler-eqns}) could also be found by
directly transforming the Euler-Lagrange equations (\ref{Eul-Lag.eqn}) from
the Lagrangian, to the Eulerian description by using the basic relations
(\ref{Basic-def}). This would avoid the step of transforming Hamilton's
principle into the Eulerian description. And this is the point: the
equivalence of the Euler-Lagrange equations (\ref{Eul-Lag-eqn}) and the
Euler-Poincar\'e equations (\ref{EPeqn}) facilitates easy, immediate
transitions from one description to the other that are helpful in developing
approximate fluid models and interpreting their solution properties. In this
paper, we shall take both the Lagrangian, and the Eulerian viewpoints,
switching from one to the other whenever it facilitates our purpose in
the development of these models. In the end, we shall be writing Eulerian
representations of approximate fluid models, found by taking either Lagrangian,
or Eulerian means of Hamilton's principle and then applying the Euler-Poincar\'e
theory.


\section{Averaged Lagrangians and Euler-Lagrange equations}
\label{avg-L-sec}

  
\subsection{Lagrangian fluid trajectory fluctuations}
The trajectory of a Lagrangian fluid parcel 
$\mathbf{X}^\xi(\mathbf{a},t\,;\omega)$
may be decomposed into its mean and fluctuating parts as
\begin{equation}\label{traj-xi}
\mathbf{X}^\xi(\mathbf{a},t\,;\omega) 
= \mathbf{X}(\mathbf{a},t) 
+ \boldsymbol\xi (\mathbf{X}(\mathbf{a},t),t\,;\omega)\;.
\end{equation}
This is the {\bfi Reynolds decomposition} of a Lagrangian fluid trajectory.
Here $\boldsymbol\xi= \mathbf{X}^\xi - \mathbf{X}$ is a rapid (or perhaps
random) vector field of fluctuations defined along the Lagrangian fluid
trajectory. The independent variable $\omega$ in equation
(\ref{traj-xi}) denotes either rapid time variation, or random fluctuations.
So the variable $\omega$ is allowed a dual interpretation. In its first
interpretation, $\omega$ is regarded as a short time scale associated with
rapid fluctuations. In its second interpretation, $\omega$ may be regarded
as a random variable associated with a stochastic process obeyed by the
fluctuations and defined along the Lagrangian fluid trajectory $\mathbf{X}$ with
label $\mathbf{a}$ and slow time variation $t$. This variability may be regarded
equally well as being either intrinsic, or extrinsic. We may assume the the
fluid trajectory is fluctuating intrinsically; or, we may assume the fluid is
subjected to a random forcing, which introduces a stochastic component into its
acceleration, resulting eventually in the decomposition (\ref{traj-xi}) of
the Lagrangian fluid trajectory. 

We denote by $\langle\boldsymbol\cdot\rangle$ the averaging
procedure performed at constant Lagrangian label
$\mathbf{a}$ and time $t$. This averaging is performed either over the rapid
time $\omega$, 
\begin{equation}\label{avg-brkt-def1}
\langle{f}\rangle
\equiv
\lim_{T\to\infty}\frac{1}{T}\int_{-T}^T f(\omega) \,d\omega\,,
\end{equation}
or, alternatively, with respect to a certain probability distribution ${\cal
P}(\omega)$ associated with the random event $\omega$,
\begin{equation}\label{avg-brkt-def2}
\langle{f}\rangle
\equiv
\int f(\omega) {\cal P}(\omega)\,d\omega\,,
\quad\hbox{with}\quad
\int {\cal P}(\omega)\,d\omega = 1
\,.
\end{equation}
Such an average taken while holding Lagrangian labels
$\mathbf{a}$ fixed is called a {\bfi Lagrangian mean}. The fluctuations are
assumed to have zero Lagrangian mean, $\langle\boldsymbol\xi\rangle=0$; so
the quantity $\mathbf{X}(\mathbf{a},t)$ in equation (\ref{traj-xi}) is the
mean Lagrangian fluid trajectory, since
$\langle\mathbf{X}^\xi(\mathbf{a},t\,;\omega)\rangle
=\mathbf{X}(\mathbf{a},t)$.  Thus, the quantity $\boldsymbol\xi$
describes the fluctuating displacement of the fluid trajectory, relative to
its Lagrangian mean. In the presence of these rapid (or random) fluctuations,
the Lagrangian $L$ in equation (\ref{Lag-Lag}) appearing in
Hamilton's principle (\ref{Ham-princ}) contains both slowly varying mean
variables and rapidly fluctuating, or random variables. Thus, the
Lagrangian (\ref{Lag-Lag}) becomes
\medskip\boxeq{2}
\begin{equation}\label{Lag-xi}\hspace{-.1in}
L(\omega) = \int d^{\,3}a\left\{\frac{1}{2} 
\left|\mathbf{\dot{X}}^\xi(\mathbf{a},t\,;\omega)\right|^2 
+ P(\mathbf{X}^\xi(\mathbf{a},t\,;\omega),t) 
\Big[\det(\mathbf{X}^{\xi\,\prime}_a)(\mathbf{a},t\,;\omega) - 1
\Big]\right\},\hspace{-.1in}
\end{equation}
where superscript $\xi$ denotes the Reynolds decomposed Lagrangian fluid
trajectory in equation (\ref{traj-xi}).

  
\subsection{Induced Eulerian velocity fluctuations} 
The decomposition $\mathbf{X}^\xi$ in (\ref{traj-xi}) of the Lagrangian fluid
trajectory into its mean and fluctuating components implies a corresponding
decomposition of the associated Eulerian velocity field. Via the basic
kinematic relations for the fluid velocity,
\begin{equation}\label{vel-xi}
\mathbf{U}\,(\mathbf{x} + \boldsymbol\xi (\mathbf{x},t\,;\omega),t)
= \mathbf{\dot{X}}^\xi (\mathbf{a},t\,;\omega)
\quad{\rm for}\quad
\mathbf{x} + \boldsymbol\xi (\mathbf{x},t\,;\omega)
= \mathbf{X}^\xi (\mathbf{a},t\,;\omega)\;, 
\end{equation}
we have
\begin{eqnarray}\label{vel-xi-dot-relation}
\hspace{-.1in}
\mathbf{U}\,(\mathbf{x} + \boldsymbol\xi (\mathbf{x},t\,;\omega),t)
=
\frac{d}{dt} \Big(\mathbf{x} 
+ \boldsymbol\xi (\mathbf{x},t\,;\omega)\Big)
&=& 
\mathbf{u}\,(\mathbf{x},t) 
+ \frac{d}{dt}\boldsymbol\xi (\mathbf{x},t\,;\omega),
\\
{\rm where}\quad
\frac{d}{dt} 
&\equiv& 
\frac{\partial}{\partial t} 
+ \mathbf{u}\,(\mathbf{x},t)\boldsymbol{\,\cdot\nabla},
\label{lag-u-advection}
\end{eqnarray}
defines the advective time derivative.
Since the displacement fluctuations of the fluid trajectory have zero
Lagrangian mean, $\langle\boldsymbol\xi\rangle=0$, the corresponding velocity
fluctuations will also have zero Lagrangian mean, so that 
$\langle d\boldsymbol\xi/dt\rangle=0$, as well. 
Therefore, upon taking the Lagrangian mean of equation
(\ref{vel-xi-dot-relation}) we find the relation
\begin{equation}
\mathbf{u}\,(\mathbf{x},t)
= \langle\mathbf{U}\,(\mathbf{x}+\boldsymbol\xi,t)\rangle
= \mathbf{\dot{X}}(\mathbf{a},t)\,.
\end{equation}
Thus, the quantity $\mathbf{u}\,(\mathbf{x},t)$ appearing equations in
(\ref{vel-xi-dot-relation}) and (\ref{lag-u-advection}) is the {\bfi
Lagrangian mean fluid velocity} at the position $\mathbf{x} =
\mathbf{X}(\mathbf{a},t)$ along the fluid trajectory with Lagrangian label
$\mathbf{a}$.

We use the basic relations in (\ref{vel-xi}) to transform variables
in the Lagrangian (\ref{Lag-xi}) from labels $\mathbf{a}$, to position
along the mean fluid trajectory $\mathbf{x} = \mathbf{X}(\mathbf{a},t)$,
yielding
\begin{eqnarray}\label{Lag-xi1}
L(\omega) &=& \int D\,d^{\,3}x \bigg\{\frac{1}{2} 
\big|\mathbf{U}(\mathbf{x} + \boldsymbol\xi
(\mathbf{x},t\,;\omega),t)\big|^2 
\nonumber\\
&& \hspace{.7in}
+\, P(\mathbf{x} + \boldsymbol\xi
(\mathbf{x},t\,;\omega),t) 
\Big[\det(\mathbf{X}^{\xi\,\prime}_X)\det(\mathbf{X}^{\,\prime}_a) - 1\Big]
\bigg\}\;,
\end{eqnarray}
where the volume element $D$ satisfies 
$\det(\mathbf{X}^{\,\prime}_a)=D^{-1}(\mathbf{x},t)$. 
Consequently, the time dependence of the product of determinants splits into
two factors,
\begin{equation}\label{det-def}
\det(\mathbf{X}^{\xi\,\prime}_X)\det(\mathbf{X}^{\,\prime}_a)
  = \underbrace{D^{-1}(\mathbf{x},t)}_{\hbox{\large {\bf slow}}}
\
   \underbrace{\det \big(I + \boldsymbol{\nabla\xi}
(\mathbf{x},t\,;\omega)\big)}_{\hbox{\large {\bf fast (or random)}}}\,.
\end{equation}
Hence we may rewrite equation (\ref{Lag-xi1}) for the Lagrangian
$L(\omega)$ in the Eulerian description equivalently as
\medskip\boxeq{3}
\begin{equation}\label{Lag-xi2}\
L(\omega ) = \int d^{\,3}x \left\{\frac{D}{2} 
\big|\mathbf{U} (\mathbf{x} 
+ \boldsymbol\xi (\mathbf{x},t\,;\omega),t)\big|^2 
+ P(\mathbf{x} + \boldsymbol\xi (\mathbf{x},t\,;\omega),t) 
\Big[\det (I + \boldsymbol{\nabla\xi}) - D\Big]\right\}.
\end{equation}
%

  
\subsection{A Taylor series approximation to order $O(|\boldsymbol\xi|^2)$}
\label{T-series-approx} Until this point in the development, we have made no
approximations. Now, though, we shall assume that the magnitude
$|\boldsymbol\xi|$ of the rapid fluctuations is small enough to allow 
meaningful Taylor expansions  of the unapproximated Eulerian velocity
$\mathbf{U}$ and pressure $P$ as,
\begin{eqnarray}\label{u-lag-split}
\mathbf{U} (\mathbf{x} + \boldsymbol\xi (\mathbf{x},t\,;\omega),t) 
&=& \mathbf{u}(\mathbf{x},t) + \frac{d\boldsymbol\xi}{dt}
\\
&=& \mathbf{U}(\mathbf{x},t) 
+ \boldsymbol\xi (\mathbf{x},t\,;\omega)\boldsymbol{\cdot\nabla} 
\mathbf{U}(\mathbf{x},t)
+ O(|\boldsymbol\xi|^2)
\,.\label{Taylor-exp-U}
\end{eqnarray}
Taking averages and setting $\langle\boldsymbol\xi\rangle=0$ and
$\langle{d}\boldsymbol\xi/dt\rangle=0$ in the two expressions for $\mathbf{U}
(\mathbf{x} + \boldsymbol\xi,t)$ in (\ref{u-lag-split}) and (\ref{Taylor-exp-U})
gives
\begin{equation}\label{Taylor-exp-avg-u} 
\langle\mathbf{U}(\mathbf{x} 
+ \boldsymbol\xi (\mathbf{x},t\,;\omega),t)\rangle
\equiv
\mathbf{u}(\mathbf{x},t) 
=
\mathbf{U}(\mathbf{x},t)
+ O(|\boldsymbol\xi|^2)
\,,
\end{equation}
where $\mathbf{u}(\mathbf{x},t)$ is again the Lagrangian mean velocity.
Thus, the slowly varying velocities $\mathbf{u}(\mathbf{x},t)$ and
$\mathbf{U}(\mathbf{x},t)$ differ only at order $O(|\boldsymbol\xi|^2)$. 

A similar Taylor expansion result holds for pressure,
\begin{equation}\label{Taylor-exp-P}
P (\mathbf{x} + \boldsymbol\xi (\mathbf{x},t\,;\omega),t) 
=
P (\mathbf{x},t) 
+ \boldsymbol\xi (\mathbf{x},t\,;\omega)\boldsymbol{\cdot\nabla} 
P(\mathbf{x},t)
+ O(|\boldsymbol\xi|^2)
\,.
\end{equation}
Taking averages then gives a relation for the {\bfi Lagrangian mean pressure},
$p$,
\begin{equation}\label{Taylor-exp-avg-p} 
\langle{P}(\mathbf{x} 
+ \boldsymbol\xi (\mathbf{x},t\,;\omega),t)\rangle
\equiv
{p}(\mathbf{x},t) 
=
P(\mathbf{x},t)
+ O(|\boldsymbol\xi|^2)
\,.
\end{equation}
Hence, the pressures $p(\mathbf{x},t)$ and $P(\mathbf{x},t)$ 
also differ at order $O(|\boldsymbol\xi|^2)$.
We shall discuss the physical meaning of equations (\ref{Taylor-exp-avg-u})
and (\ref{Taylor-exp-avg-p}) further and model their order
$O(|\boldsymbol\xi|^2)$ terms in detail later, in Section
\ref{phys-interp-sec}. Here, we only emphasize that the present approach
ascribes {\it all} of the rapid variation in velocity
$\mathbf{U} (\mathbf{x} + \boldsymbol\xi,t)$ and
pressure $P(\mathbf{x} + \boldsymbol\xi,t)$ to the {\it Lagrangian} trajectory
fluctuation displacement $\boldsymbol\xi (\mathbf{x},t\,;\omega)$; whereas there
could be an additional {\it Eulerian} fluctuation in the functions $\mathbf{U}$
and $P$, themselves, e.g., one could take $\mathbf{U} (\mathbf{x} +
\boldsymbol\xi,t\,;\omega)$. We shall return to the issue of properly
representing the velocity and pressure fluctuations in Section
\ref{phys-interp-sec}, when we compare their Lagrangian and Eulerian
representations.

Taking the difference between equations (\ref{u-lag-split}) and
(\ref{Taylor-exp-U}) and using equation (\ref{Taylor-exp-avg-u}) now yields
an equation for the modulational dynamics of $\boldsymbol\xi
(\mathbf{x},t\,;\omega)$ in slow time, valid to order
$O(|\boldsymbol\xi|^2)$,\bigskip
\boxeq{2}
\begin{equation}\label{xi-eqn}
\frac{d\boldsymbol\xi}{dt}
= \frac{\partial\boldsymbol\xi}{\partial t} 
+ \mathbf{u}(\mathbf{x},t)\boldsymbol{\cdot\nabla} \boldsymbol\xi
= \boldsymbol\xi\boldsymbol{\cdot\nabla} \mathbf{u}(\mathbf{x},t)\,.
\end{equation}
This equation is also equivalent to the vector field commutation relation,
\begin{equation}\label{xi-commutator}
\Big[\frac{d}{dt}\ ,\ \boldsymbol\xi\boldsymbol{\cdot\nabla}\Big]
= \Big[\frac{\partial}{\partial t} 
+ \mathbf{u}(\mathbf{x},t)\,\boldsymbol{\cdot\nabla}\ ,\
\boldsymbol\xi\boldsymbol{\cdot\nabla}\Big] = 0\,.
\end{equation}
Hence, we find the remarkable relation,
\begin{equation}\label{xi-commutator-imp}
\frac{d}{dt}\
\boldsymbol\xi\boldsymbol{\cdot\nabla}\mathbf{A}(\mathbf{x},t) 
= \boldsymbol\xi\boldsymbol{\cdot\nabla}
  \frac{d}{dt}\, \mathbf{A}(\mathbf{x},t)\,,
\end{equation}
for {\it any vector} $\mathbf{A}(\mathbf{x},t)$, provided relation
(\ref{xi-eqn}) holds. We shall see in a moment that this is an exact
relation, provided the fluctuation field $\boldsymbol{\xi}$ in the
decomposition (\ref{traj-xi}) satisfies $\boldsymbol\xi
(\mathbf{X}(\mathbf{a},t),t\,;\omega) =(\boldsymbol{\tilde\xi}
(\mathbf{a}\,;\omega)\boldsymbol{\cdot}
\frac{\partial}{\partial\mathbf{a}})\mathbf{X}(\mathbf{a},t)$, in which
the dependences on fast and slow time variables are {\it separated
(factored)}.
%

  
\subsection{Remarks about advected quantities and Taylor's hypothesis}
We shall show that, geometrically, equation (\ref{xi-eqn}) means that the
slow evolution, or modulation, of the fluctuation vector field
$\boldsymbol\xi(\mathbf{x},t;\omega)$ is invariant under the flow of the
Lagrangian mean velocity, $\mathbf{u}(\mathbf{x},t)$. In other words, this
vector field is an advected quantity, ``frozen'' into the flow of
the Lagrangian mean velocity. This situation is analogous to either the
vorticity stretching equation for the 3D incompressible ideal Euler
equations, or the ratio of magnetic field intensity to mass density in
ideal magnetohydrodynamics,~\cite{HMR[1998a]}. Following the solution
attributed to Cauchy~\cite{Cauchy[1863?]} for the 3D
vorticity stretching equation, the solution of the advection condition
(\ref{xi-eqn}) may be expressed in components in separated form as
\begin{equation}\label{xi-lag}
  \xi^i\,(\mathbf{X}(\mathbf{a},t;\omega),t\,) = 
\underbrace{F^i_A(\mathbf{a},t)}_{\hbox{\large {\bf slow}}}\
\underbrace{\tilde{\xi}^A(\mathbf{a}\,;\omega)}_{\hbox{\large {\bf fast}}}
\,,
\quad\hbox{with}\quad
F^i_A=(\mathbf{X}^{\,\prime}_a)^i_A
=\frac{\partial{X^i(\mathbf{a},t)}}{\partial{a}^A}\,.
\end{equation}
To verify this solution, one may substitute it into
$\partial\tilde{\xi}^A(\mathbf{a}\,;\omega)/\partial{t}|_{\mathbf{a}}=0$, 
where $\tilde{\xi}^A=(F^{-1})^A_i{\xi}^i$ and $(F^{-1})^A_i$ is the matrix
inverse of $F^i_A$, so that $(F^{-1})^A_iF^i_B=\delta^A_B$ and
$F^i_A(F^{-1})^A_j=\delta^i_j$. The existence of these matrix inverse
relations is guaranteed by preservation of fluid volume,
\begin{equation}\label{vol-pres}
\det(\mathbf{X}^{\,\prime}_a) = \det (F^i_A) \equiv
\det{F}=1\,.
\end{equation}
We also note the relation,
\begin{equation}\label{grad-u-lag}
\dot{F}^k_C\
(F^{-1})^C_j (\mathbf{a},t)
=
\frac{\partial{u^k}}{\partial{x^j}}(\mathbf{x},t)
\,, 
\quad{\rm for}\quad
\mathbf{x} = \mathbf{X} (\mathbf{a},t)
\,.
\end{equation}
Rearranging equation (\ref{xi-lag}) implies
\begin{equation}\label{xi-dot-grad}
  \boldsymbol{\xi\cdot\nabla}
=  \xi^i(\mathbf{x},t\,;\omega)\frac{\partial}{\partial x^i}
=  \tilde{\xi}^A(\mathbf{a}\,;\omega)\frac{\partial}{\partial a^A}
= \boldsymbol{\tilde\xi\,\cdot\,}
\frac{\partial}{\partial\mathbf{a}}
\quad{\rm for}\quad
\mathbf{x} = \mathbf{X} (\mathbf{a},t)\,, 
\end{equation}
where $\boldsymbol{\tilde\xi}(\mathbf{a}\,;\omega)$ may be taken as the
initial value in slow time of the fluctuation
$\boldsymbol{\xi}(\mathbf{x},t\,;\omega)$. Thus, along the Lagrangian
mean fluid trajectory the vector field
$\boldsymbol{\xi\cdot\nabla}$ with parameter $\omega$ may be expressed
solely in terms of the Lagrangian labels. This means
$\boldsymbol{\xi\cdot\nabla}$ is an advected quantity,
invariant in slow time along the Lagrangian mean fluid trajectory, just
as indicated by the commutation relation (\ref{xi-commutator}). Furthermore,
to order $O(|\boldsymbol\xi|^2)$ the decomposition (\ref{traj-xi}) may be
expressed using relation (\ref{xi-dot-grad}) as
\begin{equation}\label{traj-xi-approx}
\mathbf{X}^\xi(\mathbf{a},t\,;\omega) 
=\mathbf{X} + \boldsymbol\xi (\mathbf{X},t\,;\omega)
= \Big(1 + 
  \tilde{\xi}^A(\mathbf{a}\,;\omega)\frac{\partial}{\partial a^A}\Big)
\mathbf{X}(\mathbf{a},t) 
\,.
\end{equation}

Conversely, we may show that advection of the Lagrangian vector field
$\boldsymbol{\tilde\xi} (\mathbf{a}\,;\omega)\boldsymbol{\cdot}
\frac{\partial}{\partial\mathbf{a}}$ implies the advection
condition (\ref{xi-eqn}) for the slow time evolution of the Eulerian
vector field $\boldsymbol{\xi\cdot\nabla}$, since (in analogy to the
calculation (\ref{conteqn-der}) for the Lagrangian invariance of the
volume element),
\begin{eqnarray}\label{invar-xi}
0 \ =\ \frac{\partial}{\partial t}\Big|_{\mathbf{a}}\
\bigg(\boldsymbol{\tilde\xi} (\mathbf{a}\,;\omega)\boldsymbol{\cdot}
\frac{\partial}{\partial\mathbf{a}}\bigg)
&=& \bigg(\frac{\partial}{\partial t} + \pounds_{\mathbf{u}}\bigg) 
    (\boldsymbol{\xi\cdot\nabla})
\nonumber\\
  &=& \bigg[
\frac{\partial\boldsymbol\xi}{\partial t} 
+ \mathbf{u}\boldsymbol{\cdot\nabla} \boldsymbol\xi
- \boldsymbol\xi\boldsymbol{\cdot\nabla} \mathbf{u}\bigg]
\boldsymbol{\cdot\nabla}\,,
\end{eqnarray}
where $\pounds_{\mathbf{u}}$ denotes Lie derivative with respect to the
{\it Lagrangian mean} fluid velocity, $\mathbf{u}(\mathbf{x},t)$.  Vanishing
of the term in brackets recovers the Eulerian invariance equation
(\ref{xi-eqn}), which we now understand is equivalent to assuming the
separated form 
$\boldsymbol\xi (\mathbf{X}(\mathbf{a},t),t\,;\omega)
=(\boldsymbol{\tilde\xi} (\mathbf{a}\,;\omega)\boldsymbol{\cdot}
\frac{\partial}{\partial\mathbf{a}})\mathbf{X}(\mathbf{a},t)$ in the
decomposition (\ref{traj-xi}). 

\paragraph{\bf Remark.} The advection of the fluctuating
vector field $\boldsymbol{\xi\cdot\nabla}$ expressed in relation
(\ref{xi-eqn}), or its equivalent commutator form (\ref{invar-xi}) is an
extended form of the {\bfi Taylor hypothesis}, i.e., that small rapid
fluctuation fields are swept downstream with the mean flow
\cite{Taylor[1938]},~\cite{Hinze[1975]}. Relation (\ref{xi-eqn}) extends
the traditional interpretation of Taylor's hypothesis to apply to advection
of the rapid fluctuations of a vector quantity by the Lagrangian mean flow.
The Taylor hypothesis is usually interpreted
\cite{Hinze[1975]},~\cite{Dahm[1997]} as meaning that turbulent fluctuations
satisfy $\partial/\partial{t}=-\bar{U}\partial/\partial{x}$ for a
sufficiently short time interval, where $\bar{U}$ is the (constant) Eulerian
mean flow velocity in the $x$-direction. The extended form of Taylor's
hypothesis we introduce here asserts that the advection relation (\ref{xi-eqn})
holds along each mean Lagrangian fluid trajectory and accounts for the vector
nature of the displacement fluctuation, $\boldsymbol\xi$. According to its
derivation in the steps leading to equation (\ref{xi-eqn}), this extended
Taylor's hypothesis is valid to order $O(|\boldsymbol\xi|^2)$. When
$\mathbf{u}$ is replaced by $(\bar{U},0,0)$ with constant $\bar{U}$ the
advection relation (\ref{xi-eqn}) reduces to the traditional form of
Taylor's hypothesis, which is usually applied to scalar
quantities,~\cite{Dahm[1997]}.

  
\subsection{Averaged approximate Lagrangians} 
After the approximations leading to the Taylor hypothesis relation
(\ref{xi-eqn}) with its geometrical interpretation (\ref{invar-xi}), our
Lagrangian $L(\omega)$ in (\ref{Lag-xi2}) becomes
\begin{eqnarray}\label{Lag-xi3}
L(\omega) &=& \int d^{\,3}x\ \bigg\{ D\ \frac{1}{2} 
\big|\big(1+\boldsymbol\xi (\mathbf{x},t\,;\omega) 
\boldsymbol{\cdot\nabla}\big)
\mathbf{u}(\mathbf{x},t)\big|^2
\nonumber\\
&&+\  \Big[\big(1+\boldsymbol\xi (\mathbf{x},t\,;\omega) 
\boldsymbol{\cdot\nabla}\big)p(\mathbf{x},t) \Big] 
\Big[\det (I + \boldsymbol{\nabla\xi}) - D(\mathbf{x},t)\Big]\bigg\}\;,
\end{eqnarray}
in which $p$ is the Lagrangian mean pressure.
Using relation (\ref{xi-dot-grad}) transforms this formula to the Lagrangian
picture as, cf. equations (\ref{Lag-Lag}) and (\ref{Lag-xi}),
\begin{eqnarray}\label{Lag-xi4}
L(\omega) &=& \int d^{\,3}a\ \bigg\{\frac{1}{2} 
\bigg|\bigg(1+\tilde{\xi}^A(\mathbf{a}\,;\omega)
\frac{\partial}{\partial a^A}\bigg)
\mathbf{\dot{X}}(\mathbf{a},t)\bigg|^2 
\\
&&+\ \bigg[\bigg(1+\tilde{\xi}^A(\mathbf{a}\,;\omega)
\frac{\partial}{\partial a^A}\bigg)\,
p(\mathbf{X}(\mathbf{a},t),t)\bigg]
\bigg[\det(\mathbf{X}^{\xi\,\prime}_a)(\mathbf{a},t\,;\omega) - 1
\bigg]\bigg\}.
\nonumber
\end{eqnarray}
Therefore, we may take the Lagrangian mean of this approximate form of 
$L(\omega)$ to find the following averaged approximate Lagrangian (using
$\langle\boldsymbol{\tilde\xi}\,\rangle=0$)
\begin{eqnarray}\label{mean-Lag-Lag}
\langle L\rangle &=& \int d^{\,3}a \bigg\{\frac{1}{2}
\Big[\,\big|\mathbf{\dot{X}}(\mathbf{a},t)\big|^2  +
\langle\tilde{\xi}^A\tilde{\xi}^B\rangle 
\Big(\frac{\partial\mathbf{\dot{X}}}{\partial a^A}\boldsymbol\cdot
\frac{\partial\mathbf{\dot{X}}}{\partial a^B}\Big)\Big] 
\\
&&\quad+\  p(\mathbf{X}(\mathbf{a},t),t) 
\Big[\Big\langle\det\Big(I + \boldsymbol
{\frac{\partial\xi}{\partial\mathbf{X}}}\Big)\Big\rangle
\det(\mathbf{X}^{\,\prime}_a) - 1\Big] 
\nonumber\\
&&\quad +\ \det(\mathbf{X}^{\,\prime}_a)
\Big(\Big\langle\boldsymbol\xi \det\Big(I + \boldsymbol
{\frac{\partial\xi}{\partial\mathbf{X}}}\Big)\Big\rangle
\boldsymbol
{\cdot\,\frac{\partial}{\partial\mathbf{X}}}
\Big) p(\mathbf{X}(\mathbf{a},t),t) 
\bigg\}.\nonumber
\end{eqnarray}
In the Eulerian description, this averaged approximate Lagrangian becomes
\begin{eqnarray}\label{mean-Lag-Eul}
\langle L\rangle &=& \int d^{\,3}x \bigg\{\frac{D}{2} \Big[|\mathbf{u}\,|^2 
+ \langle\xi^k\xi^l\rangle 
\Big(\mathbf{u}_{,k}\boldsymbol\cdot\mathbf{u}_{,l}\Big)\Big] 
\\
&&+\  p \Big[\langle\det(I + \boldsymbol{\nabla\xi})\rangle - D\Big] 
+ \Big(\langle\boldsymbol\xi \det(I + \boldsymbol{\nabla\xi})
\rangle\boldsymbol{\cdot\nabla}\Big) p \bigg\},\nonumber
\end{eqnarray}
where partial spatial derivatives are denoted by subscript comma, e.g.,
$\mathbf{u}_{,k} = \partial\mathbf{u}/\partial x^k =
\partial_k\mathbf{u}$, for
$k = 1,2,3$.  In preparation for using the Euler-Poincar\'e theorem, we
compute the following  the following variational derivatives of the
averaged approximate Lagrangian,
\begin{eqnarray}\label{mean-Lag-der}
\frac{1}{D} \frac{\delta\langle L\rangle }{\delta \mathbf{u}} 
&=& \mathbf{u} 
 - \frac{1}{D}\Big( \partial_k\,
D\langle\xi^k\xi^l\rangle \partial_l \Big) \mathbf{u},
\nonumber\\
\frac{\delta\langle L\rangle }{\delta D} 
&=& - p  
+ \frac{1}{2}|\mathbf{u}\,|^2 
+ \frac{1}{2}\langle\xi^k\xi^l\rangle 
\Big(\mathbf{u}_{,k}\boldsymbol\cdot\mathbf{u}_{,l}\Big)
\equiv -P_{tot}, 
\\
\frac{\delta\langle L\rangle }{\delta p} 
&=& \langle\det (I + \boldsymbol{\nabla\xi})\rangle - D 
- \boldsymbol{\nabla\,\cdot}\,
  \langle\boldsymbol\xi \det(I + \boldsymbol{\nabla\xi})\rangle\;.
\nonumber
\end{eqnarray}
The natural boundary condition $\mathbf{u}\boldsymbol\cdot\mathbf{\hat{n}}=0$ at
a fixed boundary with unit normal vector $\mathbf{\hat{n}}$ is imposed on the
Lagrangian mean velocity in the course of deriving the Euler-Poincar\'e
equations. In addition, when taking the variations in (\ref{mean-Lag-der}),
we assume in integrating by parts that the fluctuations do not penetrate a
fixed boundary, so that
$\boldsymbol{\xi\cdot}\mathbf{\hat{n}}=0$ must be satisfied at such a
boundary. Consequently, we have
$\hat{n}_k\langle\xi^k\xi^l\rangle = 0$ at fixed boundaries, as well. We
shall discuss variations with respect to $\langle\xi^k\xi^l\rangle$ after
making further approximations in $\langle L\rangle$.

  
\subsection{Further approximations} 
Stationarity of the averaged approximate Lagrangian $\langle L\rangle$
in (\ref{mean-Lag-Eul}) and the last equation in (\ref{mean-Lag-der}) imply
\begin{equation}\label{Lag-der-D}
D = \langle \det(I + \boldsymbol{\nabla\xi})\rangle 
- \boldsymbol{\nabla\,\cdot}\, \langle\boldsymbol\xi 
\det(I + \boldsymbol{\nabla\xi})\rangle\;.
\end{equation}
We shall approximate this relation using the identity,
\begin{equation}\label{det-id}
\langle\det (I + \boldsymbol{\nabla\xi})\rangle 
= 1 
+ \boldsymbol{\nabla\cdot} \langle\boldsymbol\xi\rangle  
+ \frac{1}{2}\boldsymbol{\nabla\cdot}
  \langle\boldsymbol{\xi\nabla\cdot\xi}
- \boldsymbol{\xi\cdot\nabla\xi}\rangle
+\langle\det(\boldsymbol{\nabla\xi})\rangle\;.
\end{equation}
Since we assume $|\boldsymbol\xi|$ to be small (at least typically, or in an
average  sense), the relation (\ref{Lag-der-D}) can be given a simple and
concise approximate form by neglecting terms of order
$O(|\boldsymbol\xi|^3)$, or higher, involving
$\langle\det(\boldsymbol{\nabla\xi})\rangle$ and
$\boldsymbol{\nabla\cdot}\langle
\boldsymbol{\xi}\det(\boldsymbol{\nabla\xi})\rangle$.
Namely,
\smallskip\boxeq{2}
\begin{equation}\label{Lag-der-D-approx}
D = 
1 - \frac{1}{2}\boldsymbol{\nabla\cdot}\big[
\langle\boldsymbol{\xi\cdot\nabla\xi}\rangle
+ 
\langle
\boldsymbol\xi
(\boldsymbol{\nabla\cdot\xi})\rangle\big]
= 1 - \frac{1}{2}\langle\xi^k\xi^l\rangle_{,k\,l}\;.
\end{equation}
This order $O(|\boldsymbol\xi|^2)$ compressibility due to the 
presence of the fluctuations is consistent with the findings of 
generalized Lagrangian mean (GLM)
theory,~\cite{Andrews-McIntyre[1978a]},
\cite{Andrews-McIntyre[1978b]},~\cite{GH[1996]}.

Upon substituting the order $O(|\boldsymbol\xi|^2)$ approximations yielding
relation (\ref{Lag-der-D-approx}) into the Lagrangian (\ref{mean-Lag-Lag}),
we find the following {\bfi averaged approximate Lagrangian}, 
\smallskip\boxeq{2}
\begin{equation}\label{mean-Lag-approx}
\langle L\rangle = \int d^{\,3}x \bigg\{\frac{D}{2} \Big[|\mathbf{u}\,|^2 
+ \langle\xi^k\xi^l\rangle 
\Big(\mathbf{u}_{,k}\boldsymbol\cdot\mathbf{u}_{,l}\Big)\Big] 
+\  p \Big[1 - D
- \frac{1}{2}\langle\xi^k\xi^l\rangle_{,k\,l}\Big]\bigg\}.
\end{equation}
This approximate $\langle L\rangle$ returns to the Lagrangian picture as,
cf. equations (\ref{Lag-Lag}), (\ref{Lag-xi}) and (\ref{mean-Lag-Lag}),
\smallskip\boxeq{7}
\begin{eqnarray}\label{mean-Lag-Lag-approx}
\langle L\rangle &=& \int d^{\,3}a \bigg\{\frac{1}{2}
\Big[\,\big|\mathbf{\dot{X}}(\mathbf{a},t)\big|^2  
+
\langle\tilde{\xi}^A\tilde{\xi}^B\rangle 
\Big(\frac{\partial\mathbf{\dot{X}}}{\partial a^A}\boldsymbol\cdot
\frac{\partial\mathbf{\dot{X}}}{\partial a^B}\Big)\Big] 
\nonumber\\
&&+\ 
p(\mathbf{X}(\mathbf{a},t),t) 
\Big[\det(\mathbf{X}^{\,\prime}_a) - 1\Big]
\\
&&-\ 
\frac{1}{2}
\det(\mathbf{X}^{\,\prime}_a)
\langle\tilde\xi^A\tilde\xi^B\rangle
\bigg[
\frac{\partial^2 p}{\partial a^A\partial a^B}
\,-
\frac{\partial^2 X^i}{\partial a^A\partial a^B}
\big(F^{-1}\big)^C_i
\frac{\partial p}{\partial a^C} 
\bigg]
\bigg\}.\nonumber
\end{eqnarray}
\smallskip

The two equivalent forms (\ref{mean-Lag-approx}) and
(\ref{mean-Lag-Lag-approx}) of the averaged approximate Lagrangian possess
order $O(|\boldsymbol\xi|^2)$ corrections involving higher order derivatives
of the velocity, or pressure, that are contracted using
the metric provided by the Lagrangian mean
covariance of the fluctuations, either in its Eulerian form,
$\langle\xi^k\xi^l\rangle$, or in its Lagrangian form,
$\langle\tilde{\xi}^A\tilde{\xi}^B\rangle$. According to relation
(\ref{xi-lag}), these two representations of the covariance are related by
\begin{equation}\label{metric-transf}
\langle\tilde{\xi}^A\tilde{\xi}^B\rangle(\mathbf{a}) 
= \langle\xi^k\xi^l\rangle(\mathbf{X},t)
(F^{-1})^A_k(F^{-1})^B_l 
\,.
\end{equation}
This relation and the {\bfi Piola identity},
\begin{equation}\label{Piola-id}
\frac{\partial}{\partial a^A}\Big(\det{F}\big(F^{-1}\big)^A_k\Big) = 0\,,
\end{equation}
provide the transformation laws required in converting between the two
expressions (\ref{mean-Lag-approx}) and (\ref{mean-Lag-Lag-approx}) for the
averaged approximate Lagrangian.

\paragraph{Remark on the covariance determinant.}
Equation (\ref{metric-transf}) for the transformation of the covariance
tensor $\boldsymbol{\langle\tilde{\xi}\tilde{\xi}\rangle}$ as it
advects along a Lagrangian fluid trajectory also implies the following   
relation for its determinant,
\begin{equation}\label{det-rel}
\det\boldsymbol{
\langle\tilde{\xi}\tilde{\xi}\rangle}(\mathbf{a}) 
= 
\frac{
\det\boldsymbol{\langle\xi\xi\rangle}(\mathbf{X},t)}
{(\det F)^2}
\,.
\end{equation}
Consequently, the product $D^2\det\boldsymbol{\langle\xi\xi\rangle}$
is {\it preserved} along fluid trajectories, 
\boxeq{2}
\begin{equation}\label{cons-det}
\frac{d}{dt}(D^2
\det\boldsymbol{\langle\xi\xi\rangle})
= 0\,.
\end{equation}
Therefore, the covariance of the fluctuations cannot vanish in the course of
the motion, provided it is initially nonzero. Thus, one may regard the
principle axes of the symmetric tensor $\boldsymbol{\langle\xi\xi\rangle}$ as
describing an ellipsoid, that is carried along with each fluid parcel and
represents an additional ``internal'' degree of freedom for the fluid,
associated with the Lagrangian mean covariance of the advecting fluctuations.
This covariance ellipsoid may deform and change orientation as it follows the
course of the fluid motion, but its volume (the determinant
$\det\boldsymbol{\langle\xi\xi\rangle}$) cannot either vanish, or become
unbounded, as long as the fluid density $D$ is finite.

\paragraph{Eulerian representation of the covariance dynamics.}
The time derivative of the relation (\ref{metric-transf}) and the
basic relation (\ref{grad-u-lag}) imply
\begin{equation}\label{covariance-dynamics-calc}
0
=
\frac{\partial}{\partial{t}}\Big|_{\mathbf{a}}
\langle\tilde{\xi}^A\tilde{\xi}^B\rangle(\mathbf{a}) 
=
\Big(
\frac{d}{dt}\langle\xi^k\xi^l\rangle
- 
\langle\xi^k\xi^j\rangle{u}^l_{,j}
-
u^k_{,j}\langle\xi^j\xi^l\rangle
\Big)
(F^{-1})^A_k(F^{-1})^B_l
\,.
\end{equation}
Hence, the Eulerian dynamics for the Lagrangian mean covariance
$\langle\xi^k\xi^l\rangle$ may be expressed as
\boxeq{2}\smallskip
\begin{equation}\label{covariance-dynamics-index}
\frac{d}{dt}\langle\xi^k\xi^l\rangle
= 
\langle\xi^k\xi^j\rangle{u}^l_{,j}
+
u^k_{,j}\langle\xi^j\xi^l\rangle
\,,
\end{equation}
or, in vector form,
\smallskip

\noindent
\boxeq{2}
\begin{equation}\label{covariance-dynamics-vector}
\frac{d}{dt}\boldsymbol{\langle\xi\xi\rangle}
= \boldsymbol{\langle\xi\xi\rangle\cdot\nabla\mathbf{u}}
+ \boldsymbol{\nabla\mathbf{u}}^{\rm T}
\boldsymbol{\cdot\langle\xi\xi\rangle}
\,.
\end{equation}

  
\subsection{The Euler-Lagrange equations with order
$O(|\boldsymbol\xi|^2)$ compressibility}

The Euler-Lagrange equations of the averaged approximate Lagrangian
(\ref{mean-Lag-Lag-approx}),
\begin{equation}\label{Eul-Lag-eqn-approx}
\frac{\partial}{\partial t}\bigg|_{\mathbf{a}}
\frac{\delta\langle L\rangle}{\delta\mathbf{\dot{X}}}
- \frac{\delta\langle L\rangle}{\delta\mathbf{X}} = 0
\,,
\end{equation}
are given by (dropping endpoint contributions),
\begin{equation}\label{Eul-Lag.eqn-approx}
\Big(1-\frac{\partial}{\partial a^A}
\langle\tilde{\xi}^A\tilde{\xi}^B\rangle
\frac{\partial}{\partial a^B}\Big)\boldsymbol{\ddot\mathbf{X}} 
- \frac{\delta\langle L\rangle}{\delta\mathbf{X}} = 0
\,.
\end{equation}
We shall defer our discussion of $\delta\langle L\rangle/\delta\mathbf{X}$
until later. Stationarity of $\langle L\rangle$ in (\ref{mean-Lag-Lag-approx})
under variations in $p$ implies  order $O(|\boldsymbol\xi|^2)$ compressibility, 
\begin{equation}\label{det-F-comp}
\det{F} = 1 
+ 
\frac{1}{2}\Big(\det{F}
\langle\tilde{\xi}^A\tilde{\xi}^B\rangle
\Big)_{,AB}
+ 
\frac{1}{2}\det{F}
\big(F^{-1}\big)^C_i
\frac{\partial}{\partial a^C}
\Big(
\langle\tilde{\xi}^A\tilde{\xi}^B\rangle
X^i_{,AB}
\Big)
\,.
\end{equation}
Equivalently, by using equations (\ref{metric-transf}) and
(\ref{Piola-id})  we find the expected relation,
\begin{equation}\label{det-D-comp}
1 = D + 
\frac{1}{2}\langle\xi^k\xi^l\rangle_{,kl}
\,.
\end{equation}
To write the Eulerian form of the motion equation (\ref{Eul-Lag.eqn-approx}) we
must transform the operator,
\begin{equation}\label{Delta-op-def}
\tilde\Delta\equiv\frac{\partial}{\partial a^A}
\langle\tilde{\xi}^A\tilde{\xi}^B\rangle
\frac{\partial}{\partial a^B}\,,
\end{equation}
from the Lagrangian, to the Eulerian description. For this, we
use the metric transformation formula (\ref{metric-transf}), the
definition (\ref{vol-pres}) and the Piola identity (\ref{Piola-id})
to find the {\bfi transformation law} for $\tilde\Delta$,
\begin{eqnarray}\label{Delta-op-trans}
\tilde\Delta
&\equiv&
\frac{\partial}{\partial a^A}
\langle\tilde{\xi}^A\tilde{\xi}^B\rangle
\frac{\partial}{\partial a^B}
= 
\frac{\partial}{\partial a^A}
\langle\xi^k\xi^l\rangle
(F^{-1})^A_k(F^{-1})^B_l
\frac{\partial}{\partial a^B}
\nonumber\\
&=&
\det{F}(F^{-1})^A_k\frac{\partial}{\partial a^A}
\frac{\langle\xi^k\xi^l\rangle}{\det{F}}
(F^{-1})^B_l\frac{\partial}{\partial a^B}
\nonumber\\
&=&
\det{F} \frac{\partial}{\partial x^k}
\frac{\langle\xi^k\xi^l\rangle}{\det{F}}
\frac{\partial}{\partial x^l}
\ =\
D^{-1}\frac{\partial}{\partial x^k}\
D\, \langle\xi^k\xi^l\rangle
\frac{\partial}{\partial x^l}
\equiv\tilde\Delta_D
\,.
\end{eqnarray}
Consequently, we also have the remarkable commutation relation between the
operators $d/dt$ and $\tilde\Delta_D$,
\begin{equation}\label{remarkable-comm-rel}
\frac{ d}{dt}\Big(D^{-1}\frac{\partial}{\partial x^k}\
D\, \langle\xi^k\xi^l\rangle
\frac{\partial}{\partial x^l}\Big)
 - \Big(D^{-1}\frac{\partial}{\partial x^k}\
D\, \langle\xi^k\xi^l\rangle
\frac{\partial}{\partial x^l}\Big)\frac{ d}{dt}
\equiv\bigg[\frac{d}{dt},\tilde\Delta_D\bigg]=0\,.
\end{equation}
This relation allows us to transform the acceleration term in equation
(\ref{Eul-Lag.eqn-approx}) from the Lagrangian, to the Eulerian description, as
\begin{equation}\label{Eul-Lag.eqn-approx-Euler-form1}
\Big(\frac{\partial}{\partial t} 
+ \mathbf{u}\boldsymbol{\,\cdot\nabla}\Big)
\Big(1-\tilde\Delta_D\Big)\mathbf{u} 
- \frac{\delta\langle L\rangle}{\delta\mathbf{X}}
= 0
\,,
\end{equation}
or, equivalently,
\begin{equation}\label{Eul-Lag.eqn-approx-Euler-form2}
\Big(1-\tilde\Delta_D\Big)
\Big(\frac{\partial}{\partial t} 
+ \mathbf{u}\boldsymbol{\,\cdot\nabla}\Big)\mathbf{u} 
- \frac{\delta\langle L\rangle}{\delta\mathbf{X}}
= 0\,.
\end{equation}
In both forms of this motion equation, the divergence of the Lagrangian mean
velocity is given by
\begin{equation}\label{div-u-Lag-2nd-order}
\boldsymbol{\nabla\cdot}\mathbf{u}
=
-\,\frac{1}{D}\frac{dD}{dt}
=
\frac{1}{2-\langle\xi^k\xi^l\rangle_{,kl}}\
\frac{d}{dt}\
\langle\xi^k\xi^l\rangle_{,kl}
=
\frac{1}{2}\frac{d}{dt}\
\langle\xi^k\xi^l\rangle_{,kl}
+
O(|\boldsymbol\xi|^4)
\,.
\end{equation}
In its first form (\ref{Eul-Lag.eqn-approx-Euler-form1}) the presence of the
operator $(1-\tilde\Delta_D)$ in this motion equation smoothes the Lagrangian
mean transport velocity $\mathbf{u}$ relative to the momentum, or circulation
velocity $\mathbf{v}=(1-\tilde\Delta_D)\mathbf{u}$. In its second form
(\ref{Eul-Lag.eqn-approx-Euler-form2}) the operator $(1-\tilde\Delta_D)$ acts to
smooth the generalized force $\delta\langle L\rangle/\delta\mathbf{X}$ in an
adaptive fashion depending on the Lagrangian mean covariance
$\boldsymbol{\langle\xi\xi\rangle}$, which in turn depends on the velocity
shear $\boldsymbol{\nabla\mathbf{u}}$. At the moment, the generalized force
$\delta\langle L\rangle/\delta\mathbf{X}$ is still a Lagrangian quantity. We
shall see in Section \ref{2pt Eul closure-sec} how to complete the expression of
these equations by computing $\delta\langle L\rangle/\delta\mathbf{X}$ in purely
Eulerian form in terms of a fluctuation stress tensor.

Since the averaged Lagrangian in equation (\ref{mean-Lag-Lag-approx}) has no
explicit time dependence, Noether's theorem implies {\bfi conservation of
energy},
\begin{eqnarray}\label{erg-approx-lag}
E &=& 
\frac{1}{2}\int d^{\,3}a\bigg\{ 
\Big|\mathbf{\dot{X}}(\mathbf{a},t)\Big|^2 
+ \langle\tilde{\xi}^A\tilde{\xi}^B\rangle
\frac{\partial\mathbf{\dot{X}}}{\partial a^A}
\boldsymbol\cdot
\frac{\partial\mathbf{\dot{X}}}{\partial a^B}\bigg\}\,,
\nonumber\\
&=&
\frac{1}{2}\int d^{\,3}a\bigg\{ 
\Big|\mathbf{\dot{X}}(\mathbf{a},t)\Big|^2 
+ \langle\xi^k\xi^l\rangle(\mathbf{X},t)
\big(F^{-1}\big)^A_k\big(F^{-1}\big)^B_l
\frac{\partial\mathbf{\dot{X}}}{\partial a^A}
\boldsymbol\cdot
\frac{\partial\mathbf{\dot{X}}}{\partial a^B}\bigg\}\,,
\nonumber\\
&=&
\frac{1}{2} \int d^{\,3}x \ D \Big[\ |\mathbf{u}\,|^2 
+ \langle\xi^k\xi^l\rangle 
\Big(\mathbf{u}_{,k}\boldsymbol\cdot\mathbf{u}_{,l}\Big)\,\Big]\,,
\nonumber\\
&=&
\frac{1}{2} \int d^{\,3}x \ 
\Big(1-\frac{1}{2}\langle\xi^i\xi^j\rangle_{,ij}\Big) 
\Big[\ |\mathbf{u}\,|^2 
+ \langle\xi^k\xi^l\rangle 
\Big(\mathbf{u}_{,k}\boldsymbol\cdot\mathbf{u}_{,l}\Big)\,\Big]\,,
\end{eqnarray}
where we have used the advection solution (\ref{metric-transf}) for
the Lagrangian mean covariance in transforming between the first and second
lines, and then transformed into the Eulerian description in the last line, cf.
equation (\ref{mean-Lag-approx}). The conserved energy $E$ in
(\ref{erg-approx-lag}) is the total kinetic energy of the Lagrangian mean
motion equation (\ref{Eul-Lag.eqn-approx}), including the mean contribution
from the covariance of the rapid (or random) fluctuations. We note that $E$
contains a term of order $O(|\boldsymbol\xi|^4)$.

  
\subsection{Restoring incompressibility in the order
$O(|\boldsymbol\xi|^2)$ model}

We now simplify matters by restoring incompressibility of the
Lagrangian mean fluid velocity, thereby producing a {\bfi simplified
averaged approximate Lagrangian}, cf. equation (\ref{mean-Lag-approx})
\begin{equation}\label{mean-Lag-approx-inc}
\langle L\rangle = \int d^{\,3}x \bigg\{\frac{D}{2} \Big[|\mathbf{u}\,|^2 
+ \langle\xi^k\xi^l\rangle 
\Big(\mathbf{u}_{,k}\boldsymbol\cdot\mathbf{u}_{,l}\Big)\Big] 
+\  p \Big[1 - D\Big]\bigg\}.
\end{equation}
Here, to regain some simplicity in the analysis below, we dropped terms
in the pressure constraint (\ref{Lag-der-D-approx}) of order
$O(|\boldsymbol\xi|^2)$, but kept terms of the {\it same order} in the kinetic
energy piece of the mean Lagrangian in (\ref{mean-Lag-Eul}). The approximate
$\langle L\rangle$ in (\ref{mean-Lag-approx-inc}) so obtained returns to the
Lagrangian picture as, cf. equations (\ref{Lag-Lag}), (\ref{Lag-xi}),
(\ref{mean-Lag-Lag}) and (\ref{mean-Lag-Lag-approx}),
\smallskip\boxeq{5}
\begin{eqnarray}\label{mean-Lag-Lag-approx-inc}
\langle L\rangle = \int d^{\,3}a \bigg\{\frac{1}{2}
\Big[\,\big|\mathbf{\dot{X}}(\mathbf{a},t)\big|^2  
\!\!\!&+&\!\!\!
\langle\tilde{\xi}^A\tilde{\xi}^B\rangle 
\Big(\frac{\partial\mathbf{\dot{X}}}{\partial a^A}\boldsymbol\cdot
\frac{\partial\mathbf{\dot{X}}}{\partial a^B}\Big)\Big] 
\\
\!\!\!&+&\!\!\!\  p(\mathbf{X}(\mathbf{a},t),t) 
\Big[\det(\mathbf{X}^{\,\prime}_a) - 1\Big] 
\bigg\}.\nonumber
\end{eqnarray}
\smallskip

\noindent
The particular averaged approximate Lagrangian
$\langle L\rangle$ in (\ref{mean-Lag-Lag-approx}) also could have been
obtained by substituting the approximate decomposition in equation
(\ref{traj-xi-approx}) directly into the kinetic energy term of the original
Lagrangian (\ref{Lag-Lag}) and then averaging.

Relative to the starting Lagrangian (\ref{Lag-Lag}) for Euler's
incompressible fluid equations, our procedure of decomposition,
approximation and averaging, followed by restoring incompressibility of
$\mathbf{u}$ has merely introduced an additional term into the kinetic
energy of the averaged approximate Lagrangian
$\langle L\rangle$ in (\ref{mean-Lag-Lag-approx}). This additional term
involves higher order derivatives of the velocity, that are contracted using
the metric defined by the pullback of the advected Lagrangian mean
covariance of the fluctuations, namely,
$\langle\tilde{\xi}^A\tilde{\xi}^B\rangle$, as given in equation
(\ref{metric-transf}).


\subsection{The Euler-Lagrange equations for the LMM model}

The Euler-Lagrange equations (\ref{Eul-Lag-eqn}) for the simplified averaged
approximate Lagrangian (\ref{mean-Lag-Lag-approx-inc}) are
immediately obtained as
\boxeq{3}
\begin{equation}\label{Eul-Lag.eqn-approx1}
\Big(1-\frac{\partial}{\partial a^A}
\langle\tilde{\xi}^A\tilde{\xi}^B\rangle
\frac{\partial}{\partial a^B}\Big)\boldsymbol{\ddot\mathbf{X}} 
\ +\ \det(\mathbf{X}^{\,\prime}_a)\
\boldsymbol{\nabla}p\
= 0\,,
\end{equation}
and stationarity under variations in $p$
implies volume preservation, $\det(\mathbf{X}^{\,\prime}_a) = 1$.
To write the Eulerian form of this equation we must transform the
operator $\tilde\Delta$ as done in equation (\ref{Delta-op-trans}).
When volume preservation is imposed, this gives
\begin{equation}\label{Delta-op-id}
\tilde\Delta
=
\frac{\partial}{\partial a^A}
\langle\tilde{\xi}^A\tilde{\xi}^B\rangle
\frac{\partial}{\partial a^B}
= 
\frac{\partial}{\partial x^k}\
\, \langle\xi^k\xi^l\rangle
\frac{\partial}{\partial x^l}
= \tilde\Delta_D\Big|_{D=1}
\quad\hbox{for}\quad 
D=1
\,.
\end{equation}
This relation allows us to transform equation (\ref{Eul-Lag.eqn-approx})
from the Lagrangian, to the Eulerian description, as
\boxeq{6}
\begin{equation}\label{LMM-mot-Euler-form1}
\Big(\frac{\partial}{\partial t} 
+ \mathbf{u}\boldsymbol{\,\cdot\nabla}\Big)
\Big(1-\tilde\Delta\Big)\mathbf{u} 
\ +\ 
\boldsymbol{\nabla}\,p\
= 0
\,,\quad\hbox{with}\quad
\boldsymbol{\nabla\cdot}\mathbf{u} = 0 \,,
\end{equation}
$\quad$or, equivalently,
\begin{equation}\label{LMM-mot-Euler-form2}
\Big(1-\tilde\Delta\Big)
\Big(\frac{\partial}{\partial t} 
+ \mathbf{u}\boldsymbol{\,\cdot\nabla}\Big)\mathbf{u} 
\ +\ 
\boldsymbol{\nabla}\,p\
= 0
\,,\quad\hbox{with}\quad
\boldsymbol{\nabla\cdot}\mathbf{u} = 0\,.
\end{equation}
These are two equivalent forms of the {\bfi Lagrangian mean motion (LMM)
equation}. In its first form, the presence of the Helmholtz operator
$(1-\tilde\Delta)$ in this motion equation smoothes the Lagrangian mean
transport velocity $\mathbf{u}$ relative to the momentum, or circulation
velocity $\mathbf{v}=(1-\tilde\Delta)\mathbf{u}$. In its second form, the
Helmholtz operator $(1-\tilde\Delta)$ acts to smooth the pressure gradient in
an adaptive fashion depending on the covariance
$\boldsymbol{\langle\xi\xi\rangle}$, which in turn depends on the velocity
shear, $\boldsymbol{\nabla\mathbf{u}}$.

Since the averaged Lagrangian in equation (\ref{mean-Lag-Lag-approx-inc})
has no explicit time dependence, Noether's theorem implies {\bfi
conservation of energy}, cf. equation (\ref{erg-approx-lag})
\begin{eqnarray}\label{erg-approx-lag-inc}
E &=& 
\frac{1}{2}\int d^{\,3}a\bigg\{ 
\Big|\mathbf{\dot{X}}(\mathbf{a},t)\Big|^2 
+ \langle\tilde{\xi}^A\tilde{\xi}^B\rangle
\frac{\partial\mathbf{\dot{X}}}{\partial a^A}
\boldsymbol\cdot
\frac{\partial\mathbf{\dot{X}}}{\partial a^B}\bigg\}\,,
\nonumber\\
&=&
\frac{1}{2}\int d^{\,3}a\bigg\{ 
\Big|\mathbf{\dot{X}}(\mathbf{a},t)\Big|^2 
+ \langle\xi^k\xi^l\rangle(\mathbf{X},t)
\big(F^{-1}\big)^A_k\big(F^{-1}\big)^B_l
\frac{\partial\mathbf{\dot{X}}}{\partial a^A}
\boldsymbol\cdot
\frac{\partial\mathbf{\dot{X}}}{\partial a^B}\bigg\}\,,
\nonumber\\
&=&
\frac{1}{2} \int d^{\,3}x \ D \Big[\ |\mathbf{u}\,|^2 
+ \langle\xi^k\xi^l\rangle 
\Big(\mathbf{u}_{,k}\boldsymbol\cdot\mathbf{u}_{,l}\Big)\,\Big]
\,,
\end{eqnarray}
and $D=1$ for the incompressible LMM case.

  
\section{Physical interpretations of $\mathbf{u}$ and $\mathbf{v}$
as the Lagrangian and Eulerian mean fluid velocities}
\label{phys-interp-sec}

In this section we use the extended Taylor hypothesis
(\ref{xi-eqn}) to interpret the velocities $\mathbf{u}$ and $\mathbf{v}$   
in the Lagrangian mean motion (LMM) model by comparing the results of Section
\ref{avg-L-sec} with the results of the traditional
Reynolds fluid velocity decomposition. 

The interpretations are as follows: $\mathbf{u}=\langle\mathbf{U}\rangle^L$ is
{\it defined} as the Lagrangian mean fluid velocity (denoted as
$\langle\mathbf{U}\rangle^L$ in this section {\it only}) and
$\mathbf{v}=(1-\tilde\Delta_D)\mathbf{u}$ is {\it approximately} the Eulerian
mean fluid velocity $\langle\mathbf{U}\rangle^E$. Here, $\tilde\Delta_D$ defined
in equation (\ref{Delta-op-trans}) is the Laplacian operator whose
metric is given by the Lagrangian mean displacement covariance
$\langle\xi^k\xi^l\rangle$ for the advected fluctuations.  The difference
$\mathbf{u} - \mathbf{v} \equiv \langle\mathbf{U}\rangle^S$ is called the
{\bfi Stokes mean drift velocity}. We shall show that
$\langle\mathbf{U}\rangle^S$ satisfies the closure relation
$\langle\mathbf{U}\rangle^S=\tilde\Delta_D\mathbf{u}$. Thus, we shall identify
$\mathbf{v}=\mathbf{u}-\tilde\Delta_D\mathbf{u}$ as the Eulerian mean fluid
velocity $\langle\mathbf{U}\rangle^E
=\langle\mathbf{U}\rangle^L -\langle\mathbf{U}\rangle^S$. 

This section is meant to be self contained. The     
notation we use here for averaged quantities such as
$\langle\mathbf{U}\rangle^L$,
$\langle\mathbf{U}\rangle^E$ and $\langle\mathbf{U}\rangle^S$
is intended to be as self-explanatory as possible.

  
\subsection{Stokes mean drift closure relation}
The Reynolds decomposition of the Lagrangian fluid trajectory
$ \mathbf{X}^{\xi}$ introduced in equation (\ref{traj-xi}) implies
relation (\ref{vel-xi-dot-relation}) for the Lagrangian mean fluid velocity
$\mathbf{u}$, denoted in this section as
$\langle\mathbf{U}\rangle^L$,
\begin{equation}
\mathbf{U}(\mathbf{x}+\boldsymbol{\xi}(\mathbf{x},t;\omega),t)
\equiv 
\langle\mathbf{U}\rangle^L(\mathbf{x},t) 
+ \mathbf{U}^{\ell}(\mathbf{x},t;\omega)
\,,\quad\hbox{with}\quad
\mathbf{U}^{\ell} \equiv \frac{d\boldsymbol\xi}{dt}
(\mathbf{x},t;\omega)
\,.
\label{vel-def-Lag-mean}
\end{equation}
Note that the slow-time advective derivative, 
\begin{equation}\label{d/dt-def}
\frac{d}{dt}
= \frac{\partial}{\partial t} 
+ \langle\mathbf{U}\rangle^L(\mathbf{x},t)
  \boldsymbol{\cdot\nabla}
\,,
\end{equation}
acts only on $(\mathbf{x},t)$ dependence and does not act on
$\omega$ dependence. As before, the angle brackets as in
$\langle\mathbf{U}\rangle^L$ denote average over dynamical
behavior in $\omega$ that is either rapid in time, or random. 
In comparison, the traditional {\bfi Reynolds fluid velocity decomposition} 
is expressed at a given position $\mathbf{x}$ in terms of the {\bfi Eulerian 
mean fluid velocity}, $\langle\mathbf{U}\rangle^E$ as
\begin{equation}
\mathbf{U}(\mathbf{x},t;\omega)
\equiv 
\langle\mathbf{U}\rangle^E(\mathbf{x},t) 
+ \mathbf{U}^{\,\prime}(\mathbf{x},t;\omega)
\,.
\label{vel-def-Eul-mean-Reynolds}
\end{equation}
We note that the rapid $\omega$-dynamics appears in {\it different functional
forms} on the left hand sides of equations (\ref{vel-def-Lag-mean}) and
(\ref{vel-def-Eul-mean-Reynolds}) in the definitions of $\langle\mathbf{U}\rangle^L$ and
$\langle\mathbf{U}\rangle^E$. The Reynolds fluid velocity decomposition in
(\ref{vel-def-Eul-mean-Reynolds}) describes fluctuations in velocity without reference to
a Lagrangian parcel trajectory. In principle, such a trajectory could be
obtained by integrating this decomposition from an initial reference
configuration. Rather than developing the analysis relating these velocity
formulas from a Lagrangian viewpoint, we shall instead follow
\cite{Andrews-McIntyre[1978a]} in presenting an heuristic argument relating
these velocities as Eulerian quantities by using Taylor series approximations
and asymptotics in the magnitude $|\boldsymbol\xi|$ of the fluctuations.

Taylor expansion of the Reynolds fluid velocity decomposition
(\ref{vel-def-Eul-mean-Reynolds}) yields
\begin{eqnarray}\label{Taylor-exp-U-3}
\mathbf{U} (\mathbf{x} + \boldsymbol\xi,t\,;\omega)
&=&
\mathbf{U} (\mathbf{x},t\,;\omega)
+ \boldsymbol{\xi\cdot\nabla}\, \mathbf{U} (\mathbf{x},t\,;\omega)
+ \frac{1}{2}\xi^k\,\xi^l
\mathbf{U}_{,k\,l}(\mathbf{x},t\,;\omega)
+ O(|\boldsymbol\xi|^3)
\nonumber\\
&=& 
\langle\mathbf{U}\rangle^E(\mathbf{x},t)
+ 
\boldsymbol{\xi\cdot\nabla}
\langle \mathbf{U}\,\rangle^E(\mathbf{x},t)
+\
\frac{1}{2}\xi^k\,\xi^l
\langle\mathbf{U}\rangle^E_{,k\,l}(\mathbf{x},t)
\nonumber\\
&&
+\
\mathbf{U}^{\,\prime}(\mathbf{x},t\,;\omega)
+
\boldsymbol{\xi\cdot\nabla}\,
\mathbf{U}^{\,\prime}(\mathbf{x},t\,;\omega)
+ O(|\boldsymbol\xi|^3)
\,.
\end{eqnarray}
Upon {\it assuming} we may equate the different
functional forms of velocity arising in the two decompositions as
$\mathbf{U}(\mathbf{x} + \boldsymbol{\xi}(\mathbf{x},t;\omega),t)
= \mathbf{U} (\mathbf{x} + \boldsymbol\xi,t\,;\omega)$,
comparison of the two formulas (\ref{vel-def-Lag-mean})
and (\ref{Taylor-exp-U-3}) gives
\begin{eqnarray}
O(1):\quad 
\langle\mathbf{U}\,\rangle^L
&=& 
 \langle\mathbf{U}\,\rangle^E
\,;\label{order-one}
\\
O(|\boldsymbol\xi|):\qquad
\frac{d\boldsymbol\xi}{dt}
&=&
\boldsymbol{\xi\cdot\nabla} \langle\mathbf{U}\,\rangle^E
+
\mathbf{U}^{\,\prime}(\mathbf{x},t\,;\omega)
\,;\label{order-xi}
\\
O(|\boldsymbol\xi|^2):\quad 
\langle\mathbf{U}\rangle^S
&\equiv&
\langle\mathbf{U}\rangle^L - \langle\mathbf{U}\rangle^E
\nonumber
\\
&=&
\langle\boldsymbol{\xi\cdot\nabla}\,
\mathbf{U}^{\,\prime}\rangle(\mathbf{x},t)
+\
\frac{1}{2}\langle\xi^k\,\xi^l\rangle
\langle\mathbf{U}\rangle^E_{,k\,l}(\mathbf{x},t)
\,.\label{order-xixi}
\end{eqnarray}
There is some risk in making this order by order comparison of
Taylor expansions, because of the possibility of double counting the 
fluctuations when comparing their Eulerian and Lagrangian representations.
However, there are two cases of the order $O(|\boldsymbol\xi|)$ relation
(\ref{order-xi}) that are unequivocal:
\begin{enumerate}
\item All the fluctuation is modeled in the Lagrangian fluid trajectory. (This
is the case studied here.) Then $\mathbf{U}^{\,\prime}=0$ and (\ref{order-xi})
implies  $d\boldsymbol\xi/dt = \boldsymbol{\xi\cdot\nabla}\, \mathbf{u}$. This
is the {\bfi Taylor hypothesis} in (\ref{xi-eqn}).
\item  All the fluctuation is modeled in the Eulerian fluid velocity. Then 
$\mathbf{U}^{\ell} = d\boldsymbol\xi/dt = 0$ and we obtain
$\mathbf{U}^{\,\prime} =  - \boldsymbol{\xi\cdot\nabla}
\langle\mathbf{U}\,\rangle^E$ from equation (\ref{order-xi}). The relation
$d\boldsymbol\xi/dt = 0$ is the {\bfi Taylor-like hypothesis} introduced in
\cite{HKMRS[1998]}. We shall investigate this situation in
Section \ref{EMM-sec}, when we deal with Eulerian mean fluid theories.
\end{enumerate}
For the {\bfi Stokes mean drift velocity} 
$\langle\mathbf{U}\rangle^S
\equiv
\langle\mathbf{U}\rangle^L - \langle\mathbf{U}\rangle^E$
in the present situation with purely Lagrangian fluctuations (so that
$\mathbf{U}^{\,\prime} = 0$) we shall follow~\cite{Holm[1996]-CL}
and take
\begin{eqnarray}
\langle\mathbf{U}\rangle^S (\mathbf{x},t) 
&=& \langle\,\boldsymbol{\xi\cdot\nabla}\mathbf{U}^{\ell}\rangle
\nonumber\\
&=&
\big(\boldsymbol{\nabla\cdot}
\boldsymbol{\langle\,\xi\xi\rangle}
\boldsymbol{\cdot\nabla}\big)
\langle\mathbf{U}\rangle^L
\nonumber\\
&&
-\
\langle
(\boldsymbol{\nabla\cdot\xi})\,
\boldsymbol\xi
\rangle
\boldsymbol{\cdot\nabla}
  \langle\mathbf{U}\rangle^L\
+\
\frac{1}{2}\,\langle\xi^k\,\xi^l\rangle
\langle\mathbf{U}\rangle^L_{,k\,l}
+ 
O(|\boldsymbol\xi|^3)
\nonumber\\
&=&
\big(\boldsymbol{\nabla\cdot}
\boldsymbol{\langle\,\xi\xi\rangle}
\boldsymbol{\cdot\nabla}\big)
\langle\mathbf{U}\rangle^L
+ 
o(|\boldsymbol\xi|^2)
\,,
\quad\hbox{({\bfi Stokes drift})}
\label{vel-def-Stokes}\,.
\end{eqnarray}
We drop the latter two terms in this expansion, by arguing that 
\begin{equation}\label{neglected-terms}
\langle\boldsymbol{\nabla\cdot\xi}\rangle 
= 
o(|\boldsymbol\xi|)
\quad\hbox{and}\quad
\langle\mathbf{U}\rangle^L_{,k\,l}
=
o(1)\,.
\end{equation}
Hence, we arrive at the approximate closure relation for
$\langle\mathbf{U}\rangle^S$, to order $o(|\boldsymbol\xi|^2)$,
\smallskip
\boxeq{2}
\begin{equation}\label{ubar-S-eqn}
\langle\mathbf{U}\rangle^S
= 
\boldsymbol{\nabla\cdot}\,\big(
\boldsymbol{\langle\,\xi\xi\rangle}
\boldsymbol{\cdot\nabla}
\langle\mathbf{U}\rangle^L\big)
=
\tilde\Delta\langle\mathbf{U}\rangle^L
\,,
\end{equation}
\smallskip

\noindent
where the operator $\tilde\Delta$ is defined by 
$\tilde\Delta = (\boldsymbol{\nabla\cdot}
\boldsymbol{\langle\,\xi\xi\rangle}
\boldsymbol{\cdot\nabla}) = \tilde\Delta_D + O(|\boldsymbol\xi|^4)$, so we need
not distinguish between $\tilde\Delta$ and $\tilde\Delta_D$ here.

\paragraph{Identification of $\mathbf{v}$ as the Eulerian mean velocity.}
Subject to the assumptions (\ref{neglected-terms}) and accepting the definition
of $\langle\mathbf{U}\rangle^S$ as in (\ref{vel-def-Stokes}), the momentum, or
circulation velocity $\mathbf{v}$ may be identified as the Eulerian mean
velocity $\langle\mathbf{U}\rangle^E$, by following the chain of relations,
\begin{equation}\label{ubar-E-vee-eqn}
\langle\mathbf{U}\rangle^E
= 
\langle\mathbf{U}\rangle^L - \langle\mathbf{U}\rangle^S
=
(1-\tilde\Delta_D)\langle\mathbf{U}\rangle^L
=
(1-\tilde\Delta_D)\mathbf{u}
=
\mathbf{v}.
\end{equation}
This establishes the physical interpretation of the
quantity $\mathbf{v}$ as the Eulerian mean velocity.

\paragraph{Lagrangian mean, and Eulerian mean pressures.}
A similar Taylor expansion may be performed to determine the relation between
the Lagrangian mean, and the Eulerian mean pressures. Namely,
\begin{equation}\label{press-exp}
P(\mathbf{x} + \boldsymbol\xi)
=
P(\mathbf{x}) + \boldsymbol{\xi\cdot\nabla}P(\mathbf{x})
+\
\frac{1}{2}\,\langle\xi^k\,\xi^l\rangle
P_{,k\,l}(\mathbf{x})
+ 
O(|\boldsymbol\xi|^3)
\,.
\end{equation}
Hence, we obtain
\begin{equation}\label{press-Eul-Lag}
\langle{P}\rangle^E(\mathbf{x})
=
\langle{P}\rangle^L(\mathbf{x})
\ 
-\
\frac{1}{2}\,\langle\xi^k\,\xi^l\rangle
\langle{P}\rangle^L_{,k\,l}(\mathbf{x})
+ 
O(|\boldsymbol\xi|^3)
\,,
\end{equation}
which relates the Eulerian mean pressure $\langle{P}\rangle^E$ to the
Lagrangian mean pressure $\langle{P}\rangle^L$ to order $O(|\boldsymbol\xi|^3)$.

\paragraph{Remarks on boundary conditions.}
With the interpretation given in (\ref{ubar-E-vee-eqn}), the Eulerian
mean velocity $\mathbf{v} =
\langle\mathbf{U}\rangle^E$ naturally must be tangent to fixed
boundaries, so that
\begin{equation}\label{v-bdy-cond}
\mathbf{v}\cdot\hat{\mathbf{n}}=0\,,
\quad\hbox{at a fixed boundary.}
\end{equation}
The Lagrangian mean velocity $\mathbf{u} =
\langle\mathbf{U}\rangle^L$ must satisfy boundary conditions that
allow determination of
$\mathbf{u}$ from $\mathbf{v}$ by inverting the dynamical Helmholtz
operator, $1-\tilde\Delta$. For this, we shall choose Dirichlet
boundary conditions on the Lagrangian mean velocity,
\begin{equation}\label{u-bdy-cond}
\mathbf{u}=0\,,
\quad\hbox{at the boundary.}
\end{equation}
We note that $\hat{\mathbf{n}}\cdot
\boldsymbol{\langle\,\xi\xi\rangle} = 0$ at the boundary, as well,
since the fluctuations should not penetrate the boundary.
These remarks provide the rational for choosing the boundary
conditions (\ref{LMM-bc}) for the ideal LMM model.

\paragraph{Relation to the Monin-Yaglom formula.}
An alternative formula for the Stokes mean drift velocity due to
Monin and Yaglom~\cite{Monin-Yaglom} introduces the diffusivity
tensor $\boldsymbol\kappa$, according to 
\begin{equation}\label{ubar-S-M&Y}
\langle\mathbf{U}\rangle^S
=
\boldsymbol{\nabla\cdot\kappa}\,.
\end{equation}
Consistency of this formula with relation (\ref{ubar-S-eqn})
allows one to identify the diffusivity tensor
$\boldsymbol\kappa$ as
\begin{equation}\label{ubar-kappa-rel}
\boldsymbol\kappa
=
\boldsymbol{\langle\,\xi\xi\rangle}
\boldsymbol{\cdot\nabla}
\langle\mathbf{U}\rangle^L
\,.
\end{equation}
We decompose this tensor into its symmetric and
antisymmetric components, as $\boldsymbol\kappa =
\boldsymbol\kappa_S + \boldsymbol\kappa_A$, where,
\begin{eqnarray}\label{ubar-kappa-rel-cont}
\boldsymbol\kappa_S
&=&
\frac{1}{2}\Big(
\boldsymbol{\langle\,\xi\xi\rangle}
\boldsymbol{\cdot\nabla}
\langle\mathbf{U}\rangle^L
+
(\boldsymbol\nabla
\langle\mathbf{U}\rangle^L)^{\rm T}
\boldsymbol{\cdot\langle\,\xi\xi\rangle}
\Big)\,,
\\
\boldsymbol\kappa_A
&=&
\frac{1}{2}\Big(
\boldsymbol{\langle\,\xi\xi\rangle}
\boldsymbol{\cdot\nabla}
\langle\mathbf{U}\rangle^L
-
(\boldsymbol\nabla
\langle\mathbf{U}\rangle^L)^{\rm T}
\boldsymbol{\cdot\langle\,\xi\xi\rangle}
\Big)\,.
\end{eqnarray}
We note from equation (\ref{covariance-dynamics-vector}) that
\begin{equation}\label{kappa-sym}
\boldsymbol\kappa_S
=
\frac{1}{2}\,\frac{d}{dt}
\boldsymbol{\langle\,\xi\xi\rangle}\,,
\end{equation}
and by direct manipulation we find
\begin{equation}\label{kappa-asym}
\boldsymbol{\nabla\cdot\kappa}_A
=
-\,\frac{1}{2}\,{\rm curl}\,
\boldsymbol{\langle\,\xi\times}
(\boldsymbol{\xi\cdot\nabla}\,)
\langle\mathbf{U}\rangle^L
\boldsymbol\rangle
\,.
\end{equation}
Relation (\ref{kappa-sym}) for $\boldsymbol\kappa_S$ is
precisely the definition of the {\bfi Taylor diffusivity}
\cite{Taylor[1921]},~\cite{Bennett[1996]} and relation
(\ref{kappa-asym}) implies that
$\boldsymbol\kappa_A$ does not contribute to the divergence of
$\langle\mathbf{U}\rangle^S$. Thus, we find
\begin{equation}\label{div-ubar-S-M&Y}
\boldsymbol{\nabla\cdot}\langle\mathbf{U}\rangle^S
=
\boldsymbol{\nabla\cdot\kappa}_S
=
\boldsymbol{\nabla\cdot}
\Big(\!
\boldsymbol{\nabla\cdot}
\frac{1}{2}\,\frac{d}{dt}
\boldsymbol{\langle\,\xi\xi\rangle}\Big)
\,.
\end{equation}
In combination with the LMM equations, these identifications
provide {\it dynamical equations} for the diffusivity in the Monin-Yaglom
formula which are of potential use in modeling geophysical fluid dynamics
\cite{Dukowicz}.

\paragraph{Comparison with the Craik-Leibovich
equations~\cite{Holm[1996]-CL},~\cite{CL[1976]}.} The explicit expression
(\ref{ubar-S-eqn}) for the Stokes mean drift velocity
$\langle\mathbf{U}\rangle^S = \mathbf{u}-\mathbf{v}$ allows us to write the LMM
motion equation (\ref{LMM-mot-Euler-form1}) in the notation of this Section as
\begin{eqnarray}\label{ideal-LMM-mot-eqn}
&&\frac{\partial}{\partial t}\langle\mathbf{U}\rangle^E
+ \langle\mathbf{U}\rangle^E\boldsymbol{\,\cdot\nabla}
\langle\mathbf{U}\rangle^E
+ \langle\mathbf{U}\rangle^S \boldsymbol{\,\cdot\nabla}
\langle\mathbf{U}\rangle^E
+ \boldsymbol\nabla\langle{P}\rangle^L 
= 0
\,,
\\
&& \hspace{.25in}
\hbox{with}\quad
\langle\mathbf{U}\rangle^S
=
\tilde\Delta\langle\mathbf{U}\rangle^L
\,,
\quad\hbox{and}\quad
\boldsymbol{\nabla\cdot}
\langle\mathbf{U}\rangle^L = 0 \,.
\label{u-div-eqn}
\end{eqnarray}
Thus, the Stokes mean drift velocity introduces an {\bfi additional transport}
of Eulerian mean velocity into the ideal LMM motion equation
(\ref{ideal-LMM-mot-eqn}). The corresponding Kelvin-Noether circulation theorem
is
\begin{equation}\label{KelThm-LMM}
\frac{ d}{dt}\oint_{\gamma(\langle\mathbf{U}\rangle^L)}
\langle\mathbf{U}\rangle^E
\boldsymbol{\cdot}d\mathbf{x} 
= 
\int\int_{S(\langle\mathbf{U}\rangle^L)}
\boldsymbol{\nabla}\langle{U}\rangle^E_j
\boldsymbol{\times\nabla}\langle{U}\rangle^S_j
\boldsymbol\cdot
d\mathbf{S}
\,,
\end{equation}
for any surface $S(\langle\mathbf{U}\rangle^L)$ whose boundary is the  
closed curve $\gamma(\langle\mathbf{U}\rangle^L)$ moving with the 
Lagrangian mean fluid velocity $\langle\mathbf{U}\rangle^L$. Thus,
circulation of Eulerian mean velocity will be generated according to these
equations whenever the gradients of the Stokes drift velocity and Eulerian mean
velocity are not colinear. In this circumstance, the Stokes drift velocity
generates convective circulation of the Eulerian mean velocity that tends to
reduce the gradient of the Lagrangian mean velocity.

The motion equation (\ref{ideal-LMM-mot-eqn}) of the LMM model is reminiscent of
(but different from) the ideal Craik-Leibovich (CL)
equations~\cite{Holm[1996]-CL},~\cite{CL[1976]}.
In the CL theory, the rapidly oscillating waves at a free surface
are assumed to be unaffected by the more slowly changing currents below. The
effect of the waves on the Eulerian mean velocity is parameterized in the
CL theory by introducing into the Euler equations a ``vortex force," expressed
in terms of a prescribed Stokes drift velocity
$\langle\mathbf{U}\rangle^S(\mathbf{x},t)$. 
The CL equations are given by,
\begin{eqnarray}
&&
\frac{\partial}{ \partial t} \langle\mathbf{U}\rangle^E 
+ \langle\mathbf{U}\rangle^E 
\boldsymbol{\,\cdot\nabla} \langle\mathbf{U}\rangle^E 
-
\langle\mathbf{U}\rangle^S \times {\rm curl}\, \langle\mathbf{U}\rangle^E
+ \boldsymbol\nabla \varpi
= 0
\, ,
\label{CL1}\\
&&
\boldsymbol{\nabla\,\cdot\,} \langle\mathbf{U}\rangle^E = 0 \, ,
\quad \hbox{and} \quad
\varpi 
= 
P + \frac{1}{2} |\langle\mathbf{U}\rangle^E 
+ \langle\mathbf{U}\rangle^S|^2 - \frac{1}{2} |\langle\mathbf{U}\rangle^E|^2 \, .
\nonumber
\end{eqnarray}
Here $\varpi$ is a modified pressure term that includes the Eulerian mean
pressure $P$ as well as the increase of the kinetic energy of the fluid due to
the waves. The term $\langle\mathbf{U}\rangle^S \times {\rm curl} \,
\langle\mathbf{U}\rangle^E$  is the ``vortex force" of the CL theory of Langmuir
circulation. The Eulerian mean fluid velocity $\langle\mathbf{U}\rangle^E$ is assumed to
be divergenceless and is required to be tangential to fixed boundaries of the
domain of flow. 

The first difference from the CL equations
is that the ideal LMM model has a {\it dynamical equation} for the Stokes
mean drift velocity, while this is a {\it prescribed function} in the
Craik-Leibovich model. Also, the Stokes mean drift velocity appears
as a transport term $(\langle\mathbf{U}\rangle^S \boldsymbol{\,\cdot\nabla}
\langle\mathbf{U}\rangle^E)$ in the LMM motion equation
(\ref{ideal-LMM-mot-eqn}), while it appears as a ``vortex force''
$(-\langle\mathbf{U}\rangle^S
\times{\rm curl}\, \langle\mathbf{U}\rangle^E)$ in the CL model.
Both of these are substantial differences between the LMM model and the
Craik-Leibovich theory.

  
\subsection{An order $O(|\boldsymbol\xi|^2)$ model with
div$\langle\mathbf{U}\rangle^E=0$} 
Rather than the divergenceless condition 
$\boldsymbol{\nabla\cdot\,}
\langle\mathbf{U}\rangle^L = 0$ appearing in the
LMM equations (\ref{ideal-LMM-eqns}), the condition
$\boldsymbol{\nabla\cdot\,}
\langle\mathbf{U}\rangle^E = 0$ is implied by the
Eulerian mean of the original divergence free condition
$\boldsymbol{\nabla\cdot}\mathbf{U} = 0$, when combined with the
velocity decomposition (\ref{vel-def-Eul-mean-Reynolds}) and the assumption
that the Eulerian mean velocity fluctuation has zero Eulerian mean,
$\langle\mathbf{U}^{\,\prime}\rangle^E=0$, where
$\langle\,\boldsymbol{\cdot}\rangle^E$ denotes average over the rapid dependence
at fixed Eulerian position. 

If this were so, then preservation of the condition
$\boldsymbol{\nabla\cdot\,} \langle\mathbf{U}\rangle^E = 0$ would determine the
pressure $p$ in the LMM motion equation (\ref{ideal-LMM-mot-eqn})
by solving the Poisson equation $-\Delta{p} = \boldsymbol{\nabla\cdot\,}
(\langle\mathbf{U}\rangle^L \boldsymbol
{\,\cdot\nabla}\langle\mathbf{U}\rangle^E)$ with Neumann boundary conditions
obtained by taking the normal component of the motion equation
(\ref{ideal-LMM-mot-eqn}) and using
$\langle\mathbf{U}\rangle^E \cdot\hat{\mathbf{n}}=0$ at the
boundary. This situation is appealing because of the clear
physical interpretation of the velocity variables and their
boundary conditions. However, it requires
$\boldsymbol{\nabla\cdot\,}\langle\mathbf{U}\rangle^L =
O(|\boldsymbol\xi|^2)$. Hence, to explore this possibility we must
begin by restoring the $O(|\boldsymbol\xi|^2)$ compressibility induced by the
fluctuations and calculated earlier in equation
(\ref{Lag-der-D-approx}) as 
\begin{equation}\label{2nd-order-compress}
D = 1 - \frac{1}{2}\langle\xi^k\xi^l\rangle_{,kl}\,.
\end{equation}
The arrangements in Hamilton's principle needed for the pressure
to impose this relation as a dynamically consistent constraint are
discussed in Section \ref{avg-L-sec} leading to the Lagrangian in equation
(\ref{mean-Lag-approx}). It would be convenient if these arrangements were also
to imply that $\boldsymbol{\nabla\cdot\,}\langle\mathbf{U}\rangle^E =
o(|\boldsymbol\xi|^2) \approx 0$, i.e., that the Eulerian mean
velocity were incompressible to a certain approximation. We shall now
investigate this possibility.

In the notation of the present section, the continuity equation for
$D$ as in equation (\ref{2nd-order-compress}) implies, to order
$O(|\boldsymbol\xi|^4)$
\begin{equation}\label{2nd-order-div-u}
 \boldsymbol{\nabla\cdot\,}
 \langle\mathbf{U}\rangle^L 
=
 \frac{-1}{D}\frac{dD}{dt}
=
 \frac{1}{2}\frac{d}{dt}\langle\xi^k\xi^l\rangle_{,kl}
=
\frac{1}{2}\frac{d}{dt}(
\boldsymbol{\nabla\cdot\nabla\cdot\,}
\boldsymbol{\langle\,\xi\xi\rangle})
\,.
\end{equation}
Combining this relation with equation (\ref{ubar-S-eqn}) and
the divergence of equation (\ref{ubar-E-vee-eqn}) gives
a formula for the divergence of the Eulerian mean velocity,
\begin{eqnarray}\label{2nd-order-div-u-eul}
\boldsymbol{\nabla\cdot}\langle\mathbf{U}\rangle^E
&=& 
\boldsymbol{\nabla\cdot}\langle\mathbf{U}\rangle^L 
-
\boldsymbol{\nabla\cdot}\langle\mathbf{U}\rangle^S
=
\boldsymbol{\nabla\cdot}\langle\mathbf{U}\rangle^L 
-
\boldsymbol{\nabla\cdot}
\big(\boldsymbol{\nabla\cdot}
\boldsymbol{\langle\,\xi\xi\rangle}
\boldsymbol{\cdot\nabla}
\langle\mathbf{U}\rangle^L\big)
\nonumber\\
&=&
\frac{1}{2}\frac{d}{dt}\Big(
\boldsymbol{\nabla\cdot\nabla\cdot\,}
\boldsymbol{\langle\,\xi\xi\rangle}\Big)
-
\frac{1}{2}
\boldsymbol{\nabla\cdot}\Big[
\boldsymbol{\nabla\cdot}\Big(
\frac{d}{dt}
\boldsymbol{\langle\,\xi\xi\rangle}\Big)\Big]
\nonumber\\
&=&
\frac{1}{2}\Big[\frac{d}{dt}\,, 
\boldsymbol{\nabla\cdot}
\boldsymbol{\nabla\cdot}\Big]
\boldsymbol{\langle\,\xi\xi\rangle}
=
\frac{1}{2}\Big[
\langle\mathbf{U}\rangle^L
\boldsymbol{\cdot\nabla}\,, 
\boldsymbol{\nabla\cdot}
\boldsymbol{\nabla\cdot}\Big]
\boldsymbol{\langle\,\xi\xi\rangle}
\nonumber\\
&=&
O\big(|\boldsymbol{\nabla}
\boldsymbol{\nabla} \langle\mathbf{U}\rangle^L\,|\big)
O\big(|\boldsymbol\xi|^2\big)
\,.
\end{eqnarray}
Thus, $\boldsymbol{\nabla\cdot}\langle\mathbf{U}\rangle^E
= o(|\boldsymbol\xi|^2)\approx0$,
provided $|\boldsymbol{\nabla}
\boldsymbol{\nabla} \langle\mathbf{U}\rangle^L| = o(1)$, 
as we argued earlier in equation (\ref{neglected-terms}).
Since the Lagrangian mean velocity
$\langle\mathbf{U}\rangle^L$ is smoothed by Helmholtz inversion,
this assumption may be plausible in certain flow regimes. If this
asumption is made, the approximate equations that result are
\boxeq{11}
\begin{equation}\label{approx-2nd-order-mot-eqn}
\Big(\frac{\partial}{\partial t} 
+ \langle\mathbf{U}\rangle^L
  \boldsymbol{\cdot\nabla}\Big)
 \langle\mathbf{U}\rangle^E
+
\boldsymbol{\nabla}p = 0\,,
\quad
\boldsymbol{\nabla\cdot}\langle\mathbf{U}\rangle^E \approx 0
\,,
\end{equation}
$\quad$ where
\begin{equation}\label{u-lag-def-redux}
\langle\mathbf{U}\rangle^L
= \big(1-\tilde\Delta\big)^{-1}
\langle\mathbf{U}\rangle^E\,,
\quad
\tilde\Delta = (\boldsymbol{\nabla\cdot}
\boldsymbol{\langle\,\xi\xi\rangle}
\boldsymbol{\cdot\nabla})
\,,
\end{equation}
$\quad$ and
\begin{equation}\label{xi-xi-bold-alt}
\Big(\frac{\partial}{\partial t} 
+ \langle\mathbf{U}\rangle^L
  \boldsymbol{\cdot\nabla}\Big)
\boldsymbol{\langle\xi\xi\rangle}
= \boldsymbol{\langle\,\xi\xi\rangle}
\boldsymbol{\cdot\nabla}
\langle\mathbf{U}\rangle^L
+
(\boldsymbol\nabla
\langle\mathbf{U}\rangle^L)^{\rm T}
\boldsymbol{\cdot\langle\,\xi\xi\rangle}
\,.
\end{equation}
$\quad$ The boundary conditions for the system
(\ref{approx-2nd-order-mot-eqn}) -- (\ref{u-lag-def-redux}) are 
\begin{equation}\label{alt-LMM-bc}
\langle\mathbf{U}\rangle^E\boldsymbol{\cdot\,\hat{n}}=0\,,
\quad
\langle\mathbf{U}\rangle^L=0\,,
\quad\hbox{and}\quad
\boldsymbol{\langle\xi\xi\rangle\cdot\hat{n}}=0
\quad\hbox{on a fixed boundary.}
\end{equation}
The energy for this system is,
\begin{equation}\label{alt-LMM-cons-erg}
E =  \frac{1}{2}\int d^{\,3}x \
\Big(1 - \frac{1}{2}\langle\xi^k\xi^l\rangle_{,kl}\Big)
\langle\mathbf{U}\rangle^E\boldsymbol{\cdot\,}
\langle\mathbf{U}\rangle^L
\,.
\end{equation}
This energy will be conserved {\it exactly} for 
$\boldsymbol{\nabla\cdot}\langle\mathbf{U}\rangle^E$ satisfying
equation (\ref{2nd-order-div-u-eul}) without approximation, as we
found in the second order model in Section \ref{avg-L-sec}. And it
will be conserved {\it approximately} for
$\boldsymbol{\nabla\cdot}\langle\mathbf{U}\rangle^E=0$.

\paragraph{Summary of Lagrangian mean models with order
$O(|\boldsymbol\xi|^2)$ compressibility.} There are two contending
theories with order $O(|\boldsymbol\xi|^2)$ compressibility: one has exactly
divergenceless Eulerian mean velocity
$\langle\mathbf{U}\rangle^E$ and only approximate energy
conservation; and the other has small divergence of
$\langle\mathbf{U}\rangle^E$ and exact energy conservation. In
fact, the conserved energy in the latter case contains a term of
order $O(|\boldsymbol\xi|^4)$, which of course is higher order than the
validity of the equations. The choice between these two theories probably should
be made on the basis of their performance in practice. The divergenceless model
defined in equations (\ref{approx-2nd-order-mot-eqn}) -- (\ref{alt-LMM-bc}) may
have the advantage of being easier to implement numerically than the
nondivergent theory, which consists of equations
(\ref{approx-2nd-order-mot-eqn}) -- (\ref{alt-LMM-bc}) with
$\boldsymbol{\nabla\cdot}\langle\mathbf{U}\rangle^E \approx 0$ in
equation (\ref{approx-2nd-order-mot-eqn}) replaced by equation
(\ref{2nd-order-div-u-eul}).

Under mild conditions on $\boldsymbol{\langle\,\xi\xi\rangle}$, the
viscous version of the model in equations
(\ref{approx-2nd-order-mot-eqn}) -- (\ref{alt-LMM-bc}) will
dissipate the energy (\ref{alt-LMM-cons-erg}), if the motion
equation is modified to introduce viscosity, as
\begin{equation}\label{diss-2nd-order-mot-eqn}
\Big(\frac{\partial}{\partial t} 
+ \langle\mathbf{U}\rangle^L
  \boldsymbol{\cdot\nabla}\Big)
 \langle\mathbf{U}\rangle^E
+
\boldsymbol{\nabla}p = \nu\tilde\Delta\langle\mathbf{U}\rangle^E\,,
\quad
\boldsymbol{\nabla\cdot}\langle\mathbf{U}\rangle^E \approx 0
\,.
\end{equation}
If needed, an additional dissipative modification of the Lagrangian mean
covariance dynamics could also be proposed. Namely, we propose
\boxeq{4}
\begin{eqnarray}\label{xi-xi-bold-alt-dis}
\Big(\frac{\partial}{\partial t} 
+ \langle\mathbf{U}\rangle^L
  \boldsymbol{\cdot\nabla}\Big)
\boldsymbol{\langle\xi\xi\rangle}
&=& \boldsymbol{\langle\,\xi\xi\rangle}
\boldsymbol{\cdot\nabla}
\langle\mathbf{U}\rangle^L
+
(\boldsymbol\nabla
\langle\mathbf{U}\rangle^L)^{\rm T}
\boldsymbol{\cdot\langle\,\xi\xi\rangle}
\nonumber\\
&& -\
\frac{1}{\tau}\,(\boldsymbol{\langle\,\xi\xi\rangle}
- \alpha^2\boldsymbol{I})
+\
\lambda\,\tilde\Delta
\boldsymbol{\langle\,\xi\xi\rangle}
\,.
\end{eqnarray}
In this equation, $\tau$ is a relaxation time, $\alpha$ is a length
scale below which the effects of fluctuations on the mean flow
should be suppressed and $\lambda$ is a diffusivity
that suppresses gradients of $\boldsymbol{\langle\,\xi\xi\rangle}$.
The quantities $\tau$, $\alpha$ and $\lambda$ may all be taken as constant
parameters, with perhaps $\lambda\sim\alpha^2/\tau$. For 
$\tau^{-1}\ge |\boldsymbol\nabla
\langle\mathbf{U}\rangle^L|$, the
additional dissipation terms in equation (\ref{xi-xi-bold-alt-dis}) will cause
the covariance $\boldsymbol{\langle\,\xi\xi\rangle}$ to approach the
isotropic, homogeneous conditions represented by the VCHE model,
which will be discussed briefly in Section \ref{1pt closure-sec}. At
boundaries one may take $\boldsymbol{\langle\,\xi\xi\rangle}$ to
satisfy equation (\ref{xi-xi-bold-alt-dis}) with the $\lambda$ term
absent.


\section{One point closure equations} 
\label{1pt closure-sec}

  
\subsection{Euler-Poincar\'e equation for the approximate $\langle
L\rangle$} The Euler-Poincar\'e equation (\ref{EPeqn}) can be written for
the approximate $\langle L\rangle$ in equation (\ref{mean-Lag-approx-inc}) for
incompressible flow as
\begin{equation}\label{vee-eqn}
\frac{\partial}{\partial t}\,\mathbf{v} 
+ (\mathbf{u}\boldsymbol{\,\cdot\nabla})\,\mathbf{v} 
+ v_j \boldsymbol\nabla u^j +
\boldsymbol\nabla P_{tot} = 0\,,\quad \boldsymbol{\nabla\cdot}\mathbf{u} = 0,
\end{equation}
where the momentum conjugate to the velocity $\mathbf{u}$ is given by
\begin{equation}\label{vee-def}
\mathbf{v} \equiv  \frac{1}{D} 
\frac{\delta\langle L\rangle}{\delta\mathbf{u}}\bigg|_{D=1} =\ \mathbf{u} 
- \Big(\partial_k\,\langle\xi^k\xi^l\rangle\partial_l\Big)
\mathbf{u}\,.
\end{equation}
The total pressure $P_{tot}$ in equation (\ref{vee-eqn}) is defined as, cf.
equation (\ref{mean-Lag-der}),
\begin{equation}\label{Pee-def}
P_{tot} \equiv  p  
- \frac{1}{2}|\mathbf{u}\,|^2 
- \frac{1}{2}\langle\xi^k\xi^l\rangle 
\big(\mathbf{u}_{,k}\boldsymbol\cdot\mathbf{u}_{,l}\big)\,.
\end{equation}
In this Section, the fluctuation covariance $\langle\xi^k\xi^l\rangle$ is
taken as being independently {\it prescribed}, and thus is not varied in
Hamilton's principle. Equations (\ref{vee-eqn}) with definitions
(\ref{vee-def}) and (\ref{Pee-def}) are generalizations of the
$n$-dimensional Camassa-Holm (CH) equations derived
in~\cite{HMR[1998a]},~\cite{HMR[1998b]}. The latter equations are
recovered when the isotropy conditions
\begin{equation}\label{isotropy-cond}
\langle\xi^k\xi^l\rangle =
\alpha^2\delta^{kl}\;, 
\end{equation}
hold and, moreover, the statistics are homogeneous, so that $\alpha ^2$ is
constant. As we shall see in Section \ref{EMM-sec}, there is a natural
extension of the CH model for {\it Eulerian mean} fluid dynamics, in which
$\mathbf{u}$ is the Eulerian mean fluid velocity and $\mathbf{v}$ is its
Lagrangian mean counterpart.

\paragraph{Background.}
The derivation~\cite{HMR[1998a]},~\cite{HMR[1998b]}, of the CH equation
(\ref{vee-eqn}) with definitions (\ref{vee-def}) and
(\ref{Pee-def}), and homogeneous isotropic Eulerian fluctuation statistics
satisfying (\ref{isotropy-cond}) with constant $\alpha^2$ generalizes a
one dimensional integrable nonlinear dispersive shallow water
model~\cite{CH[1993]},~\cite{CHH[1994]}, to the
$n$-dimensional situation and provides the interpretation of $\alpha$ as
the typical Eulerian mean amplitude of the fluctuations. 
See~\cite{HKMRS[1998]} for the extension of that derivation to Riemannian
manifolds and discussions of alternative boundary conditions for the case
of homogeneous statistics. See also~\cite{S1[1998]} for further analysis
of the Riemannian case of the $n$-dimensional CH equations.

Holm, Marsden and Ratiu~\cite{HMR[1998a]},~\cite{HMR[1998b]}
note that the conditions of isotropy (\ref{isotropy-cond}) and homogeneity
(constant $\alpha^2$) need to be modified near fixed boundaries, due to
the physical requirement that $\boldsymbol{\xi\cdot}\mathbf{\hat{n}}=0$ be
satisfied, where $\mathbf{\hat{n}}$ is the unit vector normal to the
boundary. However, the condition
\begin{equation}\label{bdy-conds}
\hat{n}_k\langle\xi^k\xi^l\rangle = 0 \quad\hbox{on the boundary,} 
\end{equation}
implied by this physical requirement, cannot be satisfied for constant
$\alpha^2$ in equation (\ref{isotropy-cond}). Chen et al.~\cite{Chen
etal[1998a]}--~\cite{Chen etal[1998c]}, overcame this
difficulty by allowing spatial variation of $\alpha^2$ near fixed boundaries in
straight pipe and channel geometries.

  
\subsection{Relation to one point turbulence closure models}
We note that the velocity $\mathbf{v}$ defined in equation
(\ref{vee-def}) is the momentum conjugate to the velocity $\mathbf{u}$ (or dual
to $\mathbf{u}$, in the sense of variational derivative of the kinetic energy). 
On that basis, Chen et al.~\cite{Chen etal[1998a]}--~\cite{Chen etal[1998c]},
proposed the following viscous variant of (\ref{vee-eqn}), in which viscosity
acts to diffuse the  momentum $\mathbf{v}$
\boxeq{2}
\begin{equation}\label{VCHE}
\frac{\partial}{\partial t} \mathbf{v} 
+ (\mathbf{u}\,\boldsymbol{\cdot\nabla})\mathbf{v} 
+ v_j \boldsymbol\nabla u^j
= \nu \Delta \mathbf{v} - \boldsymbol\nabla P_{tot}\;,\qquad 
\boldsymbol{\nabla\cdot}\mathbf{u} = 0\;.
\end{equation}
\smallskip

\noindent
In their case, Chen et al. defined the momentum $\mathbf{v}$ and the
modified pressure $P$ by
\begin{equation}\label{vee-def2}
\mathbf{v} = \mathbf{u} 
- \big(\boldsymbol{\nabla\,\cdot}\langle\boldsymbol\xi\rangle\big)
\, \mathbf{u} 
- \Big(\partial_k\,\alpha^2\partial_k\Big) \mathbf{u}\,,
\end{equation}
and 
\begin{equation}\label{Pee-def2}
P_{tot} \equiv  p  
- \frac{1}{2}|\mathbf{u}\,|^2 
- \frac{\alpha^2}{2} 
\big(\mathbf{u}_{,k}\boldsymbol\cdot\mathbf{u}_{,k}\big)\,.
\end{equation}
Chen et al.~\cite{Chen etal[1998a]}--~\cite{Chen
etal[1998c]}, allowed for spatial variation of $\alpha^2$, particularly in flow
regions near boundaries. They also allowed for the mean
$\langle\boldsymbol\xi\rangle$ to be nonzero near boundaries to account
for the anisotropy there. Chen et al. referred to  equation (\ref{VCHE})
with definition (\ref{vee-def2}) as the viscous Camassa-Holm equations
(VCHE), although this model is also known as the {\bfi Navier-Stokes alpha
model} because it reduces to the Navier-Stokes equations when alpha is absent.
They proposed a one point closure model for turbulent flows in pipes and
channels by comparing equation (\ref{VCHE}) with the Reynolds averaged
Navier-Stokes equations in those geometries and identifying corresponding terms.
They then verified the predictions of this closure model by comparison with
experimental data at high Reynolds numbers.

Chen et al.~\cite{Chen etal[1998a]}--~\cite{Chen
etal[1998c]}, also gave a continuum mechanical interpretation to their VCHE or
NS-$\alpha$ closure model, by rewriting (\ref{VCHE}) (in the case
where the isotropy conditions (\ref{isotropy-cond}) hold, with $\alpha ^2 \equiv$
constant) in the equivalent {\bfi constitutive form},
\begin{equation}\label{constit-eqn}
\frac{d\mathbf{u}}{dt} = {\rm
div}\hbox{\bfi T}\;,\;\hbox{\bfi T} =  -p\hbox{\bfi I} + 2\nu (1 -
\alpha ^2\Delta)\hbox{\bfi D} + 2\alpha ^2 {\bf\dot{\hbox{\bfi D}}}\;,
\end{equation}
with $\boldsymbol\nabla\cdot\mathbf{u}=0, \hbox{\bfi D} = (1/2) 
(\boldsymbol\nabla\mathbf{u} + \boldsymbol\nabla\mathbf{u}^T),\ 
{\boldsymbol{\Omega}} =
(1/2) (\boldsymbol\nabla\mathbf{u} - \boldsymbol\nabla\mathbf{u}^T)$, and 
co-rotational (Jaumann) derivative given by ${\bf\dot{\hbox{\bfi D}}} =
d\hbox{\bfi D}/dt  + \hbox{\bfi D}\,{\boldsymbol{\Omega}} - 
{\boldsymbol{\Omega}} \hbox{\bfi D}$, with $d/dt = 
\partial/\partial t + \mathbf{u}\boldsymbol{\cdot\nabla}$.  In this
form, one recognizes the constitutive form of VCHE or NS-$\alpha$ as a
variant of the rate-dependent incompressible homogeneous {\bfi fluid of second
grade}~\cite{Dunn-Fosdick[1974]},~\cite{Dunn-Rajagopal[1995]}, whose
viscous dissipation, however, is {\it modified} by the Helmholtz operator
$(1 - \alpha^2\Delta)$.  There is a tradition at least since
Rivlin~\cite{Rivlin[1957]} of modeling turbulence by using continuum
mechanics principles such as objectivity and material frame indifference
(see also~\cite{Chorin[1988]}).  For example, this sort of approach is
taken in deriving Reynolds stress algebraic equation
models~\cite{Shih-Zhu-Lumley[1995]}. Rate-dependent closure models of
mean turbulence such as the VCHE or NS-$\alpha$ closure model have also been
obtained by the two-scale DIA approach~\cite{Yoshizawa[1984]} and by the
renormalization group methods~\cite{Rubinstein-Barton[1990]}. We shall see in
Section \ref{2pt Eul closure-sec} that the covariance
$\boldsymbol{\langle\xi\xi\rangle}$ contributes a fluctuation shear stress term
in the total stress tensor for the order $O(|\boldsymbol\xi|^2)$ compressible
Lagrangian mean motion (LMM) model.

  
\subsection{Comparison of VCHE or NS-$\alpha$ with LES and RANS
models.}  Reynolds-averaged Navier-Stokes (RANS) models of turbulence are part
of the  classic theoretical development of the
subject~\cite{Hinze[1975]},~\cite{Townsend[1967]},~\cite{Lumley-Tennekes[1972]}.
The related Large Eddy Simulation (LES) turbulence modeling
approach~\cite{WReynolds[1987]},~\cite{Piomelli[1993]},~\cite{L&M[1996]}, 
provides an operational definition of the intuitive idea of Eulerian resolved
scales of motion in turbulent flow. In this approach a filtering function ${\cal
F}(\mathbf{r})$ is introduced and the Eulerian velocity field $\mathbf{U}_E$ is
filtered in an integral sense, as
\begin{equation} \label{u-fltrd}
\bar{\mathbf{u}}(\mathbf{r})\equiv \int_{\mathbb{R}^3} d^3{r}^{\,\prime}\,
   \,{\cal F}(\mathbf{r}-\mathbf{r}^{\,\prime})
   \,\mathbf{U}_E\,(\mathbf{r}^{\,\prime})\,.
\end{equation}
This convolution of $\mathbf{U}_E$ with ${\cal F}$ defines the large scale, 
resolved, or filtered velocity, $\bar{\mathbf{u}}$. The corresponding small
scale, or subgrid scale velocity,
$\mathbf{u}^{\,\prime}$, is then defined as the difference,
\begin{equation}\label{u-small}
 \mathbf{u}^{\,\prime}(\mathbf{r})\equiv
\mathbf{U}_E\,(\mathbf{r})-\bar{\mathbf{u}}\,(\mathbf{r})\,. 
\end{equation}
When this filtering operation is applied to the Navier-Stokes system, the
following dynamical equation is obtained for the filtered velocity,
$\bar{\mathbf{u}}$, cf. equation (\ref{constit-eqn}),
\begin{equation} \label{NS-fltrd}
    \frac{\partial}{\partial t}\bar{\mathbf{u}}
   + \bar{\mathbf{u}}\boldsymbol{\cdot\nabla}\bar{\mathbf{u}}
 = -\,{\rm div}\overline{\hbox{{\bfi T}}}
   -\,\boldsymbol\nabla
   \bar{p}+\nu\,\Delta \bar{\mathbf{u}}\,, 
   \quad \boldsymbol{\nabla\cdot\bar{\mathbf{u}}}=0\,,
\end{equation}
in which $\bar{p}$ is the filtered pressure field (required to maintain
$\boldsymbol{\nabla\cdot\bar{\mathbf{u}}}=0$) and the tensor difference
\begin{equation} \label{stress-dev}
  \overline{\hbox{{\bfi T}}} =\overline{
(\mathbf{U}_E\mathbf{U}_E)}-\bar{\mathbf{u}}\bar{\mathbf{u}}\,,
\end{equation}
represents the subgrid scale stress due to the turbulent eddies.  This subgrid
scale stress tensor appears in the same form as the Reynolds stress tensor
obtained from Reynolds averaging the Navier-Stokes equation.

The results of Chen et al.~\cite{Chen etal[1998a]}--~\cite{Chen etal[1998c]}, 
may be given either an LES, or RANS interpretation simply by comparing the
constitutive form of the VCHE or NS-$\alpha$ closure model in
(\ref{constit-eqn}) term by term with equation (\ref{NS-fltrd}), provided one
may ignore the difference between Eulerian mean, and Lagrangian mean velocities
as being of higher order. Additional LES interpretations, discussions and
numerical results for forced-turbulence simulations of the VCHE model will be
presented  elsewhere~\cite{Chen etal to.appear}.

  
\subsection{Comparison of VCHE or NS-$\alpha$ to Leray's
equation}  

The Leray regularization of the Navier-Stokes equations is given
by~\cite{Leray[1934]}, 
\begin{equation} \label{Leray-eqn}
    \frac{\partial}{\partial t}\mathbf{U}
   + \langle\mathbf{U}\,\rangle_{\ell}\boldsymbol{\cdot\nabla}\mathbf{U}
 = \nu\Delta\mathbf{U}   -\,\boldsymbol\nabla{p}\,, 
   \quad \boldsymbol{\nabla\cdot\mathbf{U}}=0
= \boldsymbol{\nabla\cdot}\,\langle\mathbf{U}\,\rangle_{\ell}\,,
\end{equation}
in which the velocity field $\mathbf{U}$ is transported by the spatially
filtered velocity
\begin{equation} \label{vel-filt-def}
\langle\mathbf{U}\,\rangle_{\ell}(\mathbf{r})
= {\ell}^{-3}\int_{\mathbb{R}^3}
g\big(\ell^{-1}(\mathbf{r}-\mathbf{r}')\big)\,
\mathbf{U}(\mathbf{r}')\,d^3\mathbf{r}'
\,.
\end{equation}
Here $g\in C^{\infty}(\mathbb{R}^3)$ is a smooth positive function that
vanishes outside a finite sphere and is normalized to unity,
$\int_{\mathbb{R}^3} g = 1$. Thus, the spatial scales in
$\langle\mathbf{U}\,\rangle_{\ell}$ smaller than $\ell$ have been smoothly
filtered out. The spatially filtered velocity
$\langle\mathbf{U}\,\rangle_{\ell}$ satisfies the important inequality
\begin{equation} \label{vel-filt-ineqlty}
|\langle\mathbf{U}\,\rangle_{\ell}(\mathbf{r})|
\le \max_{\mathbf{r}'} |\mathbf{U}(\mathbf{r}')|
\,,
\end{equation}
obtained by taking the sup norm of its definition (\ref{vel-filt-def}) and
using the normalization of $g$. Solutions of the Leray equation
(\ref{Leray-eqn}) satisfy various regularity properties that are also shared by
solutions of the VCHE~\cite{Foias-Holm-Titi[1998]}. The only difference
between the {\it forms} of these two equations is that the VCHE or NS-$\alpha$
closure model contains the additional term $v_j\boldsymbol\nabla{u^j}$, stemming
from its derivation as an Euler-Poincar\'e equation. This additional term
ensures the {\bfi Kelvin-Noether circulation theorem} for the VCHE. That is, the
VCHE or NS-$\alpha$ closure model satisfies 
\begin{equation}\label{KelThm-VCHE}
\frac{ d}{dt}\oint_{\gamma(\mathbf{u})}\mathbf{v}
\boldsymbol{\cdot}d\mathbf{x} 
= \oint_{\gamma(\mathbf{u})}\Big[
\frac{\partial\mathbf{v}}{\partial t}
+\mathbf{u}\boldsymbol{\cdot\nabla}\mathbf{v}
+ v_j \boldsymbol{\nabla} u^j\Big]
\boldsymbol{\cdot}d\mathbf{x}
= \nu \oint_{\gamma(\mathbf{u})}
\Delta\mathbf{v}\boldsymbol{\cdot}d\mathbf{x}
\,,
\end{equation}
for any closed curve ${\gamma(\mathbf{u})}$ that moves with the 
Eulerian mean fluid velocity $\mathbf{u}$. In comparison, the Leray
equation satisfies 
\begin{equation}\label{KelThm-Leray-eqn}
\frac{ d}{dt}\oint_{\gamma(\langle\mathbf{U}\,\rangle_{\ell})}\mathbf{U}
\boldsymbol{\cdot}d\mathbf{x} 
= 
\oint_{\gamma(\langle\mathbf{U}\,\rangle_{\ell})}
\mathbf{U}\boldsymbol{\cdot}d\langle\mathbf{U}\,\rangle_{\ell}
+ \nu \oint_{\gamma(\langle\mathbf{U}\,\rangle_{\ell})}
\Delta\mathbf{U}\boldsymbol{\cdot}d\mathbf{x}
\,,
\end{equation}
for any closed curve ${\gamma(\langle\mathbf{U}\,\rangle_{\ell})}$ that
moves with the {\it filtered} fluid velocity
$\langle\mathbf{U}\,\rangle_{\ell}$. Thus, the Leray equation has an
additional source of circulation arising from the difference between its
filtered and unfiltered velocities, while the VCHE or NS-$\alpha$ closure model
does not have such a term. 

This difference reappears in the vorticity dynamics for the two theories.
Namely, upon using incompressibility
$\boldsymbol{\nabla\cdot\mathbf{u}}=0$ we have
\begin{equation} \label{vortex-stretching-VCHE}
\frac{\partial\mathbf{q}}{\partial t}
+ \boldsymbol{\mathbf{u}\cdot\nabla\mathbf{q}}
= \boldsymbol{\mathbf{q}\cdot\nabla\mathbf{u}} 
\,+ \nu\Delta\mathbf{q}\,,
\quad \hbox{where} \quad
\mathbf{q}\equiv{\rm curl}\,\mathbf{v}\,,
\quad \hbox{for VCHE},
\end{equation}
and
\begin{eqnarray} \label{vortex-stretching-Leray}
\frac{\partial\boldsymbol{\omega}}{\partial t}
+ \langle\mathbf{U}\,\rangle_{\ell}
\boldsymbol{\cdot\nabla\omega}
&=& \boldsymbol{\omega\cdot\nabla}\langle\mathbf{U}\,\rangle_{\ell} 
\,+ \nu\Delta\boldsymbol{\omega}
\quad \hbox{for Leray's equation}
\nonumber\\
&&+\, \Big[\boldsymbol{\nabla}\,U_j
\times\boldsymbol{\nabla}\,\langle{U}^j\,\rangle_{\ell}\Big]\,,
\ \hbox{where} \
\boldsymbol{\omega}\equiv{\rm curl}\,\mathbf{U}\,,
\end{eqnarray}
and we have used incompressibility of the filtered velocity,
$\langle\mathbf{U}\,\rangle_{\ell}$. Thus, the right hand side of the
vorticity dynamics for Leray's equation contains an additional source term,
compared to the curl of the VCHE or NS-$\alpha$ equation.

\paragraph{Outlook for the remainder of the paper.}
We shall apply the results of Holm, Marsden and
Ratiu~\cite{HMR[1998a]},~\cite{HMR[1998b]}, to recast the {\bfi Lagrangian mean
motion} (LMM) model and the order $O(|\boldsymbol\xi|^2)$ compressible model
into the Euler-Poincar\'e framework. Euler-Poincar\'e systems are the Lagrangian
version of Lie-Poisson Hamiltonian systems. Reformulating the LMM equations this
way facilitates their Eulerian analysis, e.g., by providing their Kelvin-Noether
circulation theorem, as well as energy and momentum conservation as part of a
general framework. We shall then use the equations of the LMM model in making a
natural adaptation and development of the turbulence modeling results of Chen et
al.~\cite{Chen etal[1998a]}--~\cite{Chen etal[1998c]}.
Namely, we shall formulate a {\bfi second moment closure model for turbulence}
based on adding a certain viscosity term to the ideal LMM equations. This
Lagrangian mean turbulence closure model will then be adapted to include
rotation and stratification for potential applications in geophysical problems.
Lower-dimensional examples of  Lagrangian mean theories will also be considered
and then we shall turn our attention to developing  {\bfi Eulerian mean
theories} using the Euler-Poincar\'e framework, as well.


\section{Second moment closure equations} 
\label{2pt Eul closure-sec}

  
\subsection{Euler-Poincar\'e formulation}  

We recall that the variables in the Lagrangian mean theory with averaged
approximate Lagrangian $\langle L\rangle$ in equation (\ref{mean-Lag-approx})
are the Lagrangian mean velocity $\mathbf{u}$ and the advected, or ``frozen-in''
quantities $D$ and $\langle\xi^k\xi^l\rangle$: the volume element and the
Lagrangian mean covariance of the fluctuating displacement. Such quantities
satisfy a certain Lie-derivative relation~\cite{HMR[1998a]}, such as the
continuity equation (\ref{conteqn}) for the volume element $D$,
\begin{equation}\label{Dee-eqn}
0 = \frac{\partial}{\partial t}\Big|_{\mathbf{a}}(d^{\,3}a)
= \bigg(\frac{\partial}{\partial t}+\pounds_{\mathbf{u}}\bigg)(Dd^{\,3}x)
= \bigg(\frac{\partial D}{\partial t} 
+ \boldsymbol{\nabla\cdot}(D\mathbf{u})\bigg)(d^{\,3}x)\,,
\end{equation}
where $\pounds_{\mathbf{u}}$ denotes Lie derivative with respect to the
Lagrangian mean fluid velocity, $\mathbf{u}(\mathbf{x},t)$. There is also
the geometrical relation (\ref{xi-eqn}), or its equivalent commutator form
(\ref{invar-xi}), for the slow time evolution of the fluctuation components
$\xi^k$, for $k=1,2,3$. Namely,
\begin{equation}\label{xi-eqn2}
0 \ =\ \frac{\partial}{\partial t}\Big|_{\mathbf{a}}\
\bigg(\tilde{\xi}^A
\frac{\partial}{\partial a^A}\bigg)
= \bigg(\frac{\partial}{\partial t}+\pounds_{\mathbf{u}}\bigg)
\bigg(\xi^k\frac{\partial}{\partial x^k}\bigg)
= \bigg(\frac{\partial \xi^k}{\partial t} 
+ u^j\, \xi^k_{,j} - \xi^j\, u^k_{,j}
\bigg)\frac{\partial}{\partial x^k}\,.
\end{equation}
This equation for $\xi^k$ implies a geometrical relation of the same type
for {\it all} the statistical moments $\langle\xi^k\xi^l\dots\xi^m\rangle$.
In particular, the second moment $\langle\xi^k\xi^l\rangle$ (the Lagrangian
mean covariance of the fluctuations) satisfies, cf. equation
(\ref{covariance-dynamics-index}),
\begin{eqnarray}\label{xi-xi-eqn1}
0 &=& \frac{\partial}{\partial t}\Big|_{\mathbf{a}}\
\bigg(\langle\tilde{\xi}^A\tilde{\xi}^B\rangle
\frac{\partial}{\partial a^A}\otimes\frac{\partial}{\partial a^B}\bigg)
= \bigg(\frac{\partial}{\partial t}+\pounds_{\mathbf{u}}\bigg)
  \bigg(\langle\xi^k\xi^l\rangle
  \frac{\partial}{\partial x^k}\otimes\frac{\partial}{\partial x^l}\bigg)
\\
&=& \bigg(\frac{\partial}{\partial t} \langle\xi^k\xi^l\rangle
+ u^j\, \langle\xi^k\xi^l\rangle_{,j}
- \langle\xi^j\xi^l\rangle\,u^k_{,j}
- \langle\xi^k\xi^j\rangle\,u^l_{,j}
\bigg)
\frac{\partial}{\partial x^k}\otimes\frac{\partial}{\partial x^l}\,.
\end{eqnarray}
In vector notation this equation for the Lagrangian mean covariance dynamics is,
cf. equation (\ref{covariance-dynamics-vector}),
\boxeq{2}
\begin{equation}\label{xi-xi-bold-1}
\frac{d}{dt}\boldsymbol{\langle\xi\xi\rangle}
= \boldsymbol{\langle\xi\xi\rangle\cdot\nabla\mathbf{u}}
+ \boldsymbol{\nabla\mathbf{u}}^{\rm T}
\boldsymbol{\cdot\langle\xi\xi\rangle}\,.
\end{equation}
Thus, the Taylor hypothesis (\ref{xi-eqn}) provides an approximate equation for
the evolution of the fluctuation statistical moments; in particular, for their
Lagrangian mean covariance. The Eulerian components of this equation can be
rewritten as
\begin{equation}\label{xi-xi-eqn}
\frac{\partial}{\partial t} \langle\xi^k\xi^l\rangle
 = \Big(\!\!
- \langle\xi^k\xi^l\rangle_{,j}
+ \langle\xi^a\xi^l\rangle\,\partial_a\delta^k_j
+ \langle\xi^k\xi^a\rangle\,\partial_a\delta^l_j
\Big)u^j\,,
\end{equation}
in which the right hand side is expressed as a differential operator
acting on $u^j(\mathbf{x},t)$. This operator will reappear in the
Lie-Poisson Hamiltonian formulation of the ideal second moment equations in
Section \ref{Ham-LPB-sec}. Note that the isotropic, homogeneous initial
condition $\langle\xi^k\xi^l\rangle = \delta^{kl}$ is {\it not} invariant
under the dynamics of equation (\ref{xi-xi-eqn}) for nontrivial velocity
shear. Thus, when shear is present, the Lagrangian mean covariance of
the fluctuations will {\it not} remain isotropic and homogeneous under the LMM
dynamics, even if it were initially so.

Using results of Holm, Marsden and Ratiu~\cite{HMR[1998a]},
we compute the Euler-Poincar\'e equation for the Lagrangian $\langle
L\rangle(\mathbf{u},D,\langle\xi^k\xi^l\rangle)$ depending on the
Lagrangian mean velocity $\mathbf{u}$, and advected quantities
$D$ and $\langle\xi^k\xi^l\rangle$. This Euler-Poincar\'e equation is given
by the following extension of equation (\ref{EPeqn}),
\boxeq{5}
\begin{eqnarray}\label{EPeqn-Lbar}
0 &=& \left(\frac{\partial}{\partial t} 
+ u^j\frac{\partial}{\partial x^j}\right)
\frac{1}{D}\frac{\delta \langle L\rangle}{\delta u^i} 
+ \frac{1}{D}\frac{\delta \langle L\rangle}{\delta u^j}u^j_{,i} 
- \bigg(\frac{\delta \langle L\rangle}{\delta D}\bigg)_{,i}
\nonumber\\
&& 
+\, \frac{1}{D}\bigg[\,
\frac{\delta \langle L\rangle}{\delta \langle\xi^k\xi^l\rangle}
\langle\xi^k\xi^l\rangle_{,i}
+ \bigg(\frac{\delta \langle L\rangle}{\delta \langle\xi^i\xi^l\rangle}
\langle\xi^k\xi^l\rangle\bigg)_{,k}
+ \bigg(\frac{\delta \langle L\rangle}{\delta \langle\xi^k\xi^i\rangle}
\langle\xi^k\xi^l\rangle\bigg)_{,l}\,
\bigg]\,.
\end{eqnarray}
Thus, the additional advected quantities $\langle\xi^k\xi^l\rangle$ in
general contribute their own {\bfi reactive forces}, appearing in the second 
line of the equation of motion (\ref{EPeqn-Lbar}). We compute the following
variational derivatives of the averaged approximate Lagrangian $\langle
L\rangle$ in equation (\ref{mean-Lag-approx})
\begin{eqnarray}\label{mean-Lag-der2}
\frac{1}{D} \frac{\delta\langle L\rangle }{\delta \mathbf{u}} 
&=& \mathbf{u} 
 - \frac{1}{D} \Big(\partial_k\,
D\langle\xi^k\xi^l\rangle \partial_l\Big) \mathbf{u}
\equiv \mathbf{v},
\nonumber\\
\frac{\delta\langle L\rangle }{\delta D} 
&=& - p  
+ \frac{1}{2}|\mathbf{u}\,|^2 
+ \frac{1}{2}\langle\xi^k\xi^l\rangle 
\big(\mathbf{u}_{,k}\boldsymbol\cdot\mathbf{u}_{,l}\big)
\equiv -P_{tot}, 
\nonumber\\
\frac{\delta\langle L\rangle }{\delta p} 
&=& 1 - D \,,
\nonumber\\
\frac{\delta \langle L\rangle}{\delta \langle\xi^k\xi^l\rangle}
&=& \frac{D}{2}\big(\mathbf{u}_{,k}\boldsymbol\cdot\mathbf{u}_{,l}\big)
\,.
\end{eqnarray}
Consequently, the Lagrangian mean Euler-Poincar\'e equation (\ref{EPeqn-Lbar})
for this averaged Lagrangian (after setting $D=1$) takes the form,
\boxeq{6}
\begin{eqnarray}\label{EPeqn-Lbar-vee}
&&\hspace{-.5in}
\frac{\partial v_i}{\partial t} + u^j v_{i,j} + v_j u^j_{,i} 
\,=\, \Big( - p  
+ \frac{1}{2}|\mathbf{u}\,|^2 
+ \frac{1}{2}\langle\xi^k\xi^l\rangle 
\big(\mathbf{u}_{,k}\boldsymbol\cdot\mathbf{u}_{,l}\big)
\Big)_{,i}
\\
&& 
-\, \frac{1}{2}\,\bigg[\,
\big(\mathbf{u}_{,k}\boldsymbol\cdot\mathbf{u}_{,l}\big)
\langle\xi^k\xi^l\rangle_{,i}
+ \Big(\big(\mathbf{u}_{,i}\boldsymbol\cdot\mathbf{u}_{,l}\big)
\langle\xi^k\xi^l\rangle\Big)_{,k}
+ \Big(\big(\mathbf{u}_{,k}\boldsymbol\cdot\mathbf{u}_{,i}\big)
\langle\xi^k\xi^l\rangle\Big)_{,l}\,
\bigg]\,,
\nonumber\\
&&\hbox{where}\quad
v_i = {u}_i - 
\big(\partial_k\,\langle\xi^k\xi^l\rangle \partial_l\big) {u}_i
\quad\hbox{and}\quad
{u}^i_{,i}=0
\,.
\label{vee-redef}
\end{eqnarray}
Again one sees in the entire second line of equation
(\ref{EPeqn-Lbar-vee}) the reactive forces arising from the variations of the
Lagrangian with respect to the covariance $\langle\xi^k\xi^l\rangle$.

\paragraph{Contrasting the Euler-Poincar\'e equation
(\ref{EPeqn-Lbar-vee}) with the CH equation.} In the present case, the
advected second moments $\langle\xi^k\xi^l\rangle$ for the full system
satisfy their dynamical equation (\ref{xi-xi-bold-1}). If isotropic,
homogeneous statistics were {\it prescribed} instead, so that
$\langle\xi^k\xi^l\rangle=\alpha^2\delta^{kl}$ with constant $\alpha^2$, then
both equation (\ref{xi-xi-bold-1}) and the entire second line in the motion
equation (\ref{EPeqn-Lbar-vee}) would be {\it absent}, (since the corresponding
variations in $\langle\xi^k\xi^l\rangle$ would not be taken in this case) and
the motion equation (\ref{EPeqn-Lbar-vee}) of the Lagrangian mean model then
would return to the $n$-dimensional generalization of the CH equation, i.e., the
set (\ref{vee-eqn}) -- (\ref{isotropy-cond}) introduced
in~\cite{HMR[1998a]},~\cite{HMR[1998b]}. Note, however, that the $n$-dimensional
CH equation set is {\it not} an invariant subsystem of the Euler-Poincar\'e
system (\ref{EPeqn-Lbar-vee}), with definition (\ref{vee-redef}) and advection
law (\ref{xi-xi-bold-1}), because (as mentioned earlier) the initial condition
$\langle\xi^k\xi^l\rangle =
\delta^{kl}$ is {\it not} invariant under the dynamics of equation
(\ref{xi-xi-bold-1}) for nontrivial velocity shear.  This means that the
Lagrangian mean model is a departure from the VCHE model, rather than being an
extension of it. We shall return to this matter in Section
\ref{EMM-sec}, when we discuss Eulerian mean fluid models.


\subsection{Momentum conservation -- stress tensor formulation}

Noether's theorem guarantees there is a conserved momentum for the
Euler-Poincar\'e equations (\ref{EPeqn-Lbar-vee}), since the averaged
approximate  Lagrangian $\langle L\rangle$ in equation
(\ref{mean-Lag-approx}) has no explicit spatial dependence. Moreover, the
integrand ${\cal L}$ in this Lagrangian is a {\it polynomial} in the
Lagrangian mean velocity
$\mathbf{u}$, its gradient
$\mathbf{u}_{,k}$, and the advected quantities $D$ and
$\langle\xi^k\xi^l\rangle$. That is,
\begin{equation} \label{Lag-3d-approx}
\langle L\rangle
=\int d^{\,3}x\
{\cal L}\Big(\mathbf{u},\mathbf{u}_{,k\,},
\,D,\langle\xi^k\xi^l\rangle\Big)\,,
\end{equation}
with ${\cal L}$ a polynomial function of its arguments.
In this case, we may express the Lagrangian mean Euler--Poincar\'e equations
(\ref{EPeqn-Lbar}) in the {\bfi momentum conservation form},
\begin{equation} \label{mom-cons}
\frac{\partial m_i}{\partial t} 
= -\ \frac{\partial }{\partial x^j}T^j_i\,,
\end{equation}
with {\bfi momentum density components} $m_i$, $i=1,2,3$ defined by
\begin{equation} \label{mom-comp}
m_i \equiv \frac{\delta \langle L\rangle}{\delta u^i}
= \frac{\partial{\cal L}}{\partial u^i}
- \frac{\partial}{\partial x^k}
\left(\frac{\partial{\cal L}}{\partial
u^i_{,k}}\right),
\end{equation}
and {\bfi stress tensor} $T^j_i$ given by
\smallskip\boxeq{5}\bigskip
\begin{eqnarray} \label{stress-tens}
T^j_i &=&
\left.
m_i u^j - \frac{\partial{\cal L}}{\partial u^k_{,j}}u^k_{,i}
+ \frac{\partial{\cal L}}{\partial\langle\xi^i\xi^l\rangle}
\langle\xi^j\xi^l\rangle
+ \frac{\partial{\cal L}}{\partial \langle\xi^k\xi^i\rangle}
\langle\xi^k\xi^j\rangle
\right.
\nonumber \\ &&
\left.
+\ \delta^j_i\left({\cal L}-D\frac{\partial{\cal L}}{\partial D}
\right)
\right.\,.
\end{eqnarray}
Equation (\ref{mom-cons}) then implies conservation of the domain-integrated
momentum, $\int \mathbf{m}\, d^{\,3}x$, provided the
normal component of the stress tensor $T^j_i$ vanishes on the boundary.

In our particular case, expression (\ref{stress-tens}) for the stress
tensor $T^j_i$ simplifies remarkably, to become
\begin{equation} \label{stress-tens-eval}
T^j_i = m_iu^j + p\delta^j_i\,,
\quad\hbox{where}\quad
m_i\big|_{D=1}
= v_i \equiv {u}_i - 
\big(\partial_k\,\langle\xi^k\xi^l\rangle \partial_l\big) {u}_i
\,.
\end{equation}
Consequently, the equivalent Euler-Poincar\'e motion equation
(\ref{EPeqn-Lbar-vee}) simplifies to
\begin{equation} \label{EP-aug-motion1}
\frac{\partial v_i}{\partial t} 
= -\ \frac{\partial }{\partial x^j}(v_i u^j + p\delta^j_i)\,.
\end{equation}
%

\paragraph{The ideal Lagrangian mean motion (LMM) model.}
Using incompressibility ($u^j_{,j}=0$) reduces the expression
(\ref{EP-aug-motion1}) for momentum conservation to the {\bfi Lagrangian mean
motion} (LMM) equation for an ideal fluid, cf. equation
(\ref{LMM-mot-Euler-form1}),
\medskip\boxeq{5}\bigskip
\begin{eqnarray} \label{EP-aug-motion2}
\frac{\partial\mathbf{v}}{\partial t} 
&+& \mathbf{u}\boldsymbol{\cdot\nabla}\mathbf{v}
= -\ \boldsymbol\nabla{p}\,,
\quad \boldsymbol{\nabla\cdot\mathbf{u}}=0\,,
\\
\hbox{where}&&
\mathbf{v} \equiv  \frac{1}{D} 
\frac{\delta\langle L\rangle}{\delta\mathbf{u}}\bigg|_{D=1} =
\Big( 1-\partial_k\,\langle\xi^k\xi^l\rangle\partial_l\Big)\ \mathbf{u} 
\,.\label{LMM-vee-def}
\end{eqnarray}
The boundary conditions are recalled from equations (\ref{v-bdy-cond}),
(\ref{u-bdy-cond}) and (\ref{bdy-conds}) as 
\begin{equation} \label{bc-redux}
\mathbf{v}\boldsymbol{\cdot\hat{n}}=0\,,
\quad
\mathbf{u} = 0\,,
\quad\hbox{and}\quad
\boldsymbol{\langle\xi\xi\rangle\cdot\hat{n}}=0,
\quad\hbox{on a fixed boundary.}
\end{equation} 
In (\ref{EP-aug-motion2}) the continuity
equation, $\boldsymbol{\nabla\cdot}\mathbf{u}=0$, allows the Lagrangian mean
pressure $p$ to be determined from an elliptic equation, by virtue of the
commutation relation (\ref{remarkable-comm-rel}). The auxiliary equation
(\ref{xi-xi-bold-1}) for the slow time dynamics of the advected Lagrangian mean
covariance may be rewritten as, 
\medskip\boxeq{2}
\begin{equation}\label{xi-xi-eqn-redux}
\bigg(\frac{\partial}{\partial t} 
+ \mathbf{u}\boldsymbol{\cdot\nabla}\bigg)
 \langle\xi^k\xi^l\rangle
= \langle\xi^j\xi^l\rangle\,u^k_{,j}
+ \langle\xi^k\xi^j\rangle\,u^l_{,j}\,.
\end{equation}
This equation updates the Lagrangian mean covariance $\langle\xi^k\xi^l\rangle$
and, thus, closes the LMM system.

\paragraph{Contrasting the LMM model with the Euler equations.}
Remarkably, the reaction forces due to the fluctuations in the Euler-Poincar\'e
motion equation (\ref{EPeqn-Lbar-vee}) {\it exactly cancel} the contributions of
the fluctuations in the pressure and line-element stretching term ($v_j
u^j_{,i}$). This cancellation leaves only one main effect: the fluid parcels in
the LMM model are transported by velocity $\mathbf{u}$ instead of $\mathbf{v}$.
Had we made no approximation at all in our Lagrangian $L(\omega)$
before averaging, we would have gotten exact cancellation of all of the
fluctuational effects and returned entirely to Euler's equations.

So, instead of returning us entirely to the Euler equations, the Taylor
series approximation we made in Section \ref{T-series-approx} before
averaging the Lagrangian $L(\omega)$ produces one essential difference
between the LMM model in equations (\ref{EP-aug-motion2}) --
(\ref{xi-xi-eqn-redux}) and the original Euler equations (\ref{Euler-eqns}).
Namely, the transport velocity $\mathbf{u}$ is {\it smoothed relative to}
$\mathbf{v}$, by inversion of the {\bfi dynamical Helmholtz operator},
\begin{equation}\label{dyn.Helm.op}
\mathbf{u}=(1-\tilde\Delta )^{-1}\mathbf{v}
\quad\hbox{where}\quad
1-\tilde\Delta 
=  1-\partial_k\,\langle\xi^k\xi^l\rangle\partial_l\,,
\end{equation}
whose statistical `metric' $\langle\xi^k\xi^l\rangle$ changes and adapts
dynamically according to equation (\ref{xi-xi-eqn-redux})  as the fluid moves.
Smoothing the transport velocity to improve the mathematical properties of
the incompressible fluid equations has been studied previously, going back
at least to the work of Leray~\cite{Leray[1934]} in the 1930s, as we discussed
earlier. However, the {\bfi adaptive smoothing} introduced by inverting the
dynamical Helmholtz operator (\ref{dyn.Helm.op}) is new here, as far as we
know. Euler's equations are recovered as an invariant subsystem of
the LMM equation set for $\boldsymbol{\langle\xi\xi\rangle}$ identically
zero. 

We recall the remarkable commutation relation
(\ref{remarkable-comm-rel})
\begin{equation}\label{remark-rel}
\bigg[\frac{d}{dt},\tilde{\Delta}_D \bigg] = 0\,,
\quad\hbox{where}\quad
\tilde{\Delta}_D\equiv 
D^{-1}\partial_k\,D\langle\xi^k\xi^l\rangle\partial_l\,,
\end{equation}
which may also be verified directly from the continuity equation
(\ref{conteqn}) and the Lagrangian mean covariance equation
(\ref{xi-xi-eqn}). Consequently, since
$\tilde{\Delta}_D|_{D=1}\equiv\tilde{\Delta}$, the ideal LMM motion equation
(\ref{EP-aug-motion2}) may be rewritten alternatively as \bigskip
\boxeq{2}
\begin{equation}\label{LMM-mot-u}
\frac{\partial\mathbf{u}}{\partial t} 
+ \mathbf{u}\boldsymbol{\cdot\nabla}\mathbf{u}
= -\ (1-\tilde{\Delta})^{-1}\boldsymbol\nabla{p}\,,
\quad \boldsymbol{\nabla\cdot\mathbf{u}}=0\,.
\end{equation}
Therefore, the effect of the advected fluctuations in this alternative
representation of the LMM equation  is simply to {\it smooth the pressure
gradient} in an adaptive fashion depending on the velocity shear, through
the dynamics of the covariance $\boldsymbol{\langle\xi\xi\rangle}$. 

As guaranteed in advance by the mathematical theory developed in
in~\cite{HMR[1998a]}, the Euler-Poincar\'e equation (\ref{EP-aug-motion2})
with definition (\ref{LMM-vee-def}) for the LMM model is  equivalent to the
Euler-Lagrange equation (\ref{Eul-Lag.eqn-approx}). Because
of the commutation property (\ref{remark-rel}), there is also an equivalent
alternative form of the LMM motion equation, given in (\ref{LMM-mot-u}). This
alternative, but equivalent, form of the motion equation for LMM provides
alternative interpretations of the effects of the fluctuation covariance on the
Lagrangian mean motion.

  
\subsection{Momentum conservation for the order $O(|\boldsymbol\xi|^2)$ model}

We recall the averaged approximate Lagrangian (\ref{mean-Lag-approx}) in
Eulerian form,
\begin{equation}\label{mean-Lag-approx-Eul-form}\hspace{-.2in}
\langle L\rangle = \int d^{\,3}x \bigg\{\frac{D}{2} \Big[|\mathbf{u}\,|^2 
+ \langle\xi^k\xi^l\rangle 
\Big(\mathbf{u}_{,k}\boldsymbol\cdot\mathbf{u}_{,l}\Big)\Big] 
+  \Big( p
- \frac{1}{2}\frac{\partial^2 p}
{\partial{x^k}\partial{x^l}}\langle\xi^k\xi^l\rangle\Big) 
- pD \bigg\}.
\end{equation}
Note that $p$ is the Lagrangian mean pressure and according to equation
(\ref{press-Eul-Lag}) the quantity in parentheses,
\begin{equation} \label{eul-mean-pressure}
P_E \equiv p - \frac{1}{2}\frac{\partial^2 p}
{\partial{x^k}\partial{x^l}}\langle\xi^k\xi^l\rangle
\,,
\end{equation}
is the {\bfi Eulerian mean pressure}. The momentum density components $m_i$,
$i=1,2,3$ for this Lagrangian are
\begin{equation} \label{mom-comp-order2}
m_i \equiv \frac{\delta \langle L\rangle}{\delta u^i}
= D(1-\tilde\Delta_D)u_i
\,,
\end{equation}
where $\tilde{\Delta}_D$ satisfies (\ref{remarkable-comm-rel}) and
\begin{equation} \label{D-def-order2}
D = 1 - \frac{1}{2}\langle\xi^i\xi^l\rangle_{,k\,l}
\,.
\end{equation}
The stress tensor $T^j_i$ is obtained by using equation (\ref{stress-tens}) as
\begin{equation} \label{stress-tens-order2}
T^j_i = m_iu^j 
- \frac{1}{2}\frac{\partial^2 p}
{\partial{x^i}\partial{x^l}}\langle\xi^j\xi^l\rangle 
+ \delta^j_i \Big( p - \frac{1}{2}\frac{\partial^2 p}
{\partial{x^k}\partial{x^l}}\langle\xi^k\xi^l\rangle \Big)
\,.
\end{equation}
The off-diagonal components $- \frac{1}{2}p_{,i\,l}\langle\xi^j\xi^l\rangle$ may
be regarded as a {\bfi fluctuation stress tensor}. Using the continuity equation
(\ref{conteqn}) for $D$ transforms the momentum conservation law
(\ref{mom-cons}) into the equivalent motion equation for the {\bfi LMM model
with second order compressibility}, cf. equation
(\ref{Eul-Lag.eqn-approx-Euler-form1}),
\boxeq{4}
\begin{eqnarray} \label{LMM-motion2}
\Big(\frac{\partial}{\partial t} 
+ 
\mathbf{u}\boldsymbol{\cdot\nabla}\Big)(1-\tilde\Delta_D)u_i
&=& 
-\, \frac{1}{D}\frac{\partial P_E}{\partial x^i}
+\, \frac{1}{D}\frac{\partial}{\partial x^j}
\Big(\frac{\partial^2 p}
{\partial{x^i}\partial{x^l}}\langle\xi^j\xi^l\rangle \Big)
\,,\\
\hbox{with}\quad
\boldsymbol{\nabla\cdot\mathbf{u}}
&=&
-\, \frac{1}{D}\frac{dD}{dt}
=
\frac{1}{2}\frac{d}{dt}\langle\xi^i\xi^l\rangle_{,k\,l}
+ O(|\boldsymbol\xi|^4)
\,.\label{Dee-comp}
\end{eqnarray}
The boundary conditions for this model are the same as in
(\ref{bc-redux}), namely,
\begin{equation} \label{bc-again}
\mathbf{v}\boldsymbol{\cdot\hat{n}}=0\,,
\quad
\mathbf{u} = 0\,,
\quad\hbox{and}\quad
\boldsymbol{\langle\xi\xi\rangle\cdot\hat{n}}=0,
\quad\hbox{on a fixed boundary.}
\end{equation} 
This completes the transformation of the Lagrangian form
(\ref{Eul-Lag.eqn-approx}) of this motion equation to its equivalent Eulerian
form. As guaranteed by the Euler-Poincar\'e approach, the equivalence of
equations (\ref{Eul-Lag.eqn-approx}) and (\ref{LMM-motion2}) may also be
verified by a direct calculation.

  
\subsection{Kelvin circulation theorem for the Lagrangian mean model}

Being Euler--Poincar\'e, the incompressible LMM equation (\ref{EP-aug-motion2})
has a corresponding {\bfi Kelvin-Noether circulation theorem}. Namely,
this equation implies
\begin{equation}\label{KelThm}
\frac{ d}{dt}\oint_{\gamma(\mathbf{u})}\mathbf{v}
\boldsymbol{\cdot}d\mathbf{x} 
= \oint_{\gamma(\mathbf{u})}\Big[
\frac{\partial\mathbf{v}}{\partial t}
+\mathbf{u}\boldsymbol{\cdot\nabla}\mathbf{v}
+ v_j \boldsymbol{\nabla} u^j\Big]
\boldsymbol{\cdot}d\mathbf{x}
= \oint_{\gamma(\mathbf{u})}
\mathbf{v}\boldsymbol{\cdot}d\mathbf{u}
\,,
\end{equation}
for any closed curve ${\gamma(\mathbf{u})}$ that moves with the 
Lagrangian mean fluid velocity $\mathbf{u}$. This expression for the
Kelvin-Noether property of the Lagrangian mean motion equation in
3D is reminiscent of corresponding expressions in wave, mean-flow
interaction theory~\cite{GH[1996]}. See also the Leray equation result
(\ref{KelThm-Leray-eqn}). The main point is that the presence of the
fluctuation covariance
$\langle\xi^k\xi^l\rangle$ {\bfi creates circulation} of the total specific
momentum $\mathbf{v}=\mathbf{u}  -
(\partial_k\,\langle\xi^k\xi^l\rangle\partial_l)\mathbf{u}$.

Alternatively, from equation (\ref{LMM-mot-u}) we may write {\it another}
Kelvin-Noether circulation theorem, namely 
\begin{equation}\label{LMM-KelThm-u-thm}
\frac{ d}{dt}\oint_{\gamma(\mathbf{u})}\mathbf{u}
\boldsymbol{\cdot}d\mathbf{x}
= - \oint_{\gamma(\mathbf{u})}
\Big((1-\tilde{\Delta})^{-1}\boldsymbol{\nabla}{p}\Big)
\boldsymbol{\cdot}d\mathbf{x}
\,,
\end{equation}
which represents the circulation dynamics of $\mathbf{u}$, rather than
$\mathbf{v}$. Thus, the fluctuation covariance creates circulation of both
$\mathbf{u}$ and $\mathbf{v}$.

  
\subsection{Vortex stretching equation for the Lagrangian mean model}

In three dimensions, we may use a vector identity to
re-express the LMM equation (\ref{EP-aug-motion2})
in its equivalent ``curl'' form, as
\begin{equation}\label{3d:curl-mot.eqn}
\frac{\partial }{\partial t}\mathbf{v}
- \mathbf{u} \times \Big(\boldsymbol{\nabla} \times \mathbf{v}\Big)
+ \boldsymbol{\nabla}\,{p} 
+ u^j\boldsymbol{\nabla}\,v_j
= 0\,,
\quad \boldsymbol{\nabla\cdot\mathbf{u}}=0\,.
\end{equation}
The curl of this equation in turn yields a transport and creation equation
for the {\bfi Lagrangian mean vorticity}, $\mathbf{q}\equiv{\rm
curl}\,\mathbf{v}$,
\begin{equation} \label{vortex-stretching}
\frac{\partial\mathbf{q}}{\partial t}
+ \boldsymbol{\mathbf{u}\cdot\nabla\mathbf{q}}
= \boldsymbol{\mathbf{q}\cdot\nabla\mathbf{u}} 
\,+ \Big[\boldsymbol{\nabla}\,v_j\times\boldsymbol{\nabla}\,u^j\Big]\,,
\quad \hbox{where} \quad
\mathbf{q}\equiv{\rm curl}\,\mathbf{v}\,,
\end{equation}
and we have used incompressibility of $\mathbf{u}$. Thus, $\mathbf{u}$ is the
transport velocity for the generalized vorticity $\mathbf{q}$ and the
expected {\bfi vortex stretching} term $\mathbf{q}\cdot\nabla\mathbf{u}$ is
accompanied by an additional {\bfi vortex creation} term, 
$\boldsymbol{\nabla}\,v_j\times\boldsymbol{\nabla}\,u^j$. Of course, this
additional term is also responsible for the creation of circulation of
$\mathbf{v}$ in the Kelvin-Noether circulation theorem (\ref{KelThm}). In
particular, Stokes' theorem and equation (\ref{KelThm}) imply
\begin{equation}\label{KelThm-Stokes}
\frac{ d}{dt}\int\int_{S(\mathbf{u})}{\rm curl}\,\mathbf{v}
\boldsymbol{\cdot}d\mathbf{S} 
= \int\int_{S(\mathbf{u})}
d{v}_j{\wedge}d{u}^j
= \int\int_{S(\mathbf{u})}
\Big[\boldsymbol{\nabla}\,v_j\times\boldsymbol{\nabla}\,u^j\Big]
\boldsymbol{\cdot}d\mathbf{S}
\,,
\end{equation}
where the curve ${\gamma(\mathbf{u})}$ is the boundary of the
surface ${S(\mathbf{u})}$. One may compare this result with the Leray case,
in equation (\ref{vortex-stretching-Leray}). 

Alternatively, from equation (\ref{LMM-mot-u}) we may write {\it another}
form of the vorticity dynamics, for curl$\,\mathbf{u}$ rather than
curl$\,\mathbf{v}$, namely 
\begin{equation} \label{vortex-stretching-u}
\frac{\partial\boldsymbol{\omega}}{\partial t}
+ \mathbf{u}\,\boldsymbol{\cdot\nabla\omega}
= 
\boldsymbol{\omega\cdot\nabla}\mathbf{u}\ 
-\ 
\boldsymbol{\nabla}\times 
\Big((1-\tilde{\Delta})^{-1}\boldsymbol\nabla{p}\Big)
\ \hbox{where} \
\boldsymbol{\omega}\equiv{\rm curl}\,\mathbf{u}\,,
\end{equation}
and we have used incompressibility of the Lagrangian mean velocity
$\mathbf{u}$.

  
\subsection{Energetics of the Lagrangian mean model}

The sum of the inner products of $\mathbf{v}$ with equation
(\ref{LMM-mot-u}) and $\mathbf{u}$ with equation (\ref{3d:curl-mot.eqn}) yields
{\bfi conservation of energy},
\begin{equation}\label{cons-erg-def}
E =  \frac{1}{2}\int d^{\,3}x \Big(|\mathbf{u}\,|^2 
+ \langle\xi^k\xi^l\rangle 
\mathbf{u}_{,k}\boldsymbol\cdot\mathbf{u}_{,l}\Big)
= \frac{1}{2}\int d^{\,3}x\ \mathbf{u}\boldsymbol\cdot\mathbf{v}
\,,
\end{equation}
after integrating by parts, using incompressibility and applying the
boundary conditions (\ref{bc-redux}). Naturally, this energy is also
conserved in $n$ dimensions. Thus, the covariance of the fluctuations couples to the
gradients of the Lagrangian mean velocity. So it costs the system energy either to
increase these gradients, or to increase the covariance. Moreover, there is a direct
feedback between $\boldsymbol{\nabla\mathbf{u}}$ and the dynamics of
$\boldsymbol{\langle\xi\xi\rangle}$ given in equation (\ref{xi-xi-eqn-redux}).
We shall see in Section \ref{Ham-LPB-sec}
that Legendre transforming the  Lagrangian $\langle L\rangle$ in
(\ref{mean-Lag-approx})  gives the following {\bfi Hamiltonian} (still
expressed in terms of the velocity $\mathbf{u}$, instead of the momentum
density $\mathbf{m}= \delta \langle L\rangle/ \delta \mathbf{u}
= D \mathbf{v}$),
\begin{equation}\label{Ham-def-uv}
H = \int_{\cal M} d^{\,n}x\ \Big[\,\frac{D}{2} 
\Big(|\mathbf{u}\,|^2 
+ \langle\xi^k\xi^l\rangle 
\mathbf{u}_{,k}\boldsymbol\cdot\mathbf{u}_{,l}\Big) + p(D-1)\Big]
\,.
\end{equation}
%
\paragraph{Remark on the geodesic property of the LMM model.}
When evaluated on the constraint manifold $D=1$, the Lagrangian in
(\ref{mean-Lag-approx}) and the Hamiltonian in (\ref{Ham-def-uv}) for the
Lagrangian mean motion equation (\ref{EP-aug-motion2}) coincide in
$n$ dimensions. This is expected for a stationary principle giving
rise to {\bfi geodesic motion}. The interpretation of the LMM model as
describing geodesic motion on the volume preserving diffeomorphism group
with respect to the $H_1$ metric given by (\ref{cons-erg-def}) will be discussed
elsewhere~\cite{Holm-Shkoller-inprep}. 

\paragraph{Conservation properties of the Lagrangian mean covariance.}
We recall equation (\ref{xi-xi-bold-1}) for
$\boldsymbol{\langle\xi\xi\rangle}$,
\begin{equation}\label{xi-xi-bold}
\frac{d}{dt}\boldsymbol{\langle\xi\xi\rangle}
= \boldsymbol{\langle\xi\xi\rangle\cdot\nabla\mathbf{u}}
+ \boldsymbol{\nabla\mathbf{u}}^{\rm T}
\boldsymbol{\cdot\langle\xi\xi\rangle}\,.
\end{equation}
Since tr$(\boldsymbol{\nabla\mathbf{u}})=0$, the Lagrangian mean covariance
$\boldsymbol{\langle\xi\xi\rangle}$ must always have an instantaneously
growing direction, along at least one principal axis of the velocity shear
tensor, $\boldsymbol{\nabla\mathbf{u}}$. So,
$\boldsymbol{\langle\xi\xi\rangle}$ might be systematically
growing with time. However, there are limits to this growth. In particular,
equation (\ref{det-rel}) shows that this growth in the incompressible case must
preserve the value of the determinant $\det\boldsymbol{\langle\xi\xi\rangle}$
along flow lines. Thus, stretching and rotation may occur, but for an
incompressible flow the volume of the ellipsoid composed of the principle axes
of the symmetric covariance tensor $\boldsymbol{\langle\xi\xi\rangle}$ must be
preserved on fluid parcels. In the Eulerian representation, equation
(\ref{det-rel}) implies  
\begin{equation}\label{det-rel-Eul}
\frac{d}{dt} \Big( D^2 \det\boldsymbol{\langle\xi\xi\rangle} \Big)
=0\,,
\quad\hbox{and}\quad
\frac{dD}{dt}=0
\quad\hbox{for}\quad
\boldsymbol{\nabla\cdot}\mathbf{u}=0
\,.
\end{equation}
This, in turn, implies conservation of the following quantity, for any
function $\Phi$,
\begin{equation}\label{det-Casimir}
C_{\Phi} = \int_{\cal M} d^{\,n}x\ D\, \Phi\big( D^2
\det\boldsymbol{\langle\xi\xi\rangle} \big)\,,
\quad\forall\,\Phi
\,,
\end{equation}
in which we may set $D=1$ for incompressible flow.

Conservation of the energy in equation (\ref{cons-erg-def}) also has
some indications for controlling the Lagrangian mean covariance. Of course,
this growth cannot continue indefinitely {\it in the same direction} and still
satisfy conservation of energy $E$ in (\ref{cons-erg-def}) in regions of
nontrivial shear. So the conservation of energy must also eventually saturate
this potential growth in Lagrangian mean covariance. Thus, from the viewpoint of
energetics, while $\boldsymbol{\langle\xi\xi\rangle}$ must always have a
direction in which it is growing, this growth must also occur to preserve the
determinant $\det\boldsymbol{\langle\xi\xi\rangle}$ and to limit the velocity
shear $\boldsymbol{\nabla\mathbf{u}}$ in accordance with the available energy in
equation (\ref{cons-erg-def}). Therefore, the direction of growth (along the
instantaneous principle axes of $\boldsymbol{\nabla\mathbf{u}}$ corresponding
to its positive eigenvalues) will keep changing, because of the energetic
coupling and direct feedback between $\boldsymbol{\nabla\mathbf{u}}$ and
$\boldsymbol{\langle\xi\xi\rangle}$. A dissipative modification of the
$\boldsymbol{\langle\xi\xi\rangle}$ dynamics that allows relaxation to the
homogeneous isotropic conditions of the VCHE model was introduced in equation
(\ref{xi-xi-bold-alt-dis}). This modification may be used, in the event that the
growth of $\boldsymbol{\langle\xi\xi\rangle}$ should require additional control,
e.g., in numerical simulations.


\section{Hamiltonian structure of the Lagrangian mean model}
\label{Ham-LPB-sec}

Being Euler-Poincar\'e, the LMM system consisting of the motion equation
(\ref{EPeqn-Lbar-vee}), the continuity equation (\ref{conteqn}) and the
Lagrangian mean covariance equation (\ref{xi-xi-eqn}) must also be a
Lie-Poisson Hamiltonian system. This may be verified using standard
methods, see, e.g.,~\cite{HMR[1998a]}. The corresponding Lie-Poisson
bracket is dual to the semidirect product Lie algebra
\begin{equation}\label{LP-alg}
\mathfrak{g} = \mathfrak{u} \circledS\
(\Lambda^0 \,\oplus\, S)\,,
\end{equation}
consisting of vector fields $\eta\in\mathfrak{u}$ acting from the right 
on the direct sum of functions $f\in\Lambda^0$ and metrics $g\in S$. The
semidirect product Lie algebra bracket has the expression
\begin{equation}\label{semidirectalgebraright}
[(\eta_1, f_1, g_1), (\eta_2, f_2, g_2)]
= ([\eta_1,\eta_2],\, f_1 \eta_2 - f_2 \eta_1,\, g_1 \eta_2 - g_2 \eta_1),
\end{equation}
where we denote the induced action of $\mathfrak{u}$ on 
$(\Lambda^0 \,\oplus\, S)$ from the right by
concatenation, as in $f_1 \eta_2$. Dual coordinates are: 
$\mathbf{m}=\delta\langle L\rangle/\delta\mathbf{u}=D\mathbf{v}$ dual
to $\mathfrak{u}$; $D$ dual to $\Lambda^0$; 
and $\langle\xi^k\xi^l\rangle$ dual to
$S$. The Legendre transformation of the averaged approximate Lagrangian
$\langle L\rangle$ in (\ref{mean-Lag-approx}) produces the Hamiltonian,
\begin{equation}\label{Ham-def}
H = \int d^{\,3}x \bigg[\frac{1}{2}\mathbf{m}\boldsymbol\cdot
\Big(D-\partial_k\,D\langle\xi^k\xi^l\rangle \partial_l\Big)^{-1}\mathbf{m}
+ p (D-1)\bigg],
\end{equation}
whose proper definition requires defining the inverse of the
generalized Helmholtz operator,
$\big(D-\partial_k\,D\langle\xi^k\xi^l\rangle \partial_l\big)$ and 
using the boundary conditions (\ref{bc-redux}). The variational
derivatives of this Hamiltonian are given by
\begin{eqnarray}\label{Ham-var}
\delta H &=& \int d^{\,3}x \bigg[\mathbf{u}\boldsymbol\cdot\delta\mathbf{m}
- \Big( p  
- \frac{1}{2}|\mathbf{u}\,|^2 
- \frac{1}{2}\langle\xi^k\xi^l\rangle 
\big(\mathbf{u}_{,k}\boldsymbol\cdot\mathbf{u}_{,l}\big)
\Big)\delta D
\nonumber\\
&&\hspace{.5in}-\ \frac{D}{2}\big(\mathbf{u}_{,k}
\boldsymbol\cdot\mathbf{u}_{,l}\big)
\delta\langle\xi^k\xi^l\rangle
\,+\, (D-1)\delta p\bigg]\,,
\end{eqnarray}
where evenness of the generalized Helmholtz operator under integration by
parts is used in obtaining the first term. Finally, the Lie-Poisson bracket
defined on the dual of the algebra in equation (\ref{LP-alg}) is given
explicitly by
\begin{eqnarray}\label{LPB}
&&\hspace{-.5in}
\{F,H\}(\mathbf{m},D,\langle\xi^k\xi^l\rangle) 
\\
&& = - \, \int d^{\,3}x\ \bigg\{
\frac{\delta F}{\delta m_i}(\partial_j m_i+m_j \partial_i)
\frac{\delta H}{\delta m_j}
+\frac{\delta F}{\delta m_i}(D\partial_i)\frac{\delta H}{\delta D}
+\frac{\delta F}{\delta D}(\partial_j D)\frac{\delta H}{\delta m_j}
\nonumber\\
&&\hspace{.75in}
-\ \frac{\delta F}{\delta m_i}
\Big( \langle\xi^c\xi^d\rangle_{,i}
+ \partial_a\langle\xi^a\xi^d\rangle\,\delta^c_i
+ \partial_a\langle\xi^c\xi^a\rangle\,\delta^d_i\Big)
\frac{\delta H}{\delta\langle\xi^c\xi^d\rangle}
\nonumber\\
&&\hspace{.75in}
-\ \frac{\delta F}{\delta\langle\xi^k\xi^l\rangle}
\Big(\!\!
- \langle\xi^k\xi^l\rangle_{,j}
+ \langle\xi^a\xi^l\rangle\,\partial_a\delta^k_j
+ \langle\xi^k\xi^a\rangle\,\partial_a\delta^l_j\Big)
\frac{\delta H}{\delta m_j}
\bigg\}\,.
\nonumber
\end{eqnarray}
In the second line of this Lie-Poisson bracket we see the contributions of 
the fluctuation covariance to the motion equation, cf. (\ref{EPeqn-Lbar}),
and in the third line we see the operator in the covariance dynamics itself,
cf. (\ref{xi-xi-eqn}).

The dynamical system consisting of the motion equation
(\ref{EPeqn-Lbar-vee}), the continuity equation (\ref{conteqn}) and the
Lagrangian mean covariance equation (\ref{xi-xi-eqn}) now emerges as a
Lie-Poisson Hamiltonian system in the form 
\begin{equation}\label{mu-dot-syst}
\frac{\partial \mu}{\partial t}=\{\mu,H\},
\quad \hbox{with}\quad
\mu\in(\mathbf{m},D,\langle\xi^k\xi^l\rangle)\,,
\end{equation}
with Hamiltonian $H$ given in equation (\ref{Ham-def}) and implying
conservation of the following energy, cf. equation (\ref{erg-approx-lag-inc})
and equation (\ref{cons-erg-def}) with $\mathbf{m}\big|_{D=1}=\mathbf{v}$,
\begin{equation}\label{erg-def}
E = \frac{1}{2}\int d^{\,3}x \bigg[\mathbf{v}\boldsymbol\cdot
\Big(1-\partial_k\,\langle\xi^k\xi^l\rangle \partial_l\Big)^{-1}\mathbf{v}
\bigg]
= \frac{1}{2}\int d^{\,3}x \bigg[|\mathbf{u}\,|^2 
+ \langle\xi^k\xi^l\rangle 
\Big(\mathbf{u}_{,k}\boldsymbol\cdot\mathbf{u}_{,l}\Big)\bigg] 
\,.
\end{equation}
There is a Casimir for the Lie-Poisson bracket (\ref{LPB}); namely, the
conserved quantity $C_{\Phi}$ in equation (\ref{det-Casimir}). This quantity
satisfies the Casimir relation,
\begin{equation}\label{Casimir-rel}
\{C_{\Phi},H\} = 0\,,
\quad\forall\,\Phi
\quad \hbox{and}\quad
\forall\,{H}
\,.
\end{equation}
Thus, of course, $C_{\Phi}$ in equation (\ref{det-Casimir}) is also a constant
of motion for the Lagrangian mean model Hamiltonian in equation (\ref{Ham-def}).


\section{The 2D ideal LMM equations have no
velocity Casimirs}\label{2D-ideal-version-sec}

We recall the ideal Lagrangian mean equation of motion (\ref{EP-aug-motion2}) 
\begin{eqnarray} \label{EP-aug-motion2-redux}
\frac{\partial\mathbf{v}}{\partial t} 
&+& \mathbf{u}\boldsymbol{\cdot\nabla}\mathbf{v}
= -\ \boldsymbol\nabla{p}\,,
\quad \boldsymbol{\nabla\cdot\mathbf{u}}=0\,,
\\
\hbox{where}&&
\mathbf{v}  =\ \mathbf{u} 
- \Big(\partial_k\,\langle\xi^k\xi^l\rangle\partial_l\Big)\mathbf{u}
\equiv (1-\tilde\Delta)\mathbf{u}
\,,
\end{eqnarray}
The curl form (\ref{3d:curl-mot.eqn}) of the motion equation is
\begin{equation}\label{3d:curl-mot.eqn-redux}
\frac{\partial }{\partial t}\mathbf{v}
- \mathbf{u} \times \Big(\boldsymbol{\nabla} \times \mathbf{v}\Big)
+ \boldsymbol{\nabla}\,{p} 
+ u^j\boldsymbol{\nabla}\,v_j
= 0\,.
\end{equation}
The curl of this equation, in turn, gives the vorticity dynamics,
\begin{equation} \label{vortex-stretch}
\frac{\partial\mathbf{q}}{\partial t}
+ \boldsymbol{\mathbf{u}\cdot\nabla\mathbf{q}}
= \boldsymbol{\mathbf{q}\cdot\nabla\mathbf{u}}
\,+ \Big[\boldsymbol{\nabla}\,v_j\times\boldsymbol{\nabla}\,u^j\Big]\,,
\quad \hbox{where} \quad
\mathbf{q}\equiv{\rm curl}\,\mathbf{v}\,.
\end{equation}
In two dimensions these equations simplify because the vortex stretching term is
absent, but they remain extremely nonlinear (and essentially different from
the 2D Euler equations) because of the vorticity creation term. In
two dimensions, we may define the stream function
$\psi(\mathbf{x},t)$ satisfying
$\mathbf{u}=\boldsymbol{\hat{z}}\times\boldsymbol\nabla\psi$, so that,
\begin{equation} \label{strmfctn}\hspace{-.1in}
q = \boldsymbol{\hat{z}\cdot{q}}
=\boldsymbol{\hat{z}\,\cdot\,}{\rm curl}\,
\big( (1-\tilde\Delta)
\boldsymbol{\hat{z}}\times\boldsymbol\nabla\psi\big)
=
\partial_x(1-\tilde\Delta)\psi_x
+ \partial_y(1-\tilde\Delta)\psi_y
\equiv{\cal O}\psi\,.
\end{equation}
This equation defines the operator ${\cal O}$.
Note that partial spatial derivatives do not commute with the dynamical
Laplacian, $\tilde\Delta=\partial_k\,\langle\xi^k\xi^l\rangle\partial_l$,
because of the spatial dependence in $\langle\xi^k\xi^l\rangle$. 

In terms of the stream function $\psi$ and the operator ${\cal O}$, the
2D vorticity dynamics (\ref{vortex-stretch}) may be written as
\begin{eqnarray} \label{vortex-stretch-2D}
\frac{\partial}{\partial t}{\cal O}\psi
+ J\{\psi,{\cal O}\psi\}
= J\{v_j,u^j\}
&=& - J\{\tilde\Delta\psi_x,\psi_x\}
  - J\{\tilde\Delta\psi_y,\psi_y\}\,,
\nonumber\\
&=& -\,\frac{1}{2}\,\tilde\Delta_x|\boldsymbol\nabla\psi|^2_y
    \,+\,\frac{1}{2}\,\tilde\Delta_y|\boldsymbol\nabla\psi|^2_x\,,
\end{eqnarray}
where $J\{f,g\}=f_xg_y-f_yg_x$ is the 2D Jacobi operation, subscripts denote 
partial derivatives and, e.g., the operator 
$\tilde\Delta_x = \partial_k\,\langle\xi^k\xi^l\rangle_x \partial_l$
operates to its right. The highly nonlinear right hand side of this equation
prevents conservation of the domain integrated powers of the 2D Lagrangian mean
vorticity, $q={\cal O}\psi$. For example, the enstrophy $\int q^2 dx\,dy$ is
not conserved by this flow, and the Lagrangian mean vorticity $\int q\, dx\,dy$
is only conserved for certain boundary conditions (those for which
$\oint_{bdy}\mathbf{v}\boldsymbol\cdot d\mathbf{u}=0$). Thus, the 2D ideal
LMM equations apparently have no Casimirs that depend upon the velocity.
Of course, they do have the Casimir $C_{\Phi}$ in equation
(\ref{det-Casimir}) associated with the determinant of the Lagrangian mean
covariance.


\section{Relation to second moment turbulence closure models} 
\label{NS2pt closure-sec}

By adapting the work of Chen et al.~\cite{Chen
etal[1998a]}--~\cite{Chen etal[1998c]}, 
we form a second moment turbulence closure
model that combines the one point closure model they studied with the
results of the previous sections and yields the following system of
equations,
\boxeq{2}
\begin{equation}\label{VCHE-2pt}
\frac{\partial}{\partial t} \mathbf{v} 
+ (\mathbf{u}\,\boldsymbol{\cdot\nabla})\mathbf{v} 
= \nu \tilde\Delta \mathbf{v} 
- \boldsymbol{\nabla}p
+ \mathbf{F}
\;,\qquad 
\boldsymbol{\nabla\cdot\,}\mathbf{u} = 0\;,
\end{equation}
\smallskip\vspace{-.1in}

\noindent
where $\nu$ is a constant kinematic viscosity and in this case $\mathbf{v}$
is given by
\begin{equation}\label{vee-def3}
\mathbf{v} = \mathbf{u} 
 - \big(\partial_k\langle\xi^k\xi^l\rangle \partial_l\big)\mathbf{u}
\equiv (1-\tilde\Delta)\mathbf{u}\,.
\end{equation}
Note that the Lagrangian mean fluctuation covariance appears in the dissipation
operator $\tilde\Delta$. In the absence of the forcing $\mathbf{F}$, this
viscous LMM turbulence model dissipates the energy $E$ in equation
(\ref{cons-erg-def}) according to
\begin{equation}\label{LMM-erg-dissip1}
\frac{dE}{dt}
= -\,\nu \int d^{\,3}x
\Big[ {\rm tr}(\boldsymbol{\nabla}\mathbf{u}^T 
\boldsymbol{\cdot\,\langle\xi\xi\rangle\,\cdot}
\boldsymbol{\nabla}\mathbf{u}) +
\tilde\Delta\mathbf{u}\boldsymbol{\cdot}\tilde\Delta\mathbf{u}\,\Big]\,.
\end{equation}
This strictly negative energy dissipation law is the reason for adding viscosity
with $\tilde\Delta$, instead of using the ordinary Laplacian operator.

These equations are closed by the dissipative dynamical equation
(\ref{xi-xi-bold-alt-dis}) for the advected Lagrangian mean
covariance, rewritten as,
\begin{equation}\label{cov-dyn}
\frac{d}{dt} \langle\xi^k\xi^l\rangle
= u^k_{,j}\,\langle\xi^j\xi^l\rangle
+ \langle\xi^k\xi^j\rangle\,u^l_{,j}
-\
\frac{1}{\tau}\,\big(\langle\xi^k\xi^l\rangle
- \alpha^2\delta^{kl}\big)
+\
\lambda\frac{\alpha^2}{\tau}\,\tilde\Delta
\langle\xi^k\xi^l\rangle
\,.
\end{equation}
The boundary conditions for this {\bfi dissipative LMM model} for a fixed
boundary are $\mathbf{v}=0$, $\mathbf{u}=0$, for the velocities and either
$\boldsymbol{\langle\xi\xi\rangle\cdot\hat{n}} = 0$ when dissipation is absent
in the dynamics of $\boldsymbol{\langle\xi\xi\rangle}$, or equation
(\ref{cov-dyn}) without the diffusion term when such dissipation is present.

As mentioned before, isotropy and homogeneity of the Lagrangian mean covariance
tensor $\langle\xi^k\xi^l\rangle$ are {\it not} preserved under the
dynamics of equation (\ref{cov-dyn}) for nontrivial velocity shear.
Therefore, the VCHE or NS-$\alpha$ closure model of Chen et al.~\cite{Chen
etal[1998a]}--~\cite{Chen etal[1998c]}, is not an
invariant subsystem of the dissipative LMM model in equations
(\ref{VCHE-2pt})--(\ref{cov-dyn}).  However, the phenomenological addition of
relaxation and dissipation to the covariance dynamics does make the dissipative
LMM model relax to the VCHE or NS-$\alpha$ closure model, whose validity in
steady state has been verified by comparison with experimental data,~\cite{Chen
etal[1998a]}--~\cite{Chen etal[1998c]}. These
dissipative LMM equations comprise a 3D dynamically self-consistent
second-moment turbulence closure model, whose solution may be sought as a
dynamical systems problem, e.g., as an initial value problem for decaying
turbulence, or as motion of a forced dissipative system. 

The symmetric quantity
$\boldsymbol{\kappa}_S=d\boldsymbol{\langle\xi\xi\rangle}/dt$ in equation
(\ref{cov-dyn}) is called the {\bfi Taylor diffusivity}~\cite{Taylor[1921]} and
is sometimes used in turbulence modeling as a semi-empirical model for passive
scalar diffusion~\cite{Monin-Yaglom},~\cite{Bennett[1996]},~\cite{Smith[1998]}.
This application will be considered further in Section \ref{LM-geo-appl-sec},
when we include the effects of rotation and stratification, as well. For
now, the main effect of the fluctuations in equations (\ref{VCHE-2pt}) --
(\ref{vee-def3}) from the viewpoint of the Navier-Stokes equations for an
incompressible fluid is to smooth the transport velocity $\mathbf{u}$
relative to the circulation velocity $\mathbf{v}$, by inversion of the
Helmholtz operator $(1-\tilde\Delta)$. This smoothing tends to suppress
triad interactions at wave numbers greater in magnitude than about
$|\boldsymbol{\langle\xi\xi\rangle}|^{-1/2}$. Therefore, the cascade of the 
spectral kinetic energy density
$\hat{E}(\mathbf{k})=\frac{1}{2}(\boldsymbol{\widehat{\mathbf{u}
\cdot\mathbf{v}}})(\mathbf{k})$ as a function of wave number will be
suppressed for
$|\mathbf{k}|>|\boldsymbol{\langle\xi\xi\rangle}|^{-1/2}$. Hence, viscous
dissipation should take over at length scales smaller than the local length
scale given by
$|\boldsymbol{\langle\xi\xi\rangle}|^{1/2} (\mathbf{x},t)$ and suppress the
effects of nonlinearity at these smaller scales, just as if the LMM model in
(\ref{VCHE-2pt}) -- (\ref{cov-dyn}) were an {\bfi adaptive LES scheme}.
The LES aspects of the dissipative LMM model will be pursued elsewhere.


\section{Geophysical applications -- adding rotation and
stratification to the Lagrangian mean model}\label{LM-geo-appl-sec}

  
\subsection{Formulation of Lagrangian mean Euler-Boussinesq (LMEB) equations}
At leading order in $|\boldsymbol\xi|^2$, introduction of rotation and
stratification alters the averaged approximate Lagrangian $\langle
L\rangle$ in equation (\ref{mean-Lag-approx}) to the following form,
\begin{equation}\label{approx-mean-Lag+rot}\hspace{-.2in}
\langle L\rangle = \int d^{\,3}x \bigg\{\frac{D}{2} \Big[|\mathbf{u}\,|^2 
+ \langle\xi^k\xi^l\rangle 
\Big(\mathbf{u}_{,k}\boldsymbol\cdot\mathbf{u}_{,l}\Big)\Big] 
+ D \mathbf{R}(\mathbf{x})\boldsymbol\cdot\mathbf{u}
- g\,b\,D\,z
+\  p \Big[1 - D\Big]\bigg\}.
\end{equation}
The corresponding Euler-Poincar\'e equations are,
cf.~\cite{HMR[1998a]},~\cite{HMR[1998b]},
\smallskip\boxeq{6}
\begin{eqnarray}\label{LMEB-model}
&&\frac{d}{dt}\mathbf{v} 
- \mathbf{u}\times{\rm curl}\,\mathbf{R}(\mathbf{x})
+ \boldsymbol\nabla p + gb\,\mathbf{\hat{z}}
= 0
\,,\quad\hbox{with}\quad
\boldsymbol{\nabla\cdot}\mathbf{u} = 0 \,,
\\
&& 
\hbox{where  }
\frac{d}{dt} \equiv \Big(\frac{\partial}{\partial t} 
+ \mathbf{u}\boldsymbol{\,\cdot\nabla}\Big),
\quad
\mathbf{v} \equiv (1-\tilde\Delta)\mathbf{u}\,,
\quad
\tilde\Delta \equiv
\boldsymbol{\nabla\cdot\langle\xi\xi\rangle\cdot\nabla}\,,
\nonumber\\
&& 
\frac{db}{dt}
=0\,,
\quad\hbox{and}\quad
\frac{d}{dt}\boldsymbol{\langle\xi\xi\rangle}
= \boldsymbol{\langle\xi\xi\rangle\cdot\nabla\mathbf{u}}
+ \boldsymbol{\nabla\mathbf{u}}^{\rm T}
\boldsymbol{\cdot\langle\xi\xi\rangle}
\,,\label{LMEB-model-defs}
\end{eqnarray}
\smallskip

\noindent
with boundary conditions,
\begin{equation}\label{LMEB-mod-bc}
\mathbf{v}\boldsymbol{\cdot\hat{n}}=0
\quad
\mathbf{u}=0\,,
\quad\hbox{and}\quad
\boldsymbol{\langle\xi\xi\rangle\cdot\hat{n}}=0
\quad\hbox{on a fixed boundary.}
\end{equation}
These {\bfi Lagrangian mean Euler-Boussinesq (LMEB) equations} describe the
Lagrangian mean effects of fluctuations on the ideal motion of a stratified
incompressible fluid in a rotating reference frame, given in the Eulerian
description. (Note that the $\boldsymbol{\langle\xi\xi\rangle}$ equation in
(\ref{LMEB-model-defs}) does not refer to internal waves. See~\cite{GH[1996]}
for a discussion of the Lagrangian mean effects of internal waves from a
viewpoint similar to the one taken here.) The motion equation for this system
may be rewritten as
\begin{equation}\label{LMEB-mot-curl}
\frac{\partial}{\partial t}\mathbf{v} 
- \mathbf{u}\times{\rm curl}\,(\mathbf{v}+\mathbf{R})
+ u^j\boldsymbol\nabla{v_j} + \boldsymbol\nabla{p}
+ g b \boldsymbol\nabla{z} = 0\,.
\end{equation}
From this expression, we immediately find the {\bfi Kelvin-Noether
circulation theorem} showing how the covariance of the fluctuations interacts
with the buoyancy to  generate mean total circulation. Namely,
\begin{equation}\label{LMEB-KelThm}
\frac{ d}{dt}\oint_{\gamma(\mathbf{u})}(\mathbf{v}+\mathbf{R})
\boldsymbol{\cdot}d\mathbf{x} 
= \int\int_{S(\mathbf{u})}
\Big[\boldsymbol{\nabla}\,v_j\times\boldsymbol{\nabla}\,u^j
- \, g\, \boldsymbol{\nabla}b\boldsymbol{\times\nabla}z\Big]
\boldsymbol{\cdot}d\mathbf{S}
\,,
\end{equation}
where the closed curve ${\gamma(\mathbf{u})}$ moves with the 
Lagrangian mean fluid velocity $\mathbf{u}$ and is the boundary of the
surface ${S(\mathbf{u})}$. The terms
$[\boldsymbol{\nabla}\,v_j\boldsymbol{\times\nabla}\,u^j]$ and $[g\,
\boldsymbol{\nabla}b\boldsymbol{\times\nabla}z\,]$ are both {\bfi vorticity
creation terms} in this Lagrangian mean Kelvin-Noether circulation theorem.
Therefore, the fluctuations act together with buoyancy to create circulation
of the sum $\mathbf{v}+\mathbf{R}$ around closed fluid loops moving with
velocity $\mathbf{u}$. In particular, the vorticity creation term
$[\boldsymbol{\nabla}\,v_j\times\boldsymbol{\nabla}\,u^j]$ may cause mixing
across isopycnal (constant $b$) surfaces, even when the buoyancy is stably
stratified. According to the interpretation given in Section
\ref{phys-interp-sec}, this is the effect of the Stokes mean drift velocity.

The fluctuations and the buoyancy also act together in the {\bfi conserved
energy} for the LMEB system (\ref{LMEB-model}) -- (\ref{LMEB-mod-bc}). Namely,
cf. equation (\ref{LMM-cons-erg}),
\begin{equation}\label{LMEB-cons-erg}
E_{LMEB} = \int d^{\,3}x\ \bigg[
\frac{1}{2} \Big(|\mathbf{u}\,|^2 
+ {\rm tr}(\boldsymbol{\nabla\mathbf{u}}^{\rm T} 
\boldsymbol{\cdot\langle\xi\xi\rangle\cdot
 \nabla\mathbf{u}})
\Big)
+ g\,b\,z\,\bigg]\,.
\end{equation}
This expression shows the trade offs in energetics among translational
kinetic energy, Lagrangian mean velocity shear combined with the
fluctuations, and gravitational potential energy that may occur in the
dynamics of the LMEB model. These energetic tradeoffs should have
implications for the stability properties of the LMEB model's equilibrium
solutions. We note that the LMEB model has all the same equilibrium
solutions of the original Euler-Boussinesq equations (namely, those on the
invariant manifold $\boldsymbol{\langle\xi\xi\rangle}=0$) in addition to
others associated with critical points of the sum of the energy $E_{LMEB}$
and the conserved Casimirs for the EAB theory,
\begin{equation}\label{LMEB-Casimirs}
C_{LMEB} = \int d^{\,3}x\ D\,
\Phi\big(b, \tilde{\Delta}{b}, D^2
\det\boldsymbol{\langle\xi\xi\rangle} \big)
\quad\hbox{for an arbitrary function,}\ \Phi\,.
\end{equation}
These quantities Poisson commute with every Hamiltonian $H$: that is, under
a modification of the Lie-Poisson bracket in equation (\ref{LPB}) to
include the buoyancy variable $b$, as follows,
\begin{equation}\label{LMEB-LPB}
\{F,H\}_{LMEB} = \{F,H\}_{(\ref{LPB})} -
\int d^{\,3}x\
\bigg[\frac{\delta F}{\delta m_i}
\big( - b_{,i}\,\big)\frac{\delta H}{\delta b}
\ +\ \frac{\delta F}{\delta b}
\big( b_{,j}\big)
\frac{\delta H}{\delta m_j}\bigg]
\,,
\end{equation}
we have
$\{C_{LMEB},H\}_{LMEB}=0$, for all $H[\mathbf{m},D,
\boldsymbol{\langle\xi\xi\rangle},b]$.

  
\subsection{Passive scalar diffusion in the LMEB model.}
Following Bennett~\cite{Bennett[1996]}, we introduce the symmetric Taylor
diffusivity tensor~\cite{Taylor[1921]} as 
\begin{equation}\label{T-diff}
\kappa_S^{k\,l} 
\equiv
 \frac{1}{2} \frac{d}{dt}\langle\xi^k\xi^l\rangle\,,
\quad\hbox{or}\quad
\boldsymbol{\kappa}_S 
\equiv
 \frac{1}{2} \frac{d}{dt}\boldsymbol{\langle\xi\xi\rangle}
\,.
\end{equation}
According to the semi-empirical
theory~\cite{Monin-Yaglom},~\cite{Bennett[1996]},~\cite{Smith[1998]}, the
corresponding passive scalar equation with this diffusivity is 
\begin{equation}\label{pass-scal-diff}
\frac{db}{d t} = \boldsymbol{\nabla\cdot}
(\boldsymbol{\kappa}_S
\boldsymbol{\cdot\nabla}\,b)\,.
\end{equation}
As indicated here, we shall adopt this equation for the dissipative dynamics 
of the buoyancy, even though $b$ is not strictly passive. The present theory
provides a {\it dynamical equation} (\ref{LMEB-model-defs}) for the evolution
of the Taylor diffusivity tensor, $\boldsymbol\kappa_S$. Thus, passive scalars
in this theory advect and diffuse according to equation
(\ref{pass-scal-diff}) with tensor diffusivity given by equation
(\ref{xi-xi-bold}) as 
\begin{equation}\label{tensor-diff-xi-xi}
2\boldsymbol{\kappa}_S 
\equiv
\frac{d}{dt}\boldsymbol{\langle\xi\xi\rangle}
= \boldsymbol{\langle\xi\xi\rangle\cdot\nabla\mathbf{u}}
+ \boldsymbol{\nabla\mathbf{u}}^{\rm T}
\boldsymbol{\cdot\langle\xi\xi\rangle}\,.
\end{equation}
Hence, the passive scalar equation we find is,
\begin{equation}\label{pass-scal-diff1}
\frac{db}{d t} 
= 
\frac{1}{2}
\big(\boldsymbol{\nabla\cdot}
(\boldsymbol{\langle\xi\xi\rangle\cdot\nabla\mathbf{u}}
+ \boldsymbol{\nabla\mathbf{u}}^{\rm T}
\boldsymbol{\cdot\langle\xi\xi\rangle})
\boldsymbol{\cdot\nabla}\big)\,b\,,
\end{equation}
where the Lagrangian mean covariance $\boldsymbol{\langle\xi\xi\rangle}$ and the
velocity shear tensor $\boldsymbol{\nabla\mathbf{u}}$ are determined
self-consistently from the LMEB dynamics,  suitably modified to include
dissipation. We see that the tensor diffusivity
$\boldsymbol{\kappa}_S=
\frac{1}{2}(\boldsymbol{\langle\xi\xi\rangle\cdot\nabla\mathbf{u}}
+ \boldsymbol{\nabla\mathbf{u}}^{\rm T}
\boldsymbol{\cdot\langle\xi\xi\rangle})$ is
determined dynamically as the symmetrized product of the Lagrangian mean
covariance and the mean velocity shear of the flow. As we have discussed,
equation (\ref{tensor-diff-xi-xi}) implies that the covariance
$\boldsymbol{\langle\xi\xi\rangle}$ at any time will be increasing along
the instantaneous positive eigendirections of the velocity shear tensor
$\boldsymbol{\nabla\mathbf{u}}$. Conservation of energy (\ref{LMEB-cons-erg})
should control this tendency, in general, because the cost in energy for the
covariance to grow is highest precisely where the growth rate is highest.

The covariance dynamics following from the Taylor
hypothesis for the advection of turbulent structures by the (Lagrangian)
mean flow implies the determinant of $\boldsymbol{\langle\xi\xi\rangle}$ will be
conserved along flow lines in the incompressible case. Therefore, the covariance
will remain finite and nonzero. However, if any additional dissipation is
needed to moderate the growth of $\boldsymbol{\langle\xi\xi\rangle}$ due to
shear forcing, equation (\ref{cov-dyn}) may be used as an alternative
expression for the Taylor diffusivity.

Finally, following Chen et al.~\cite{Chen etal[1998a]}--~\cite{Chen
etal[1998c]}, we introduce viscosity into the motion equation for the Lagrangian
mean Euler-Boussinesq theory as
$\tilde\Delta$-diffusion of momentum
$\mathbf{v}$, cf. equation (\ref{VCHE-2pt}), with boundary conditions as
discussed after equation (\ref{cov-dyn}). This final step -- the semi-empirical
introduction of dissipation --  completes our formulation of the LMEB equation
set (\ref{LMEB-model-intro}) -- (\ref{LMEB-bc}) for the Lagrangian mean motion
of a rotating stratified incompressible fluid. In what follows, we shall consider
some lower dimensional subcases of this Lagrangian mean motion in Section
\ref{LMSW-eqns-sec}, then turn our attention to the corresponding Eulerian mean
theory in Sections \ref{EMM-sec} and  \ref{EM-geo-appl-sec}.

\paragraph{Remark on asymmetry of the diffusivity and other modeling steps.}
Because of the Earth's rotation, asymmetric diffusivity may be expected in
the ocean. This is an additional modeling step, which will be pursued 
elsewhere. For discussions of this aspect, see, e.g., the review by
Davis~\cite{Davis[1991]} and references therein. Other modeling steps would
address the {\it creation} of fluctuation covariance, as well as its
dissipation, or treat the covariance between fluctuations at two different times.
We do not pursue these other sub-gridscale modeling directions here. Instead, we
proceed from the Taylor hypothesis (\ref{xi-eqn}) that the Lagrangian
mean fluctuation covariance is advected by the Lagrangian mean fluid velocity.


\section{Lagrangian mean motion in fewer dimensions} 
\label{LMSW-eqns-sec}

  
\subsection{2D Lagrangian mean rotating shallow water (LMRSW) equations}

To the averaged approximated LMEB Lagrangian $\langle
L\rangle$ in equation (\ref{approx-mean-Lag+rot}) we apply the standard
shallow water approximations: neglect vertical gradients; neglect
kinetic energy of vertical motion; assume constant density; and
integrate in the vertical direction from the bottom topography at $z=-B(x,y)$ to
the free surface at $z=h(x,y,t)$.
These approximations lead to a new Lagrangian $\langle
L\rangle^L_{RSW}$ for Lagrangian mean rotating shallow water (LMRSW)
dynamics,
\begin{eqnarray}\label{LMRWS-Lag}
\langle L\rangle^L_{RSW} &=& \int dx\,dy\ \bigg\{\frac{D}{2}
\Big[|\mathbf{u}\,|^2  + \langle\xi^k\xi^l\rangle 
\Big(\mathbf{u}_{,k}\boldsymbol\cdot\mathbf{u}_{,l}\Big)\Big] 
\nonumber\\ 
&&\hspace{.75in}
+\ D \mathbf{R}(\mathbf{x})\boldsymbol\cdot\mathbf{u}
- \frac{1}{2}g\,D^2 + g\,D\,B(x,y)\bigg\}.
\end{eqnarray}
Here $D=h+B(x,y)$ is the total depth of the water. The corresponding
Euler-Poincar\'e motion equation for LMRSW dynamics in momentum conservation
form is, cf. equations (\ref{mom-cons}) -- (\ref{stress-tens}),
\begin{equation} \label{LMRSW-mom-cons}
\frac{\partial m_i}{\partial t} 
= -\ \frac{\partial }{\partial x^j}\Big(m_iu^j
 +  \frac{1}{2}g\,D^2\delta^j_i \Big)
 +  gD\,B_{,i}(x,y)
 + D u_j R^j_{,i}
\,,
\end{equation}
where one defines momentum density components $m_i$, for $i=1,2$ by
\begin{equation} \label{LMRSW-mom-comp}
m_i \equiv \frac{\delta \langle L_{SW}\rangle}{\delta u^i}
= D (u_i + R_i) - \big(D{\langle\xi^k\xi^l\rangle}u_{i,l}\big)_{,k}
\equiv D(1-\tilde\Delta_D)u_i + D R_i
\,.
\end{equation}
The depth $D$ satisfies the continuity equation,
\begin{equation}\label{LMRSW-cont-eqn}
\frac{\partial D}{\partial t} 
+ \boldsymbol{\nabla\cdot}(D\mathbf{u}) = 0\,,
\end{equation}
and $\boldsymbol{\langle\xi\xi\rangle}$ satisfies the Lagrangian mean covariance
dynamics inherited from the Taylor hypothesis (\ref{xi-eqn}),
\begin{equation}\label{LMRSW-xi-xi}
\frac{d}{dt}\boldsymbol{\langle\xi\xi\rangle}
= \boldsymbol{\langle\xi\xi\rangle\cdot\nabla\mathbf{u}}
+ \boldsymbol{\nabla\mathbf{u}}^{\rm T}
\boldsymbol{\cdot\,\langle\xi\xi\rangle}\,.
\end{equation}
In terms of the circulation velocity 
$\mathbf{v}=(1-\tilde\Delta_D)\mathbf{u}$ defined as before, but now
in 2D, we find the motion equation for $\mathbf{v}$,
\begin{equation}\label{LMRSW-mot-eqn}
\Big(\frac{\partial}{\partial t} 
+ \mathbf{u}\boldsymbol{\,\cdot\nabla}\Big)\mathbf{v} 
- \mathbf{u}\times{\rm curl}\,\mathbf{R}(\mathbf{x})
+ g\, \boldsymbol\nabla\Big(D-B(x,y)\Big)
= 0 \,.
\end{equation}
This is the same as the standard motion equation for rotating ideal shallow
water dynamics, modulo the substitution $\mathbf{u}\rightarrow\mathbf{v}$ and
the additional dynamics for the  covariance $\boldsymbol{\langle\xi\xi\rangle}$.
This motion equation for LMRSW implies the Kelvin-Noether circulation theorem,
cf. equation (\ref{KelThm}),
\begin{equation}\label{LMRSW-KelThm}
\frac{ d}{dt}\oint_{\gamma(\mathbf{u})}
(\mathbf{v} + \mathbf{R})
\boldsymbol{\cdot}d\mathbf{x} 
= \int\int_{S(\mathbf{u})}
d{v}_j{\wedge}d{u}^j
= \int\int_{S(\mathbf{u})}
\Big[\boldsymbol{\nabla}\,v_j\times\boldsymbol{\nabla}\,u^j\Big]
\boldsymbol{\cdot}d\mathbf{S}
\,.
\end{equation}
Thus, as for the LMM model, the fluctuations have the effect of generating
(total) circulation. The curl of the LMRSW motion equation yields, with
$q=\boldsymbol{{\hat z}\cdot}{\rm curl}\,(\mathbf{v} + \mathbf{R})$,
\begin{equation}\label{LMRSW-PV-eqn}
\Big(\frac{\partial}{\partial t} 
+ \mathbf{u}\boldsymbol{\,\cdot\nabla}\Big)
\Big(\frac{q}{D}\Big) 
= D^{-1}\boldsymbol{{\hat z}\cdot}
\big(\boldsymbol{\nabla}\,v_j 
\times \boldsymbol{\nabla}\,u^j\big)
= D^{-1} J\big(v_j,u^j\big)
\,.
\end{equation}
Thus, perhaps unexpectedly, the potential vorticity $q/D$ is
{\it not conserved on fluid parcels} by the LMRSW model, but instead has a
local creation term, $D^{-1} J\big(v_j,u^j\big)$. This creation of potential
vorticity is in fact the convective mechanism by which the smoothing of
velocity gradients occurs in the LMRSW model. The usual equations of rotating
shallow water theory are recovered on the invariant subsystem for
$\boldsymbol{\langle\xi\xi\rangle}=0$, on which $\mathbf{v}=\mathbf{u}$, and
the right hand side of the potential vorticity equation (\ref{LMRSW-PV-eqn})
vanishes.


\subsection{One dimensional Lagrangian mean shallow water (LMSW) equations}

Restricting the Lagrangian $\langle
L\rangle^L_{RSW}$ in equation (\ref{LMRWS-Lag}) to one dimensional motion
(without rotation) results in
\begin{equation}\label{1D-AWS-Lag}
\langle L\rangle^L_{SW} = \int dx\ \bigg[\frac{D}{2} 
\Big( u^2 + w u_x^2 \Big) 
- \frac{1}{2}g\,D^2 + g\,D\,B(x)\bigg]\,.
\end{equation}
Here we denote ${\langle\xi\xi\rangle}$ by $w$ in one dimension. We also denote
partial derivatives by subscripts and $D=h+B(x)$ is the total depth of the
water. The corresponding Euler-Poincar\'e equations in momentum conservation
form for LMSW in one dimension are, cf. equations (\ref{mom-cons})  --
(\ref{stress-tens}),
\begin{eqnarray} \label{1D-LMSW-mom-cons}
m_t &=& -\ \Big(mu
 +  \frac{1}{2}g\,D^2\Big)_x
 +\  gDB_x
\,,\\
\hbox{with $m$ defined by }\quad
m &\equiv& \frac{\delta\langle L\rangle^L_{SW}}{\delta u}
    = Du - (Dwu_x)_x
\,,\\
\hbox{where $D$ and $w$ satisfy}\quad
D_t &=& - (Du)_x
\,,\label{1D-sw-mass-eqn}\\
\hbox{and }\quad
w_t &=& -u\,w_x+2w\,u_x\,.\label{1D-sw-xi-xi-eqn}
\end{eqnarray}
The corresponding motion equation for the momentum velocity $v=m/D$ is given
by,
\begin{equation}\label{1D-AWS-motion-eqn}
v_t + u\,v_x + g \big(D-B(x)\big)_x = 0\,,
\quad\hbox{where}\quad
v = u - D^{-1}(Dwu_x)_x\,.
\end{equation}
Thus, the Lagrangian mean equation in the case of one dimensional shallow water again
makes just one simple change $v\rightarrow u$ in the advection term, and
introduces an additional equation for the dynamics of the Lagrangian
mean covariance ${\langle\xi\xi\rangle}=w$.

An explicit calculation using equations (\ref{1D-sw-mass-eqn}),
(\ref{1D-sw-xi-xi-eqn}) and the definition of the momentum velocity
$v$ in (\ref{1D-AWS-motion-eqn}) regains the remarkable relation
(\ref{remarkable-comm-rel}), expressed in one dimension as,
\begin{equation}\label{1D-AWS-uv-relation}
v_t + u\,v_x
= (1-D^{-1}\partial_x D\,w \partial_x)(u_t + u\,u_x)\,.
\end{equation}
Hence, the one dimensional Lagrangian mean shallow water equation system
may be rewritten as
\begin{eqnarray}\label{1D-asw-mot-eqn}
&&u_t + u\,u_x 
+ g\, (1-D^{-1}\partial_x D\,w \partial_x)^{-1}(D-B)_x = 0
\,,\\
&&D_t + u\,D_x + D u_x = 0
\,,\label{1D-asw-cont-eqn}\\
&&w_t + u\,w_x - 2w\,u_x = 0
\,.\label{1D-asw-xi-xi-eqn}
\end{eqnarray}
This system for $w\ne0$ is no longer hyperbolic; rather, it is conservative
and dispersive. However, when $w=0$, we again recover the usual shallow water
dynamics as an invariant subsystem. 

One may also obtain the equations for {\bfi Lagrangian mean polytropic gas
dynamics} with pressure-density relation $p=p_0(D/D_0)^{\gamma}$,  by
replacing the motion equation in the Lagrangian mean shallow water system
(\ref{1D-asw-mot-eqn}) -- (\ref{1D-asw-xi-xi-eqn}) with
\begin{equation}\label{1D-polytropic-gas-eqn}
u_t + u\,u_x 
+ \frac{p_0\gamma}{D_0^{\gamma}}\, (1-D^{-1}\partial_x D\,w
\partial_x)^{-1}D^{\gamma-2}D_x = 0\,.
\end{equation}
Thus, equations (\ref{1D-asw-cont-eqn}) -- (\ref{1D-polytropic-gas-eqn})
provide a system of {\bfi Lagrangian mean polytropic gas equations}, which is no
longer hyperbolic for $w\ne0$, but which recovers the usual
polytropic gas dynamics as an invariant subsystem when $w=0$. This system
conserves the energy,
\begin{equation}\label{1D-polytropic-gas-erg}
E = \int dx\ \bigg[\frac{D}{2} 
\Big( u^2 + w u_x^2 \Big) 
+ \frac{p_o}{\gamma-1}\frac{D^{\gamma}}{D_0^{\gamma}}\bigg]\,.
\end{equation}

The one dimensional Lagrangian mean models for shallow water and polytropic gas
dynamics given in this Section should provide ample opportunities for testing the
dynamical effects of fluctuations on one dimensional Lagrangian mean fluid
motion, without requiring the more intensive numerics and analysis needed in
higher dimensions. (The three dimensional version of the Lagrangian mean
polytropic gas equations is readily obtained at this point.)


\subsection{The Lagrangian mean Riemann (LMR) equation}

The process we apply here of fast-slow decomposition, followed by Taylor
approximation and averaging at fixed Lagrangian fluid label in Hamilton's
principle for ideal fluids produces the following situation. There are more
equations in the averaged approximate model than in the original model and they
include the original model as an invariant subsystem in which the fluctuation
covariance vanishes.

The simplest situation for which this occurs is probably the Riemann
equation,
\begin{equation}\label{R-eqn}
u_t + 3u\,u_x = 0\,,
\end{equation}
which describes one dimensional shock formation and is the Euler-Poincar\'e
equation for the Lagrangian,
\begin{equation}\label{R-lag}
L_R = \int dx\ \frac{1}{2} u^2\,.
\end{equation}
The corresponding averaged approximate Lagrangian for this problem is
(again writing ${\langle\xi\xi\rangle}=w$ for the covariance in one dimension)
\begin{equation}\label{LMR-lag}
\langle L\rangle_R  = \int dx\ \frac{1}{2} 
( u^2 + w\,u_x^2 )\,,
\end{equation}
whose Euler-Poincar\'e equation is 
\begin{equation}\label{LMR-eqn}
v_t + \Big(\frac{1}{2}u^2 + \frac{1}{2}w\,u_x^2 + u\,v \Big)_x  =0\,,
\end{equation}
with 
\begin{equation}\label{LMR-vee-def}
v = \frac{\delta\langle L\rangle_R}{\delta u} 
= u - \big(w\,u_x\big)_x
\,,
\end{equation}
and the one dimensional Lagrangian mean covariance $w$ satisfies
\begin{equation}\label{LMR-xi-xi}
w_t = -\,u\,w_x+2w\,u_x\,.
\end{equation}
The Lagrangian mean Riemann system (\ref{LMR-eqn}) -- (\ref{LMR-xi-xi}) can be
rewritten equivalently in (nonlocal) characteristic form as
\begin{eqnarray} \label{LMR-char-form}
u_t + u\,u_x &=& -\ \Big(1 - \partial_x\, w\, \partial_x \Big)^{-1}
\partial_x\,\Big(u^2 - \frac{1}{2} w\,u_x^2 \Big)
\,,\\
\hbox{and }\quad
w_t &=& -\,u\,w_x+2w\,u_x\,.
\end{eqnarray}
We compare the Lagrangian mean Riemann system (\ref{LMR-eqn}) 
-- (\ref{LMR-xi-xi})
with the Camassa-Holm (CH) equation, a one dimensional completely integrable
nonlinear dispersive shallow water model~\cite{CH[1993]},~\cite{CHH[1994]},
\begin{equation}\label{CH-eqn}
(u-u_{xx})_t + \Big(\frac{1}{2}u^2 
- \frac{1}{2}u_x^2 + u\,(u-u_{xx}) \Big)_x 
=0\,,
\end{equation}
which may also be written in (nonlocal) characteristic form closely similar
to equation (\ref{LMR-char-form}) as
\begin{equation} \label{CH-char-form}
u_t + u\,u_x = -\ \Big(1 - \partial_x^2\, \Big)^{-1}
\partial_x\,\Big(u^2 + \frac{1}{2} u_x^2 \Big)
\,.
\end{equation}
The Camassa-Holm equation (\ref{CH-eqn}) is the Euler-Poincar\'e equation
for the Lagrangian 
\begin{equation}\label{CH-lag}
L_{CH}  = \int dx\ \frac{1}{2} 
( u^2 + u_x^2 )\,.
\end{equation}
Further discussions of the solution behavior for the Lagrangian mean Riemann
system (\ref{LMR-eqn}) -- (\ref{LMR-xi-xi}), including its traveling waves
(which may be solved explicitly) will be given
elsewhere~\cite{Holm-L&EMR[1998]}. For now, we simply observe that there is a
key difference in sign between the LMR equation (\ref{LMR-char-form}) and the CH
equation (\ref{CH-char-form}).


\section{Eulerian mean theory of advected fluctuations}\label{EMM-sec}

Here we develop an alternative Euler-Poincar\'e theory of advected
fluctuations from the viewpoint of Eulerian averaging. We have the same point
of departure as for the Lagrangian mean theory, namely the Lagrangian
(\ref{Lag-Eul}) in the Eulerian description, 
\begin{equation}\label{Lag-Eul-A}
L(\omega) 
= 
\int d^{\,3}x \left\{\frac{D}{2}
|\mathbf{U}(\mathbf{x},t\,;\omega)|^2  
+ 
P(\mathbf{x},t\,;\omega) \Big(1-
D(\mathbf{x},t\,;\omega) \Big)\right\}\;.
\end{equation}
The traditional {\bfi Reynolds decomposition of fluid velocity} is
expressed at a given position $\mathbf{x}$ in terms of the {\bfi Eulerian
mean fluid velocity}, $\bar{\mathbf{U}}$ as
\begin{equation}
\mathbf{U}(\mathbf{x},t\,;\omega)
\equiv 
\bar{\mathbf{U}}(\mathbf{x},t) 
+ \mathbf{U}^{\,\prime}(\mathbf{x},t\,;\omega)
\,.
\label{Re-vel-decomp}
\end{equation}
According to equation (\ref{order-xi}) the Eulerian velocity fluctuation
$\mathbf{U}^{\,\prime}(\mathbf{x},t\,;\omega)$ is related to the Eulerian
displacement fluctuation --- denoted as $\boldsymbol\zeta
(\mathbf{x},t\,;\omega)$ --- by
\begin{equation}
\frac{\partial\boldsymbol\zeta}{\partial t}
+
\bar{\mathbf{U}}
\boldsymbol{\cdot\nabla\zeta}
=
\boldsymbol{\zeta\cdot\nabla} \bar{\mathbf{U}}
+
\mathbf{U}^{\,\prime}(\mathbf{x},t\,;\omega)
\,.\label{order-zeta-A}
\end{equation}
As discussed in Section \ref{phys-interp-sec}, for purely Eulerian
velocity fluctuations as in equation (\ref{Re-vel-decomp}), this relation
separates into two relations: the ``Taylor-like'' hypothesis
of~\cite{HKMRS[1998]},
\begin{equation}
\frac{\partial\boldsymbol\zeta}{\partial t}
+
\bar{\mathbf{U}}
\boldsymbol{\cdot\nabla\zeta}
=
0
\,;\label{Taylor-like-hypoth}
\end{equation}
and the relation
\begin{equation}
0
=
\boldsymbol{\zeta\cdot\nabla} \bar{\mathbf{U}}
+
\mathbf{U}^{\,\prime}(\mathbf{x},t\,;\omega)
\,.\label{u-prime-zeta}
\end{equation}
Hence, the Reynolds velocity decomposition (\ref{Re-vel-decomp})
separates the Lagrangian (\ref{Lag-Eul-A}) into its mean and fluctuating
pieces, as
\begin{equation}\label{Lag-Eul-A1}
L(\omega) 
= 
\int d^{\,3}x \left\{\frac{D}{2}
\big|\bar{\mathbf{U}}(\mathbf{x},t)
+
\mathbf{U}^{\,\prime}
\big|^2  
+ 
P(\mathbf{x},t) \Big(1-
D(\mathbf{x},t) \Big)\right\}
\,.
\end{equation}
No modification is needed in the pressure constraint in this Lagrangian, because
the Eulerian mean {\it preserves} the condition that the velocity be
divergenceless; hence,
$\boldsymbol{\nabla\cdot}
\bar{\mathbf{U}} = 0$. It remains only to take the Eulerian mean
of this Lagrangian, in which we assume
$\langle\boldsymbol\zeta\,\rangle^E = 0$. 
The {\bfi Eulerian mean averaging process} at fixed position $\mathbf{x}$ is
denoted $\langle\,\boldsymbol{\cdot}\,\rangle^E$ with, e.g.,
\begin{equation}\label{Eul-mean-def}
\bar{\mathbf{U}}(\mathbf{x},t)
=
\langle\mathbf{U}(\mathbf{x},t\,;\omega)\rangle^E
\equiv
\lim_{T\to\infty}\frac{1}{T}\int_{-T}^T
\mathbf{U}(\mathbf{x},t\,;\omega)\,d\omega\,.
\end{equation}
By equation (\ref{u-prime-zeta}), the Eulerian mean kinetic energy due to the
velocity fluctuation satisfies
\begin{equation}\label{Eul-mean-fluct-KE}
\langle\,|\mathbf{U}^{\,\prime}|^2\,\rangle^E 
= \langle\zeta^k\zeta^l\rangle^E 
\bar{\mathbf{U}}_{,k}\boldsymbol\cdot\bar{\mathbf{U}}_{,l}
\,.
\end{equation}
Thus, we find, the following Eulerian mean Lagrangian, (cf. equation
(\ref{mean-Lag-approx-inc}) for the corresponding Lagrangian mean form)
\boxeq{3}
\begin{equation}\label{Lag-Eul-B}
\langle{L}\rangle^E 
= 
\int d^{\,3}x \left\{\frac{D}{2}\Big[
\big|\bar{\mathbf{U}}(\mathbf{x},t)\big|^2
+
\langle\,\zeta^k\zeta^l\rangle^E
\bar{\mathbf{U}}_{,k}
\boldsymbol{\cdot}
\bar{\mathbf{U}}_{,l}
\Big]
+ 
P(\mathbf{x},t) \Big(1-
D(\mathbf{x},t) \Big)\right\}\;.
\end{equation}
The advection relation  (\ref{Taylor-like-hypoth}) implies a similar
advection of {\it each component} of the symmetric Eulerian mean covariance
tensor $\langle\,\zeta^k\zeta^l\rangle^E$. Thus, we have
\begin{equation}
\Big(
\frac{\partial}{\partial t}
+
\bar{\mathbf{U}}
\boldsymbol{\cdot\nabla}
\Big)
\langle\,\zeta^k\zeta^l\rangle^E
=
0
\,.\label{cov-dyn-E}
\end{equation}
This relation the continuity equation for the volume element
$D$, 
\begin{equation}
\frac{\partial{D}}{\partial t}
+
\boldsymbol{\nabla\cdot}
D\bar{\mathbf{U}}
=
0
\,,\label{cont-eqn-A}
\end{equation}
complete the auxiliary equations needed for deriving the equation of motion for
the Eulerian mean velocity $\bar{\mathbf{U}}$ from the averaged
Lagrangian $\langle{L}\rangle^E$ in (\ref{Lag-Eul-B}) by using the
Euler-Poincar\'e theory.

The results of Holm, Marsden and Ratiu~\cite{HMR[1998a]}, allow one to compute
the Euler-Poincar\'e equation for the Lagrangian $\langle
L\rangle^E(\bar{\mathbf{U}},D,\langle\zeta^k\zeta^l\rangle^E)$ in
(\ref{Lag-Eul-B}) depending on the Eulerian mean velocity $\bar{\mathbf{U}}$,
and advected quantities $D$ and $\langle\zeta^k\zeta^l\rangle^E$ as, cf.
equations (\ref{EPeqn}) and (\ref{EPeqn-Lbar}),
\boxeq{5}
\begin{eqnarray}\label{EPeqn-Lbar-E}
0 &=& \left(\frac{\partial}{\partial t} 
+ \bar{U}^j\frac{\partial}{\partial x^j}\right)
\frac{1}{D}\frac{\delta \langle L\rangle^E}{\delta \bar{U}^i} 
+ \frac{1}{D}\frac{\delta \langle L\rangle^E}{\delta \bar{U}^j}\bar{U}^j_{,i} 
\\&&
\qquad
-\
\frac{\partial}{\partial x^i}\, 
\frac{\delta \langle L\rangle^E}{\delta D} \
+\ 
\frac{1}{D}
\frac{\delta \langle L\rangle^E}{\delta \langle\zeta^k\zeta^l\rangle^E}\
\frac{\partial}{\partial x^i}\,
\langle\zeta^k\zeta^l\rangle^E
\,.
\nonumber
\end{eqnarray}
Thus, the Eulerian mean covariance $\langle\zeta^k\zeta^l\rangle^E$
satisfying the scalar advection relation (\ref{cov-dyn-E}) 
contributes considerably simpler reactive forces than those arising in equation
(\ref{EPeqn-Lbar}) from the Lagrangian mean covariance satisfying
the tensor advection equation (\ref{xi-xi-bold-1}). We compute the
following variational derivatives of the averaged approximate Lagrangian
$\langle L\rangle^E$ in equation (\ref{Lag-Eul-B})
\begin{eqnarray}\label{mean-Lag-der-E}
\frac{1}{D} \frac{\delta\langle L\rangle^E }{\delta \bar{\mathbf{U}}} 
&=& \bar{\mathbf{U}} 
 - \frac{1}{D} \Big(\partial_k\,
D\langle\zeta^k\zeta^l\rangle^E \partial_l\Big) \bar{\mathbf{U}}
\equiv \mathbf{V},
\nonumber\\
\frac{\delta\langle L\rangle^E }{\delta D} 
&=& - P  
+ \frac{1}{2}|\bar{\mathbf{U}}\,|^2 
+ \frac{1}{2}\langle\zeta^k\zeta^l\rangle^E 
\big(\bar{\mathbf{U}}_{,k}\boldsymbol\cdot\bar{\mathbf{U}}_{,l}\big)
\equiv -P_{tot}^E, 
\nonumber\\
\frac{\delta\langle L\rangle^E }{\delta P} 
&=& 1 - D \,,
\nonumber\\
\frac{\delta \langle L\rangle^E}{\delta \langle\zeta^k\zeta^l\rangle^E}
&=& \frac{D}{2}\big(\bar{\mathbf{U}}_{,k}
\boldsymbol\cdot\bar{\mathbf{U}}_{,l}\big)
\,.
\end{eqnarray}
Of course, these variational derivatives are in the {\it same form} as in
equation set (\ref{mean-Lag-der2}). However, the variables here are Eulerian
mean quantities and they will enter a {\it different} Euler-Poincar\'e equation,
namely, (\ref{EPeqn-Lbar-E}), instead of (\ref{EPeqn-Lbar}). This
difference arises because the Eulerian mean covariance
$\langle\zeta^k\zeta^l\rangle^E$ advects as an array of scalars under the
Eulerian mean evolution, rather than as the components of a symmetric tensor. 
Consequently, the Euler-Poincar\'e equation (\ref{EPeqn-Lbar-E})
for this averaged Lagrangian takes the form,
\boxeq{4}
\begin{eqnarray}\label{EPeqn-EMM}&&
\frac{\partial V_i}{\partial t} + \bar{U}^j V_{i,j} + V_j \bar{U}^j_{,i} 
=
-\
\frac{\partial {P_{tot}^E}}{\partial x^i} \
-\ \frac{1}{2}\,
\big(\bar{\mathbf{U}}_{,k}\boldsymbol\cdot\bar{\mathbf{U}}_{,l}\big)
\langle\zeta^k\zeta^l\rangle^E_{,i}
\,,
\\&&\hspace{-.85in}
\quad\hbox{where}\quad
V_i = \bar{U}_i - \tilde\Delta^E_D\bar{U}_i
\quad\hbox{with}\quad
\tilde\Delta^E_D
\equiv
\frac{1}{D}
\big(\partial_k\,D\,\langle\zeta^k\zeta^l\rangle^E \partial_l\big) 
\quad\hbox{and}\quad
\bar{U}^i_{,i}=0
\,.
\label{vee-redef-E}
\end{eqnarray}
The boundary conditions we shall choose for this motion equation are
\begin{equation} \label{bc-E}
\mathbf{V}\boldsymbol{\cdot\hat{n}}=0\,,
\quad
\bar{\mathbf{U}} = 0\,,
\quad\hbox{and}\quad
\boldsymbol{\hat{n}\,\cdot}
\boldsymbol{\langle\xi\xi\rangle}^E
=0,
\quad\hbox{on a fixed boundary.}
\end{equation} 
Then, provided the Helmholtz operator $1-\tilde\Delta^E_D$ for $D=1$ may be
inverted, the Eulerian mean pressure $P$ may be obtained by solving an elliptic
equation.

\paragraph{Contrasting the Euler-Poincar\'e equation
(\ref{EPeqn-EMM}) with the CH equation.}
When the Eulerian mean covariance is isotropic and homogeneous, so that 
$\langle\zeta^k\zeta^l\rangle^E=\alpha^2\delta^{kl}$, for a constant length
scale  $\alpha$, then this equation {\bfi reduces} to the $n$-dimensional
Camassa-Holm equation introduced in \cite{HMR[1998a]},~\cite{HMR[1998b]}. Thus,
the $n$-dimensional CH equation set {\bfi is an invariant subsystem} of the
Euler-Poincar\'e system (\ref{EPeqn-EMM}), with definition (\ref{vee-redef-E})
and advection law (\ref{cov-dyn-E}), because the initial condition
$\langle\zeta^k\zeta^l\rangle^E = \alpha^2\delta^{kl}$ is invariant under the
dynamics of equation (\ref{cov-dyn-E}).

\paragraph{Physical interpretation of $\mathbf{V}$ as the Lagrangian mean
velocity for EMM.} The Stokes mean drift velocity is defined
by~\cite{Andrews-McIntyre[1978a]},
\begin{equation} \label{Stokes-def}
\langle\mathbf{U}\rangle^S
\equiv
\langle\boldsymbol{\zeta\cdot\nabla}
\mathbf{U}^{\,\prime}\,\rangle^E
\,.
\end{equation} 
Hence, equation (\ref{u-prime-zeta}) implies
\begin{equation} \label{Stokes-for-EMM}
\langle\mathbf{U}\rangle^S
=
-\,
\langle\boldsymbol{\zeta\cdot\nabla}
\boldsymbol{\zeta\cdot\nabla}\rangle^E \bar{\mathbf{U}}
=
-\,
\tilde\Delta^E\bar{\mathbf{U}} 
+ o(|\boldsymbol\zeta|^2)
\,,
\end{equation} 
where 
\begin{equation} \label{delta-E-def}
\tilde\Delta^E 
\equiv
\big(\partial_k\,\langle\zeta^k\zeta^l\rangle^E \partial_l\big) 
= 
\tilde\Delta^E_D\big|_{D=1}
\,,
\end{equation} 
and we again argue that $\boldsymbol{\nabla\cdot\zeta} =
o(|\boldsymbol\zeta|^2)$. Thus, we find that $\mathbf{V}$ satisfies, to order
$o(|\boldsymbol\zeta|^2)$,
\begin{equation} \label{V is U-Lag for EMM}
\mathbf{V}
\equiv
\bar{\mathbf{U}} 
-
\tilde\Delta^E\bar{\mathbf{U}}
=
\bar{\mathbf{U}}
+
\langle\mathbf{U}\rangle^S
=
\langle\mathbf{U}\rangle^L
\,.
\end{equation} 
Therefore, to this order, $\mathbf{V}$ is the {\bfi Lagrangian mean velocity
for the EMM theory}. Thus, the duality between the Lagrangian mean velocity and
the Eulerian mean velocity is reciprocated. In Eulerian mean theories the dual
momentum is the Lagrangian mean velocity, and {\it vice versa}.

\paragraph{Contrasting the Euler-Poincar\'e equation
(\ref{EPeqn-EMM}) with the CL equation.}

The EMM motion equation (\ref{EPeqn-EMM}) may be rewritten equivalently as
\begin{equation}\label{EMM-mot-eqn1}
\Big(\frac{\partial}{\partial t} 
+ \mathbf{V}\boldsymbol{\,\cdot\nabla}\Big)
\mathbf{V} 
+ 
\underbrace{
(\mathbf{V}-\bar{\mathbf{U}})
\boldsymbol{\times}{\rm curl}\,\mathbf{V}}
_{\hbox{\large {\bf Stokes vortex force}}}
+\
 \boldsymbol\nabla P 
- \frac{1}{2}
\langle\zeta^k\zeta^l\rangle^E 
\boldsymbol\nabla
\big(\bar{\mathbf{U}}_{,k}\boldsymbol\cdot\bar{\mathbf{U}}_{,l}\big)
= 0
\,.
\end{equation}
Thus, the Stokes mean drift velocity
$\mathbf{V}-\bar{\mathbf{U}} = \tilde\Delta^E\bar{\mathbf{U}}$ contributes an
{\bfi additional vortex force} in this motion equation for the Lagrangian mean
fluid velocity $\mathbf{V}$. This equation is similar in form to
the CL equation (\ref{CL1}). However, the meaning of the velocities are
{\it reversed} in the two cases: the CL equation is for the Eulerian mean
velocity $\bar{\mathbf{U}}$; and the EMM equation is for the Lagrangian mean
velocity $\mathbf{V}$. The Stokes mean drift velocity is also {\it prescribed}
for the CL equation, rather than being determined dynamically.

  
\subsection{Kelvin circulation theorem for the Eulerian mean model}

Being Euler--Poincar\'e, the Eulerian mean model (EMM) equation
(\ref{EPeqn-EMM}) has a corresponding {\bfi Kelvin-Noether circulation
theorem}. Namely, this equation implies, cf. equation (\ref{KelThm}),
\begin{equation}\label{KelThm-E}
\frac{ d}{dt}\oint_{\gamma(\bar{\mathbf{U}})}\mathbf{V}
\boldsymbol{\cdot}d\mathbf{x} 
= 
-\,\frac{1}{2}\int\int_{S(\bar{\mathbf{U}})}
\boldsymbol{\nabla}
\big(\bar{\mathbf{U}}_{,k}\boldsymbol\cdot\bar{\mathbf{U}}_{,l}\big)
\boldsymbol{\times}
\boldsymbol{\nabla}
\langle\zeta^k\zeta^l\rangle^E
\boldsymbol{\cdot}
d\mathbf{S}
\,,
\end{equation}
for any closed curve ${\gamma(\bar{\mathbf{U}})}$ that moves with the 
Eulerian mean fluid velocity $\bar{\mathbf{U}}$ and surface $S(\bar{\mathbf{U}})$ with
boundary ${\gamma(\bar{\mathbf{U}})}$. Thus, in this Kelvin-Noether
circulation theorem the presence of the Eulerian mean fluctuation covariance
$\langle\zeta^k\zeta^l\rangle^E$ {\bfi creates circulation} of the Lagrangian
mean velocity
$\mathbf{V}=\bar{\mathbf{U}}  - \tilde\Delta^E\bar{\mathbf{U}}$.

  
\subsection{Vortex stretching equation for the Eulerian mean model}

In three dimensions, the EMM equation (\ref{EPeqn-EMM}) may be expressed
in its equivalent ``curl'' form, as
\begin{equation}\label{EPeqn-EMM-curl}
\frac{\partial }{\partial t}\mathbf{V}
- \bar{\mathbf{U}} \times \Big(\boldsymbol{\nabla} \times \mathbf{V}\Big)
+ \boldsymbol{\nabla}
\big(P_{tot}^E 
+ \bar{\mathbf{U}}\boldsymbol{\cdot}\mathbf{V}\big)
= 
-\ \frac{1}{2}\,
\big(\bar{\mathbf{U}}_{,k}\boldsymbol\cdot\bar{\mathbf{U}}_{,l}\big)
\boldsymbol{\nabla}
\langle\zeta^k\zeta^l\rangle^E\,,
\quad \boldsymbol{\nabla\cdot\bar{\mathbf{U}}}=0\,.
\end{equation}
The curl of this equation in turn yields an equation of transport and creation 
for the {\bfi Lagrangian mean vorticity}, $\mathbf{Q}\equiv{\rm
curl}\,\mathbf{V}$,
\begin{equation} \label{vortex-stretching-E}
\frac{\partial\mathbf{Q}}{\partial t}
+ \boldsymbol{\bar{\mathbf{U}}\cdot\nabla\mathbf{Q}}
= \boldsymbol{\mathbf{Q}\cdot\nabla\bar{\mathbf{U}}} 
-\ \frac{1}{2}\,
\boldsymbol{\nabla}
\big(\bar{\mathbf{U}}_{,k}\boldsymbol\cdot\bar{\mathbf{U}}_{,l}\big)
\boldsymbol{\times}
\boldsymbol{\nabla}
\langle\zeta^k\zeta^l\rangle^E
\,,
\quad \hbox{where} \quad
\mathbf{Q}\equiv{\rm curl}\,\mathbf{V}\,,
\end{equation}
and we have used incompressibility of $\bar{\mathbf{U}}$. Thus,
$\bar{\mathbf{U}}$ is the transport velocity for the generalized vorticity
$\mathbf{Q}$ and the expected {\bfi vortex stretching} term 
$\mathbf{Q}\cdot\nabla\bar{\mathbf{U}}$ is
accompanied by an additional {\bfi vortex creation} term. Of course,
this additional term is also responsible for the creation of circulation of
$\mathbf{V}$ in the Kelvin-Noether circulation theorem (\ref{KelThm-E}) and
it vanishes when the Eulerian mean covariance is homogeneous in space, thereby
recovering the corresponding result for the three dimensional CH
equation~\cite{HMR[1998a]},~\cite{HMR[1998b]}. 

  
\subsection{Energetics of the Eulerian mean model}

Noether's theorem guarantees {\bfi conservation of energy} for the
Euler-Poincar\'e equations (\ref{EPeqn-EMM}), since the Eulerian mean 
Lagrangian $\langle L\rangle^E$ in equation (\ref{Lag-Eul-B}) has no explicit
dependence on time. This constant energy is given by
\begin{equation}\label{cons-erg-def-E}
E =  \frac{1}{2}\int d^{\,3}x \Big(|\bar{\mathbf{U}}\,|^2 
+ \langle\zeta^k\zeta^l\rangle^E 
\bar{\mathbf{U}}_{,k}\boldsymbol\cdot\bar{\mathbf{U}}_{,l}\Big)
= \frac{1}{2}\int d^{\,3}x\ \bar{\mathbf{U}}\boldsymbol\cdot\mathbf{V}
\,.
\end{equation}
Thus, the total kinetic energy is the integrated product of the Eulerian mean
and Lagrangian mean velocities. In this kinetic energy, the Eulerian mean
covariance of the fluctuations couples to the gradients of the Eulerian mean
velocity. So there is a cost in kinetic energy for the system either to increase
these gradients, or to increase the Eulerian mean covariance. 

As one might expect from the analysis in Section \ref{Ham-LPB-sec}, Legendre
transforming the  Lagrangian $\langle L\rangle^E$ in (\ref{Lag-Eul-B})  gives
the following {\bfi Hamiltonian} (still expressed in terms of the velocity
$\bar{\mathbf{U}}$, instead of the momentum density $\mathbf{M}= \delta
\langle L\rangle^E/
\delta \bar{\mathbf{U}} = D \mathbf{V}$),
\begin{equation}\label{Ham-def-uv-E}
H^E = \int_{\cal M} d^{\,n}x\ \Big[\,
\frac{D}{2} 
\Big(|\bar{\mathbf{U}}\,|^2 
+ 
\langle\zeta^k\zeta^l\rangle^E 
\bar{\mathbf{U}}_{,k}\boldsymbol\cdot\bar{\mathbf{U}}_{,l}\Big) 
+ 
P(D-1)\Big]
\,.
\end{equation}
%
\paragraph{Remark on the geodesic property of the EMM model.}
When evaluated on the constraint manifold $D=1$, the Lagrangian in
(\ref{Lag-Eul-B}) and the Hamiltonian in (\ref{Ham-def-uv-E}) for the
Eulerian mean fluid motion equation (\ref{EPeqn-EMM}) coincide in
$n$ dimensions. This is expected for a stationary principle giving
rise to {\bfi geodesic motion}. The interpretation of the EMM model as
describing geodesic motion on the volume preserving diffeomorphism group
with respect to the $H_1$ metric given by (\ref{cons-erg-def-E}) is discussed
in~\cite{HKMRS[1998]} for the CH case, in which the Eulerian mean covariance
is isotropic and homogeneous. The corresponding discussion for the EMM
equations is equivalent to that given in~\cite{HKMRS[1998]}, since the
metrics are equivalent, provided the initial conditions for
$\langle\zeta^k\zeta^l\rangle^E$ are bounded away from zero.


\subsection{Momentum conservation -- stress tensor formulation}

Noether's theorem also guarantees conservation of momentum for the
Euler-Poincar\'e equations (\ref{EPeqn-EMM}), since the Eulerian mean 
Lagrangian $\langle L\rangle^E$ in equation (\ref{Lag-Eul-B}) has no explicit
spatial dependence. As before, the integrand ${\cal L}$ in this Lagrangian is
a {\it polynomial} in the Lagrangian mean velocity
$\bar{\mathbf{U}}$, its gradient
$\bar{\mathbf{U}}_{,k}$, and the advected quantities $D$ and
$\langle\zeta^k\zeta^l\rangle^E$. That is,
\begin{equation} \label{Lag-3d-approx-E}
\langle L\rangle^E
=\int d^{\,3}x\
{\cal L}\Big(\bar{\mathbf{U}},\bar{\mathbf{U}}_{,k\,},
\,D,\langle\zeta^k\zeta^l\rangle^E\Big)\,,
\end{equation}
with ${\cal L}$ a polynomial function of its arguments.
In this case, we may express the Eulerian mean Euler--Poincar\'e equations
(\ref{EPeqn-EMM}) in the {\bfi momentum conservation form},
\begin{equation} \label{mom-cons-E}
\frac{\partial M_i}{\partial t} 
= -\ \frac{\partial }{\partial x^j}{\cal T}^j_i\,,
\end{equation}
with {\bfi momentum density components} $M_i$, $i=1,2,3$ defined by
\begin{equation} \label{mom-comp-E}
M_i \equiv \frac{\delta \langle L\rangle^E}{\delta \bar{U}^i}
= \frac{\partial{\cal L}}{\partial \bar{U}^i}
- \frac{\partial}{\partial x^k}
\left(\frac{\partial{\cal L}}{\partial
\bar{U}^i_{,k}}\right),
\end{equation}
and {\bfi stress tensor} ${\cal T}^j_i$ given by, cf. equation
(\ref{stress-tens}),
\smallskip\boxeq{3}\bigskip
\begin{equation} \label{stress-tens-E}
{\cal T}^j_i =
M_i \bar{U}^j 
- \frac{\partial{\cal L}}{\partial \bar{U}^k_{,j}}\bar{U}^k_{,i}
+\ 
\delta^j_i\bigg({\cal L}-D\frac{\partial{\cal L}}{\partial D}\bigg)
\,.
\end{equation}
Equation (\ref{mom-cons-E}) then implies conservation of the total
momentum, $\int \mathbf{M}\, d^{\,3}x$, provided the
normal component of the stress tensor ${\cal T}^j_i$ vanishes on the boundary.

In our particular case, expression (\ref{stress-tens-E}) for the stress
tensor ${\cal T}^j_i$ becomes
\begin{equation} \label{stress-tens-eval-E}
{\cal T}^j_i = M_i\bar{U}^j 
- D \bar{\mathbf{U}}_{,k}\boldsymbol\cdot\bar{\mathbf{U}}_{,i}\,
\langle\zeta^k\zeta^j\rangle^E 
+ P\delta^j_i\,,
\quad\hbox{where}\quad
M_i
=
D(\bar{U}_i - \tilde\Delta^E_D \bar{U}_i)
\,.
\end{equation}
Consequently, the equivalent Euler-Poincar\'e motion equation
(\ref{EPeqn-EMM}), or (\ref{EPeqn-EMM-curl}) is also expressible as
\begin{equation} \label{EPeqn-EMM2}
\frac{\partial V_i}{\partial t} 
= -\ \frac{\partial }{\partial x^j}
\Big(V_i\, \bar{U}^j + P\delta^j_i
-  
\bar{\mathbf{U}}_{,k}\boldsymbol\cdot\bar{\mathbf{U}}_{,i}\,
\langle\zeta^k\zeta^j\rangle^E 
\Big)\,,
\quad\hbox{with}\quad
V_i
\equiv
M_i\big|_{D=1}
\,.
\end{equation}
The boundary conditions are given in equation (\ref{bc-E}). 


\subsection{A second moment turbulence closure model for EMM}

When dissipation and forcing are added to the EMM motion equation
(\ref{EPeqn-EMM}) by using the phenomenological viscosity
$\nu\tilde\Delta^E\mathbf{V}$ and forcing $\mathbf{F}$, one finds a {\bfi
second moment Eulerian mean turbulence model} given by
\begin{eqnarray}\label{EMM-2pt-eqns}
\Big(\frac{\partial}{\partial t} 
+ \bar{\mathbf{U}}\boldsymbol{\,\cdot\nabla}\Big)\mathbf{V}
+ V_j \boldsymbol\nabla \bar{U}^j
&+& \boldsymbol\nabla P^E_{tot} 
+\ \frac{1}{2}\,
\big(\bar{\mathbf{U}}_{,k}\boldsymbol\cdot\bar{\mathbf{U}}_{,l}\big)
\boldsymbol\nabla
\langle\zeta^k\zeta^l\rangle^E
\\
&=& \nu\,\tilde\Delta^E\mathbf{V} + \mathbf{F}\,,
\quad\hbox{where}\quad
\boldsymbol{\nabla\cdot}\bar{\mathbf{U}}=0\,,
\nonumber
\end{eqnarray}
with viscous boundary conditions $\mathbf{V}=0$, $\bar{\mathbf{U}}=0$ at a fixed
boundary. Note that the Eulerian mean fluctuation covariance
$\langle\zeta^k\zeta^j\rangle^E$ appears in the dissipation operator
$\tilde\Delta^E$. In the absence of the forcing $\mathbf{F}$ satisfying
appropriate regularity conditions, this viscous EMM turbulence model dissipates
the energy $E$ in equation (\ref{cons-erg-def-E}) according to
\begin{equation}\label{EMM-erg-dissip}
\frac{dE}{dt}
= -\,\nu \int d^{\,3}x
\Big[ {\rm tr}(\boldsymbol{\nabla}\bar{\mathbf{U}}^T 
\boldsymbol{\cdot\,\langle\zeta\zeta\rangle}^E
\boldsymbol{\,\cdot\nabla}\bar{\mathbf{U}}) +
\tilde\Delta^E\bar{\mathbf{U}}
\boldsymbol{\cdot}
\tilde\Delta^E\bar{\mathbf{U}}\,\Big]
\,.
\end{equation}
This negative definite energy dissipation law justifies adding viscosity
with $\tilde\Delta^E$, instead of using the ordinary Laplacian operator.


\section{Geophysical applications of the Eulerian mean model}
\label{EM-geo-appl-sec}

  
\subsection{Eulerian mean Euler-Boussinesq equations}
Introducting rotation and stratification alters the averaged approximate
Lagrangian $\langle L\rangle^E$ in equation (\ref{Lag-Eul-B}) to the following
expression,
\begin{eqnarray}\label{Eul-mean-Lag+rot}\hspace{-.2in}
\langle L\rangle^E_{EB} 
&=& \int d^{\,3}x \bigg\{\frac{D}{2} 
\Big[|\bar{\mathbf{U}}\,|^2 
+ \langle\zeta^k\zeta^l\rangle^E 
\Big(\bar{\mathbf{U}}_{,k}\boldsymbol\cdot\bar{\mathbf{U}}_{,l}\Big)\Big] 
\nonumber\\
&&\qquad 
+\
 D \mathbf{R}(\mathbf{x})\boldsymbol\cdot\bar{\mathbf{U}}
- g\,b\,D\,z
+\  P \Big[1 - D\Big]\bigg\}.
\end{eqnarray}
To account for buoyancy $b$ in the Euler-Poincar\'e equation
(\ref{EPeqn-Lbar-E}), the right hand side should have the additional 
summand, $D^{-1}b_{,i}\,\delta\langle L\rangle^E_{EB}/\delta{b}$. Then, the
Euler-Poincar\'e equation resulting from $\langle L\rangle^E_{EB}$ in
(\ref{Eul-mean-Lag+rot}) is
\begin{eqnarray}\label{EPeqn-EMM-rot}&&
\frac{\partial \mathbf{V}}{\partial t} 
+ \bar{\mathbf{U}}\boldsymbol{\cdot\nabla}\mathbf{V}
+ V_j \boldsymbol{\nabla}\,\bar{U}^j 
- \bar{\mathbf{U}}\times{\rm curl}\,\mathbf{R}(\mathbf{x})
\nonumber\\
&&\hspace{1in}
=
-\, gb\,\mathbf{\hat{z}}\,
-\boldsymbol{\nabla}{P_{tot}^E}\,
-\ \frac{1}{2}\,
\big(\bar{\mathbf{U}}_{,k}\boldsymbol\cdot\bar{\mathbf{U}}_{,l}\big)
\boldsymbol{\nabla}
\langle\zeta^k\zeta^l\rangle^E
\,,
\\&&
\quad\hbox{where}\quad
\mathbf{V} \equiv (1-\tilde\Delta^E)\bar{\mathbf{U}}\,,
\quad\hbox{with}\quad
\tilde\Delta^E
\equiv
\boldsymbol{\nabla\cdot\langle\,\zeta\zeta\rangle}^E
\boldsymbol{\cdot\nabla}\,,
\label{vee-redef-E1}
\end{eqnarray}
and the Eulerian mean flow is incompressible, so
$\boldsymbol{\nabla\cdot}\bar{\mathbf{U}} = 0$. The auxiliary equations that
complete this set are
\begin{equation}\label{EMEB-aux-eqn}
\Big(\frac{\partial}{\partial t} 
+ \bar{\mathbf{U}}\boldsymbol{\,\cdot\nabla}\Big)b = 0
\,,\quad
\Big(\frac{\partial}{\partial t} 
+ \bar{\mathbf{U}}\boldsymbol{\,\cdot\nabla}\Big)
\boldsymbol{\langle\,\zeta\zeta\rangle}^E = 0
\,,\quad
\frac{\partial D}{\partial t} 
+ \boldsymbol{\nabla\cdot\,}(D\mathbf{U}) = 0\,,
\end{equation}

These {\bfi Eulerian mean Euler-Boussinesq (EMEB) equations} describe the
Eulerian mean effects of fluctuations on the ideal motion of a stratified
incompressible fluid in a rotating reference frame. The motion equation for this
system may be rewritten in curl form, as
\begin{equation}\label{EMEB-mot-curl}
\frac{\partial}{\partial t}\mathbf{V} 
- \bar{\mathbf{U}}
\times{\rm curl}\,(\mathbf{V}+\mathbf{R})
+ \boldsymbol\nabla\big({P_{tot}^E} 
+ \bar{\mathbf{U}}\boldsymbol\cdot\mathbf{V}\big)
+ g b \boldsymbol\nabla{z} 
+\ \frac{1}{2}\,
\big(\bar{\mathbf{U}}_{,k}\boldsymbol\cdot\bar{\mathbf{U}}_{,l}\big)
\boldsymbol{\nabla}
\langle\zeta^k\zeta^l\rangle^E
= 0\,.
\end{equation}
The {\bfi Kelvin-Noether circulation theorem for the EMEB model} is given by,
\begin{equation}\label{EMEB-KelThm}
\frac{ d}{dt}\oint_{\gamma(\bar{\mathbf{U}})}(\mathbf{V}+\mathbf{R})
\boldsymbol{\cdot}d\mathbf{x} 
= - \int\int_{S(\bar{\mathbf{U}})}
\Big[
\frac{1}{2}\,
\boldsymbol{\nabla}
\big(\bar{\mathbf{U}}_{,k}\boldsymbol\cdot\bar{\mathbf{U}}_{,l}\big)
\boldsymbol{\times\nabla}
\langle\zeta^k\zeta^l\rangle^E
+ \, g\, \boldsymbol{\nabla}b\boldsymbol{\times\nabla}z\Big]
\boldsymbol{\cdot}d\mathbf{S}
\,.
\end{equation}
Here, the closed curve ${\gamma(\bar{\mathbf{U}})}$ moves with the 
Eulerian mean fluid velocity $\bar{\mathbf{U}}$ and is the boundary of the
surface $S(\bar{\mathbf{U}})$. There are two {\bfi vorticity
creation terms} in the Kelvin-Noether circulation theorem (\ref{EMEB-KelThm}).
Thus, spatial variation in the Eulerian mean covariance and
nonvertical buoyancy gradients can both create circulation of the sum
$\mathbf{V}+\mathbf{R}$ around closed fluid loops moving with velocity
$\bar{\mathbf{U}}$. Naturally, the same vorticity creation terms appear in the
dynamics of the Lagrangian mean vorticity $\mathbf{Q}={\rm curl}\,\mathbf{V}$,
which may be obtained by taking the curl of equation (\ref{EMEB-mot-curl}).

The {\bfi conserved energy for the EMEB system} (\ref{EPeqn-EMM-rot}) --
(\ref{vee-redef-E1}) is given by a formula similar to equation
(\ref{LMEB-cons-erg}) for the LMEB model, which is
\begin{equation}\label{EMEB-cons-erg}
E_{EMEB} = \int d^{\,3}x\ \bigg[
\frac{1}{2} \Big(|\bar{\mathbf{U}}\,|^2 
+ {\rm tr}(\boldsymbol{\nabla\bar{\mathbf{U}}}^{\rm T} 
\boldsymbol{\cdot\,\langle\,\zeta\zeta\rangle}^E
\boldsymbol{\cdot\nabla}\bar{\mathbf{U}})
\Big)
+ g\,b\,z\,\bigg]\,.
\end{equation}
Adding dissipation semiempirically as before yields equations
(\ref{EMEB-model-intro}) -- (\ref{EMEB-model-defs-intro}) in the Introduction.

  
\subsection{2D Eulerian mean rotating shallow water (EMRSW)}

We apply the standard shallow water approximations
to the averaged approximated EMEB Lagrangian $\langle
L\rangle^E$ in equation (\ref{Eul-mean-Lag+rot}) 
to find a new Lagrangian $\langle 
L\rangle^E_{RSW}$ for the dynamics of {\bfi Eulerian mean rotating
shallow water (EMRSW)},
\begin{eqnarray}\label{EMRWS-Lag}
\langle L\rangle^E_{RSW} &=& \int dx\,dy\ \bigg\{\frac{D}{2}
\Big[|\bar{\mathbf{U}}\,|^2  + \langle\zeta^k\zeta^l\rangle^E 
\Big(\bar{\mathbf{U}}_{,k}\boldsymbol\cdot\bar{\mathbf{U}}_{,l}\Big)\Big] 
\nonumber\\ 
&&\hspace{.75in}
+\ D \mathbf{R}(\mathbf{x})\boldsymbol\cdot\bar{\mathbf{U}}
- \frac{1}{2}g\,D^2 + g\,D\,B(x,y)\bigg\}.
\end{eqnarray}
Here $D=h+B(x,y)$ is the total depth of the water, which satisfies the
continuity equation,
\begin{equation}\label{EMRSW-cont-eqn}
\frac{\partial D}{\partial t} 
+ \boldsymbol{\nabla\cdot\,}(D\bar{\mathbf{U}}) = 0\,,
\end{equation}
and the Eulerian mean
covariance $\boldsymbol{\langle\xi\xi\rangle}^E$ satisfies the dynamics
inherited from the Taylor-like hypothesis (\ref{Taylor-like-hypoth}),
\begin{equation}\label{EMRSW-zeta-zeta}
\Big(\frac{\partial}{\partial t} 
+ \bar{\mathbf{U}}\boldsymbol{\,\cdot\nabla}\Big)
\boldsymbol{\langle\,\zeta\zeta\rangle}^E
= 0
\,.
\end{equation}
The Euler-Poincar\'e motion equation (\ref{EPeqn-Lbar-E}) for EMRSW dynamics
generated by the Lagrangian $\langle L\rangle^E_{RSW}$ is expressed as
\begin{eqnarray}\label{EMRSW-mot-eqn}
\Big(\frac{\partial}{\partial t} 
+ \bar{\mathbf{U}}\boldsymbol{\,\cdot\nabla}\Big)\mathbf{V} 
&-& \bar{\mathbf{U}}\times{\rm curl}\,\mathbf{R}(\mathbf{x})
+ g\, \boldsymbol\nabla\Big(D-B(x,y)\Big)
\nonumber\\
&=& 
-\ V_j \boldsymbol{\nabla}\,\bar{U}^j\ 
-\ \frac{1}{2}\,
\big(\bar{\mathbf{U}}_{,k}\boldsymbol\cdot\bar{\mathbf{U}}_{,l}\big)
\boldsymbol{\nabla}
\langle\zeta^k\zeta^l\rangle^E
 \,,
\end{eqnarray}
where the Eulerian mean circulation velocity
$\mathbf{V}=(1-\tilde\Delta^E_D)\bar{\mathbf{U}}$ is defined as in equation
(\ref{vee-redef-E}), but now in 2D. The left hand side of this equation is the
same as the standard motion equation for rotating ideal shallow water dynamics,
modulo the substitution $\bar{\mathbf{U}}\rightarrow\mathbf{V}$ and the
additional dynamics for the covariance
$\boldsymbol{\langle\,\zeta\zeta\rangle}^E$. The right hand side has a ``line
element stretching term'' and a ``covariance gradient term.''

This motion equation for EMRSW implies the {\bfi Kelvin-Noether circulation
theorem}, cf. equation (\ref{KelThm-E}),
\begin{equation}\label{EMRSW-KelThm}
\frac{ d}{dt}\oint_{\gamma(\bar{\mathbf{U}})}
(\mathbf{V} + \mathbf{R})
\boldsymbol{\cdot}d\mathbf{x} 
= -\ \frac{1}{2}\,\int\int_{S(\bar{\mathbf{U}})}
\Big[
\boldsymbol{\nabla}
\big(\bar{\mathbf{U}}_{,k}\boldsymbol\cdot\bar{\mathbf{U}}_{,l}\big)
\boldsymbol{\times\nabla}
\langle\zeta^k\zeta^l\rangle^E
\Big]
\boldsymbol{\cdot}d\mathbf{S}
\,,
\end{equation}
with fluid loop $\gamma(\bar{\mathbf{U}})$ and surface
$S(\bar{\mathbf{U}})$, as before. Thus, as for the EMM model, the
covariance gradient term has the effect of generating (total) circulation. The
curl of the EMRSW motion equation yields, with
$Q=\boldsymbol{{\hat z}\cdot}{\rm curl}\,(\mathbf{V} + \mathbf{R})$,
\begin{equation}\label{EMRSW-PV-eqn}
\Big(\frac{\partial}{\partial t} 
+ \bar{\mathbf{U}}\boldsymbol{\,\cdot\nabla}\Big)
\Big(\frac{Q}{D}\Big) 
= -\ \frac{1}{2D}\boldsymbol{\hat{\mathbf{z}}\,\cdot}
\boldsymbol{\nabla}
\big(\bar{\mathbf{U}}_{,k}\boldsymbol\cdot\bar{\mathbf{U}}_{,l}\big)
\boldsymbol{\times\nabla}
\langle\zeta^k\zeta^l\rangle^E
\,.
\end{equation}
Thus, the potential vorticity $Q/D$ is {\it not conserved on fluid parcels} by
the EMRSW model, but instead has a local creation term proportional to 
$\boldsymbol{\nabla} \langle\zeta^k\zeta^l\rangle^E$. However, 
conservation of circulation on fluid loops and conservation of potential
vorticity $Q/D$ on fluid parcels is {\it recovered} for the invariant subsystem
$\boldsymbol{\langle\,\zeta\zeta\rangle}^E=\alpha^2\delta^{kl}$ with constant
$\alpha^2$. For this invariant subsystem, we have
$\mathbf{V}=(1-\alpha^2\Delta)\bar{\mathbf{U}}$ (with the ordinary Laplacian
operator) and the covariance gradient $\boldsymbol{\nabla}
\langle\zeta^k\zeta^l\rangle^E$ vanishes. Hence, the right hand sides of
equations (\ref{EMRSW-KelThm}) and (\ref{EMRSW-PV-eqn}) also vanish in this
case; so that the Kelvin circulation integrals are constant and the  potential
vorticity is conserved on fluid parcels in this case.


\subsection{1D Eulerian mean shallow water model}

Restricting the Lagrangian $\langle
L\rangle^E_{RSW}$ in equation (\ref{EMRWS-Lag}) to one dimensional motion
(without rotation) results in
\begin{equation}\label{1D-EMSW-Lag}
\langle L\rangle^E_{SW} = \int dx\ \bigg[\frac{D}{2} 
\Big( \bar{U}^2 + W \bar{U}_x^2 \Big) 
- \frac{1}{2}g\,D^2 + g\,D\,B(x)\bigg]\,.
\end{equation}
Here $W$ denotes ${\langle\,\zeta\zeta\rangle}^E$ in 1D, subscripts denote
partial derivatives and $D=h+B(x)$ is the depth of the water. The
corresponding Euler-Poincar\'e equation is, from equation
(\ref{EPeqn-Lbar-E}),
\begin{equation}\label{1D-EMSW-EP-eqn}
V_t + \bar{U}V_x + V\bar{U}_x + \frac{1}{2} W_x\,\bar{U}_x^2  
+ \Big[g ( D-B(x) )
-\,\frac{1}{2}(\bar{U}^2+W\bar{U}_x^2) 
\Big]_x
=0\,.
\end{equation}
This may also be written in momentum conservation form for EMSW
in 1D as, cf. equations (\ref{mom-cons-E})  -- (\ref{stress-tens-E}),
\begin{eqnarray} \label{1D-EMSW-mom-cons}
&&\hspace{-2in}
M_t = -\ \Big(M\bar{U} - D W \bar{U}_x^2
 +  \frac{1}{2}g\,D^2\Big)_x
 +\  gDB_x
\,,\\
\hbox{with $M$ defined by }\quad
M &\equiv& \frac{\delta\langle L\rangle^E_{SW}}{\delta \bar{U}}
    = D\bar{U} - (DW\bar{U}_x)_x
\,,\\
\hbox{where $D$ and $W$ satisfy}\quad
D_t &=& -\, (D\bar{U})_x
\,,\label{1D-EMSW-mass-eqn}\\
\hbox{and }\quad
W_t &=& -\,\bar{U}\,W_x
\,.\label{1D-EMSW-W-eqn}
\end{eqnarray}
The equivalent equations (\ref{1D-EMSW-EP-eqn}) and (\ref{1D-EMSW-mom-cons})
simplify into the following motion equation for the Lagrangian mean velocity
$V=M/D$,
\begin{eqnarray}\label{1D-EMSW-motion-eqn}
V_t + \bar{U}\,V_x 
- D^{-1}(DW\bar{U}_x^2)_x
&=& 
- g \big(D-B(x)\big)_x 
\,,
\nonumber\\
\quad\hbox{where}\quad
V = \bar{U} &-& D^{-1}(DW\bar{U}_x)_x\,.
\end{eqnarray}
Thus, the Eulerian mean equation for one dimensional shallow water
makes the change $V\rightarrow \bar{U}$ in the advection term,
and introduces another term on the left hand side. Amusingly, this other term is
{\it half the commutator} of the advective derivative $d/dt$ and the Helmholtz
operator $1-\tilde\Delta^E_D$ acting on $\bar{U}$. That is,
\begin{equation}\label{1D-EMSW-com-rel}
2D^{-1}(DW\bar{U}_x^2)_x
= 
\Big[ \Big(\frac{\partial}{\partial t} 
+ \bar{U}\frac{\partial}{\partial x}\Big),
\Big(1 - D^{-1}\frac{\partial}{\partial x}DW
\frac{\partial}{\partial x} \Big) \Big]\bar{U} 
\,.
\end{equation}
Therefore, we have two equivalent forms of the 1D EMSW equations
\begin{eqnarray}\label{1D-EMSW-equivmot-eqn}
V_t + \bar{U}\,V_x 
- D^{-1}(DW\bar{U}_x^2)_x
&=& - g \big(D-B(x)\big)_x 
\,,
\\
(1-\tilde\Delta^E_D)
(\bar{U}_t + \bar{U}\,\bar{U}_x)
+ D^{-1}(DW\bar{U}_x^2)_x
&=& - g \big(D-B(x)\big)_x 
\,,
\\
\quad\hbox{where}\quad
V 
= 
(1-\tilde\Delta^E_D)\,\bar{U}
=
\bar{U} &-& D^{-1}(DW\bar{U}_x)_x\,.
\end{eqnarray}
Hence, the {\bfi one dimensional Eulerian mean shallow water system}
may be rewritten as, cf. equation (\ref{1D-asw-mot-eqn}),
\boxeq{4}
\begin{eqnarray}\label{1D-EMSW-sys}
&&\bar{U}_t + \bar{U}\,\bar{U}_x 
+ (1-\tilde\Delta^E_D)^{-1}
\big[g\, (D-B)_x 
+ D^{-1}(DW\bar{U}_x^2)_x
\big]
= 0
\,,\\
&&D_t + \bar{U}\,D_x + D \bar{U}_x = 0
\,,
\quad
W_t + \bar{U}\,W_x = 0
\,,\label{1D-EMSW-cont+W-eqn}
\\
&&(1-\tilde\Delta^E_D) 
\equiv 
(1-D^{-1}\partial_x D\,W\,\partial_x)
\,.\label{1D-EMSW-del-til-def}
\end{eqnarray}
This system for $W\ne0$ is no longer hyperbolic; rather, it is conservative
and dispersive. When $W$ is constant, we obtain the {\bfi shallow
water alpha-model} as an invariant subsystem. 

The equations for {\bfi Eulerian mean polytropic gas
dynamics} with pressure-density relation $p=p_0(D/D_0)^{\gamma}$,  are obtained
by replacing the motion equation in the Eulerian mean shallow water system
(\ref{1D-EMSW-sys}) -- (\ref{1D-EMSW-del-til-def}) with
\begin{equation}\label{1DEM-polytrope-eqn}
\bar{U}_t + \bar{U}\,\bar{U}_x 
+ 
\Big(1-\tilde\Delta^E_D\Big)^{-\,1}
\Big[\,\frac{p_0\gamma}{D_0^{\gamma}}\,D^{\gamma-2}D_x 
+ D^{-1}(DW\bar{U}_x^2)_x
\Big]
=
0\,.
\end{equation}
Thus, equations (\ref{1D-EMSW-cont+W-eqn}) -- (\ref{1DEM-polytrope-eqn})
provide a system of {\bfi Eulerian mean polytropic gas equations}, which is no
longer hyperbolic for $W\ne0$, and which recovers the {\bfi polytropic gas alpha
model} as an invariant subsystem when $W$ is constant. This system conserves the
energy,
\begin{equation}\label{1DEM-polytropic-gas-erg}
E = \int dx\ \bigg[\frac{D}{2} 
\Big( \bar{U}^2 + W \bar{U}_x^2 \Big) 
+ \frac{p_o}{\gamma-1}\frac{D^{\gamma}}{D_0^{\gamma}}\bigg]\,.
\end{equation}
%


\subsection{The Eulerian mean Riemann (EMR) equation}

For another illustration of the Eulerian mean methodology
in one dimension, we return to the Riemann equation,
\begin{equation}\label{R-eqn-redux}
\bar{U}_t + 3\bar{U}\,\bar{U}_x = 0\,.
\end{equation}
This is the Euler-Poincar\'e equation for the Lagrangian,
\begin{equation}\label{R-lag-redux}
L_R = \int dx\ \frac{1}{2} \bar{U}^2\,.
\end{equation}
The corresponding Eulerian mean approximate Lagrangian for this problem is
(again writing ${\langle\,\zeta\zeta\rangle}^E=W$ for the Eulerian
mean covariance in one dimension)
\begin{equation}\label{EMR-lag}
\langle L\rangle^E_R  = \int dx\ \frac{1}{2} 
( \bar{U}^2 + W\,\bar{U}_x^2 )\,,
\end{equation}
whose Euler-Poincar\'e equation is 
\begin{equation}\label{EMR-EP-eqn}
V_t + \bar{U}V_x + 2V\bar{U}_x + \frac{1}{2}W_x\,\bar{U}_x^2  =0\,,
\end{equation}
with 
\begin{equation}\label{EMR-vee-def}
V = \frac{\delta\langle L\rangle^E_R}{\delta \bar{U}} 
= \bar{U} - \big(W\,\bar{U}_x\big)_x
\,,
\end{equation}
where the one dimensional Eulerian mean covariance $W$ satisfies
the scalar advection law
\begin{equation}\label{EMR-W-eqn}
W_t = -\,\bar{U}\,W_x\,.
\end{equation}
The Eulerian mean Riemann system (\ref{EMR-EP-eqn}) -- (\ref{EMR-W-eqn}) can
be rewritten in {\bfi momentum conservation form} as
\begin{equation}\label{EMR-mom-eqn}
V_t + \Big(\frac{1}{2}\bar{U}^2 - \frac{W}{2}\,\bar{U}_x^2 
+ \bar{U}\,V \Big)_x  = 0\,,
\end{equation}
and in {\bfi nonlocal characteristic form} as
\begin{eqnarray} \label{EMR-char-form}
\bar{U}_t + \bar{U}\,\bar{U}_x &=& -\ 
\Big(1 - \partial_x\, W\, \partial_x \Big)^{-1}
\partial_x\,\Big(\bar{U}^2 + \frac{W}{2} \,\bar{U}_x^2 \Big)
\,,\\
\hbox{and }\quad
W_t &=& -\,\bar{U}\,W_x\,.
\end{eqnarray}
We compare the Eulerian mean Riemann system (\ref{EMR-EP-eqn}) --
(\ref{EMR-W-eqn}) with the completely integrable model of
Camassa-Holm~\cite{CH[1993]},~\cite{CHH[1994]} for nonlinearly dispersive
shallow water waves. The equation of motion for the CH model can be
written in momentum conservation form as, cf. equation (\ref{EMR-mom-eqn}),
\begin{equation}\label{CH-eqn-redux}
(\bar{U}-\bar{U}_{xx})_t + \Big(\frac{1}{2}\bar{U}^2 
- \frac{1}{2}\bar{U}_x^2 + \bar{U}\,(\bar{U}-\bar{U}_{xx}) \Big)_x 
=0\,.
\end{equation}
Moreover, the CH equation may also be written in {\it nearly the
same} (nonlocal) characteristic form as equation
(\ref{EMR-char-form}) for the EMR problem,
\begin{equation} \label{CH-char-form-redux}
\bar{U}_t + \bar{U}\,\bar{U}_x = -\ \Big(1 - \partial_x^2\, \Big)^{-1}
\partial_x\,\Big(\bar{U}^2 + \frac{1}{2} \bar{U}_x^2 \Big)
\,.
\end{equation}
The CH equation (\ref{CH-eqn-redux}) is the Euler-Poincar\'e equation
for the Lagrangian 
\begin{equation}\label{CH-lag-redux}
L_{CH}  = \int dx\ \frac{1}{2} 
( \bar{U}^2 + \bar{U}_x^2 )\,.
\end{equation}
Thus, the Eulerian mean Riemann system (\ref{EMR-EP-eqn}) -- (\ref{EMR-W-eqn})
reduces to the CH equation (\ref{CH-eqn-redux}) for $W=1$ (or for any other
nonzero constant, which can be absorbed into the spatial length scale). Hence,
the Eulerian mean Riemann system is the natural extension of the CH model to
allow for time dependence of the length scale associated with $W$, the
Eulerian mean covariance of the rapid fluctuations. More analysis of this
equation is given in~\cite{Holm-L&EMR[1998]}.


\section{Conclusions}\label{conclusions-sec}

We have considered two classes of models that describe the mean motion of a
fluid in the presence of rapid or random fluctuations. These are: 
\begin{itemize}
\item The {\bfi Lagrangian mean fluid models}, in which the fluctuation is
modeled as a displacement of the Lagrangian fluid parcel trajectory and averages
are taken holding its Lagrangian label fixed; and
\item The {\bfi Eulerian mean fluid models}, in which the fluctuation is modeled
as occurring at a fixed position, based on the traditional Reynolds
decomposition of the fluid velocity, and averages are taken at fixed spatial
position.
\end{itemize}
At linear order in a Taylor expansion in the magnitude of the
fluctuations we obtain the relation (\ref{order-xi}) that allows us to treat the
two classes of models on the same mathematical footing. We apply asymptotics and
averaging methods to Hamilton's principle for an ideal fluid and use the
Euler-Poincar\'e theory of Holm, Marsden and
Ratiu~\cite{HMR[1998a]},~\cite{HMR[1998b]} to determine the equations of motion
that result in each class of model. 

From the Euler-Poincar\'e viewpoint, the two classes of models differ from each
other primarily in the way they treat Taylor's hypothesis, that the fluctuations
should be ``frozen'' into the mean flow. We take Taylor's hypothesis to mean
that the fluctuations should ``transform'' in a certain way under the action of
the mean flow. In the Eulerian mean models, the displacement fluctuation
transforms under the action of the Eulerian mean flow as a collection of
scalars, by parallel transport. (See~\cite{HKMRS[1998]} for a mathematical
description of this action using {\it parallel transport} in a composition of
diffeomorphisms in a manifold setting.) In the Lagrangian mean models, the
displacement fluctuation transforms under the action of the Lagrangian mean flow
as a vector field, by {\it Lie transport} as in equation (\ref{xi-eqn}). 

The two different models yield different equations via the  Euler-Poincar\'e
theory. However, these equations are ``dual'' to each other in the physical
interpretations of their solutions. Namely, the momentum evolving in the
Lagrangian mean models is interpreted in Section \ref{phys-interp-sec} as the
Eulerian mean velocity. And, {\it vice versa}, the momentum evolving in the
Eulerian mean models is interpreted in Section \ref{EMM-sec} as the Lagrangian
mean velocity. The energy in both models is the total mean kinetic energy. The
formal source of this duality between the two classes of models turns out to be
their shared form of conserved total mean kinetic energy. This kinetic
energy in both cases may be interpreted as the domain-integrated product of the
Eulerian mean velocity times the Lagrangian mean velocity. These two velocities
are related in both classes of models by a dynamical Helmholtz operator whose
metric is the covariance of the fluctuations. Evenness of this Helmholtz operator
allows the kinetic energy to be written as the $H^1$ norm of whichever velocity
is being studied. Then, the variational derivative of the kinetic energy with
respect to one of these velocities summons the other one and thereby produces
the duality, in their Euler-Poincar\'e equations. 

The effect of the averaging in either case is to make the solution velocity
smoother than the momentum, or circulation velocity that it transports, via the
inversion of the Helmholtz operator that relates the two velocities. The
covariance that provides the metric appearing in this Helmholtz operator
introduces a length scale which evolves with the mean flow according to the
appropriate Taylor hypothesis for the fluctuations. The magnitude of this length
scale determines the smoothness of the solution velocity for both classes of
models. The mechanism for the smoothing is {\bfi nonlinear dispersion} in these
ideal fluid theories (before viscosity is added). The nonlinear dispersion
contains the length scale associated with the covariance and acts to suppress
the magnitude of the fluidic triad interaction at smaller length scales.
Normally, the triad interaction in fluid dynamics drives the cascade of
energy forward to smaller length scales. However, in both of the classes of
mean fluid models we study here this process is suppressed by nonlinear
dispersion. In summary, 
\begin{quote}
{\it The nonlinear dispersion in these mean fluid theories
acts to make the transport velocity smoother than the circulation or momentum
velocity, by the inversion of a Helmholtz operator whose length scale
corresponds to the fluctuation covariance.}
\end{quote}

These mean theories each provide either an extension, or a development of the
viscous Camassa-Holm equation (VCHE, or NS-$\alpha$) that has
recently been introduced as a one-point turbulence closure model,~\cite{Chen
etal[1998a]}--~\cite{Chen etal[1998c]}. The Eulerian mean models are indeed
natural extensions of the VCHE, or NS-$\alpha$ models to second order closures
for turbulence. The Eulerian mean models reduce to the VCHE, or NS-$\alpha$
model when the Eulerian mean fluctuation covariance is spatially homogeneous.
The Lagrangian mean models are another departure which also provides a second
moment closure for turbulence that is related to the VCHE, or NS-$\alpha$ model,
but does not contain it as an invariant subsystem unless the covariance vanishes
entirely. Which approach will eventually lead to an appropriate model for
climate and other long time geophysical applications remains to be seen.

We also formulated several examples of these Eulerian mean and Lagrangian mean
fluid models in fewer dimensions, in the hopes that these simpler examples will
provide additional physical insight into the dynamical behavior of these two
classes of models.


\section*{Acknowledgements} We are grateful to S.Y.~Chen, J.K.~Dukowicz,
C.~Foias, S.~Ghosal, R.~Jordan, R.H.~Kraichnan, J.E.~Marsden, J.R.~Ristorcelli,
S.~Shkoller and E.S.~Titi for constructive comments and enlightening 
discussions during the course of this work.



\begin{thebibliography}{99}

\singlespace

\bibitem{HMR[1998a]}  D.D. Holm, J.E. Marsden, T.S. Ratiu,
The Euler-Poincar\'e equations and semidirect
products with applications to continuum theories, 
{\it Adv. in Math.} {\bf 137} (1998) 1.

\bibitem{HMR[1998b]}  D.D. Holm, J.E. Marsden, T.S. Ratiu, 
Euler-Poincare models of ideal fluids with nonlinear dispersion, 
{\it Phys. Rev. Lett.} {\bf 80} (1998) 4173--4177. 

\bibitem{Chen etal[1998a]} S. Chen, C. Foias, D.D. Holm, E.
Olson, E.S. Titi, S. Wynne,
The Camassa-Holm equations as a closure
model for turbulent channel and pipe flow, 
{\it Phys. Rev. Lett.}, {\bf 81} (1998) 5338-5341.

\bibitem{Chen etal[1998b]} S. Chen, C. Foias, D.D. Holm, E.
Olson, E.S. Titi, S. Wynne,
A connection between the Camassa-Holm equations and 
turbulent flows in channels and pipes, 
{\it Phys. Fluids}, to appear.

\bibitem{Chen etal[1998c]} S. Chen, C. Foias, D.D. Holm, E.
Olson, E.S. Titi, S. Wynne,
The Camassa-Holm equations and turbulence, 
{\it Physica D}, to appear.

\bibitem{Holm-Shkoller-inprep} D.D. Holm and S. Shkoller, Mean fluid
motion on Riemannian manifolds, in preparation.

\bibitem{Holm[1998]} D.D. Holm, Isopycnal hydrostatic mean fluid
dynamics, in preparation.

\bibitem{Cauchy[1863?]} G.K. Batchelor, {\it An Introduction to Fluid
Dynamics} (Cambridge University Press, 1970), p. 276.

\bibitem{Taylor[1938]} G.I. Taylor, 
The spectrum of turbulence,
{\it Proc. Roy. Soc. A} {\bf 164}  (1938) 476.

\bibitem{Hinze[1975]} J.O. Hinze, {\it Turbulence}, (Mc-Graw-Hill:
New York, 2nd edition, 1975).

\bibitem{Dahm[1997]} W.J.A. Dahm and K.B. Southerland,
Experimental assessment of Taylor's hypothesis and its applicability to
dissipation estimates in turbulent flows,
{\it Phys. Fluids} {\bf9} (1997) 2101-2107.

\bibitem{Andrews-McIntyre[1978a]} D.G. Andrews and M.E. McIntyre,
An exact theory of nonlinear waves on a Lagrangian mean flow,
{\it J. Fluid Mech.} {\bf 89} (1978) 609-646.

\bibitem{Andrews-McIntyre[1978b]} D.G. Andrews and M.E. McIntyre,
On wave action and its relatives,  
{\it J. Fluid Mech.} {\bf 89} (1978) 647-664.
(Corrigendum {\bf 95} (1978) 796.)

\bibitem{GH[1996]}  I. Gjaja and D.D. Holm, 
Self-consistent Hamiltonian dynamics of wave
mean-flow interaction for a rotating stratified
incompressible fluid,
{\it Physica D} {\bf 98} (1996) 343--378.

\bibitem{Holm[1996]-CL} D.D. Holm,
The ideal Craik-Leibovich equations, 
{\it Physica D}, {\bf 98} (1996) 415-441.

\bibitem{CL[1976]} A.D.D. Craik, and S. Leibovich, 
A rational model for Langmuir circulations,
{\it J. Fluid Mech.} {\bf 73} (1976) 401-426.

\bibitem{Monin-Yaglom} A.S. Monin and A.M. Yaglom, 
{\it Statistical Fluid Dynamics},
vols. {\bf1},{\bf2}, (MIT Press, 1971).

\bibitem{Taylor[1921]} G.I. Taylor, Diffusion by continuous
movements, {\it Proc. London Math. Soc.} {\bf20} (1921) 196-211.

\bibitem{Bennett[1996]} A.F. Bennett, Particle displacements in
inhomogeneous turbulence, in {\it Stochastic modelling in Physical
Oceanography}, R.J. Adler, P. M\"uller and B. Rozovskii, eds.,
Birkha\"user, (1996) 1-45.

\bibitem{Dukowicz} J.K. Dukowicz and D. D. Holm, in preparation.

\bibitem{CH[1993]}  R. Camassa and D.D. Holm,  
An integrable shallow water equation with peaked solitons,
{\it Phys. Rev. Lett.} {\bf 71} (1993) 1661--1664.

\bibitem{CHH[1994]}
R. Camassa, D.D. Holm and J.M. Hyman,
A new integrable shallow water equation.
{\it Advances in Applied Mechanics}, Academic Press: Boston,
vol {\bf 31} (1994) pp 1--33.
   
\bibitem{HKMRS[1998]} D.D. Holm, S. Kouranbaeva, J.E. Marsden,
T. Ratiu and S. Shkoller,
A nonlinear analysis of the averaged Euler equations, 
{\it Fields Inst. Comm., Arnold Vol. 2}, Amer. Math. Soc., (Rhode
Island) (1998) to appear.

\bibitem{S1[1998]} S. Shkoller, Geometry and curvature of
  diffeomorphism groups with $H^1$ metric and mean hydrodynamics. 
  {\it J. Func. Anal.} {\bf160} (1998) 337-365. 

\bibitem{Dunn-Fosdick[1974]} J.E. Dunn and R.L. Fosdick,
Thermodynamics, stability, and boundedness
of fluids of complexity 2 and fluids of second grade, 
{\it Arch. Rat. Mech. Anal.} {\bf56} (1974) 191--252.

\bibitem{Dunn-Rajagopal[1995]} J.E. Dunn and K.R. Rajagopal,
Fluids of differential type: 
Critical reviews and thermodynamic analysis,
{\it Int. J. Engng. Sci.} {\bf33} (1995) 689--729.

\bibitem{Rivlin[1957]} R.S. Rivlin, 
The relation between the flow of non-Newtonian
fluids and turbulent Newtonian fluids,
{\it Q. Appl. Math.} {\bf 15} (1957) 212--215.

\bibitem{Chorin[1988]} A.J. Chorin, 
Spectrum, dimension, and polymer analogies in
fluid turbulence,
{\it Phys. Rev. Lett.} {\bf 60} (1988) 1947--1949.

\bibitem{Shih-Zhu-Lumley[1995]} T.H. Shih, J. Zhu and J.L. Lumley,  
A new Reynolds stress algebraic equation model,
{\it Comput. Methods Appl. Mech. Engng.} {\bf 125} (1995) 287--302.

\bibitem{Yoshizawa[1984]} A. Yoshizawa, 
Statistical analysis of the derivation of the
Reynolds stress from its eddy-viscosity representation,
{\it Phys. Fluids} {\bf 27} (1984) 1377--1387.

\bibitem{Rubinstein-Barton[1990]}  R. Rubinstein and J.M. Barton,  
Nonlinear Reynolds stress models and the renormalization group,
{\it Phys. Fluids A} {\bf2} (1990) 1472--1476.

\bibitem{Townsend[1967]}  A.A. Townsend, {\it The Structure of
Turbulent Flows}, (Cambridge University Press, 1967).

\bibitem{Lumley-Tennekes[1972]}  J.L. Lumley and H.
Tennekes, {\it A First Course in Turbulence}, (MIT Press, 1972).

\bibitem{WReynolds[1987]}  W. C. Reynolds, Fundamentals of turbulence for
turbulence modelling and simulation, in: {\it Lecture Notes for Von Karman
Institute}, AGARD Lecture Note Series, (NATO, New York, 1987) pp.1-66.

\bibitem{Piomelli[1993]}  U. Piomelli, Applications of large eddy simulations
in engineering: an overview, in: {\it Large-eddy Simulation of Complex
Engineering and Geophysical Flows}, ed. B. Galperin and S. A. Orszag 
(Cambridge University Press, 1993).

\bibitem{L&M[1996]} M. Lesieur and O. M\'etais, New trends in large-eddy
simulations of turbulence, {\it Annual Rev. Fluid Mech.} {\bf28} (1996) 45-82.

\bibitem{Chen etal to.appear} S. Chen, D.D. Holm, L.G. Margolin and R. Zhang,
3D DNS of the Camassa-Holm equation and its interpretation as an LES model,    
in preparation.

\bibitem{Leray[1934]} J. Leray, {\it Acta Math.} {\bf63} (1934) 193. 
Reviewed, e.g., in G. Gallavotti, Some rigorous results about 3D
Navier-Stokes, in Les Houches 1992 NATO-ASI meeting on {\it Turbulence
in Extended Systems}, eds. R. Benzi, C. Basdevant and S. Ciliberto (Nova
Science, New York, 1993) pp. 45-81. See also, P. Constantin, C. Foia\c s, 
B. Nicolaenko and R. Temam, {\it Integral manifolds and inertial manifolds
for dissipative partial differential equations}. Applied Mathematical
Sciences, {\bf70}, (Springer-Verlag, New York-Berlin, 1989). 

\bibitem{Foias-Holm-Titi[1998]} C. Foias, D.D. Holm and E.S. Titi,
Regularity of the viscous Camassa-Holm equations, in preparation.

\bibitem{Smith[1998]} R.D. Smith, The primitive equations in the
stochastic theory of adiabatic stratified turbulence, LA-UR-97-4319.
{\it J. Phys. Oceanog.}, to appear.

\bibitem{Davis[1991]} R.E. Davis, Lagrangian ocean studies, {\it Ann. Rev.
Fluid Mech.} {\bf23} (1991) 43-64.

\bibitem{Holm-L&EMR[1998]} D.D. Holm, The Eulerian and Lagrangian means of the
Riemann equation, in preparation.

\end{thebibliography}
\end{document}